%% file: main.tex
\documentclass[12pt]{article}
\usepackage{lineno}
\usepackage{amsmath} 
\usepackage{amsfonts}   
\usepackage{amssymb}    
\usepackage{graphicx}   
\usepackage{xcolor}
\usepackage{authblk}
\usepackage{bm}
\usepackage{secdot}		
\usepackage{mathptmx}
\usepackage{siunitx}
\usepackage[version=4]{mhchem}
\usepackage{float}
\usepackage[utf8]{inputenc}
\usepackage{textcomp}
\usepackage[hang,flushmargin,bottom]{footmisc} 
\usepackage{authblk}
\usepackage{pdfpages}
\usepackage{titlesec}
\usepackage{enumitem}
\usepackage{slashed}
\usepackage{tikz}
\usepackage[compat=1.1.0]{tikz-feynman}
\usepackage{xcolor,colortbl}
\usepackage{feynmf}
\usepackage{subcaption}
\usepackage[square,sort,comma,numbers]{natbib}

\titleformat{\section}{\normalsize\bfseries}{\thesection.}{1em}{}	
\titleformat*{\subsection}{\normalsize\bfseries}

\usepackage[normalem]{ulem} 
\usepackage{upgreek} 

\usepackage{tocloft}	

\usepackage{enumitem}         

\usepackage[hidelinks]{hyperref}
\hypersetup{
	colorlinks = true,
urlcolor ={blue},
citecolor = {.},
linkcolor = {.},
anchorcolor = {.},
filecolor = {.},
menucolor = {.},
runcolor = {.}
pdftitle={},
pdfsubject={},
pdfauthor={}, 
pdfkeywords={} 
}
\usepackage{epstopdf} 

\usepackage{fancyhdr, lastpage}	
\usepackage{caption} 
\newcommand{\nue}{\nu_e}

\titlespacing*{\paragraph}  {0pt}{0pt}{.5em}

\begin{document}
	\urlstyle{rm} 
\vfill

\title{Physics Opportunities at a Beam Dump Facility at PIP-II at Fermilab and Beyond}
\input{authors}

\maketitle
\thispagestyle{empty}
\section*{Abstract}
The Fermilab Proton-Improvement-Plan-II (PIP-II) is being implemented in order to support the precision neutrino oscillation measurements at the Deep Underground Neutrino Experiment, the U.S. flagship neutrino experiment.
The PIP-II LINAC is presently under construction and is expected to provide 800~MeV protons with 2~mA current.
This white paper summarizes the outcome of the first workshop on May 10 through 13, 2023, to exploit this capability for new physics opportunities in the kinematic regime that are unavailable to other facilities, in particular a potential beam dump facility implemented at the end of the LINAC.
Various new physics opportunities have been discussed in a wide range of kinematic regime, from eV scale to keV and MeV.
We also emphasize that the timely establishment of the beam dump facility at Fermilab is essential to exploit these new physics opportunities.

\begin{center}
	\tableofcontents
\end{center}


\pagebreak

\clearpage

\section{Introduction}
The precision measurements of the neutrino oscillation parameters require high intensity neutrino beams to ensure high statistics.
High intensity proton beams enable production of high flux neutrinos.
The PIP-II LINAC is designed to fulfil this task.
The capability of the accelerator which can provide up to 2~mA of protons of 800~MeV is sufficient to enable the precision required for neutrino oscillation parameter measurements, by using just $1\sim2\%$ of the total beam flux.

It is this proton driver capability that enables exploring a brand new area of physics which has been difficult to contemplate in the traditional fixed target environment.
In addition, several recent theoretical advances on the physics at the low energy scale reachable at fixed target experiment enable expansion of the scope and the complementary measurements to those experiments at other frontiers, including the energy frontier.
One such example is the potential of producing dark sector particles (DSP) which do not interact directly with the Standard Model (SM) particles but could results in SM particles in the final state via a kinetic coupling of a new U(1) gauge which couples to the SM photons. 

In order to exploit the capability of the physics reach made possible by the PIP-II facility, the first workshop on Physics Opportunity at PIP-II Beam Dump was held on May 10 through May 13, 2023 at Fermilab.  The outcome of this workshop is summarized in the following sections. 

This white paper is organized as follows.  Section~\ref{sec:pip2acc} describes the PIP-II facility as currently planned and some potential modifications to enhance the physics capabilities of the facility.
Section~\ref{sec:theory-directions} lays down the theoretical foundations and its directions, including the signal and related backgrounds.
Section~\ref{sec:exp-considerations} focuses on experimental considerations in exploring the physics in a wide ranging energy scale, eV to MeV.  
Section~\ref{sec:det-tec} presents various detector technologies and potential experiments which could utilize the PIP-II beam dump facility.
Finally, section~\ref{sec:conc} provides outlook for the physics at PIP-II beam dump facility, including a proposal for moving forward to establish such facility to fully exploit its capability.

\section{PIP-II and ACE Proton Sources} \label{sec:pip2acc}

\subsection{PIP-II Linac Capabilities}

The PIP-II CDR~\cite{PIP2_CDR} describes the planned performance and pulse structure of the PIP-II Linac. The PIP-II linac will be continuous wave (CW) capable, but will operate in a pulsed fashion if the Booster is the only user. All systems are designed to capable of CW operation with a small fraction (percent-level) of the overall project cost. The PIP-II bunch-by-bunch chopper is the only system not necessarily compatible with CW operation, but even this case upgrade options can still be considered.

With CW-capability, the ``macropulse'' structure (greater than microsecond timescale) is that the H$^{-}$ beam current of the PIP-II linac cannot exceed 2~mA and can be turned off or on a ms-timescale (by the LEBT chopper). The ``micropulse'' structure (less than a microsecond timescale) is that the H$^{-}$ beam is composed of a series of $\sim$1ns bunches (650 MHz RF) separated by 6.2 ns intervals (162.5~MHz), and bunches can be removed in any specified pattern (by the bunch-by-bunch MEBT chopper). The individual bunch charge cannot exceed $14\times 10^{7}$ H$^{-}$ particles. If every bunch is populated with $14\times 10^{7}$ H$^{-}$ particles and no bunches are removed, that would exceed the 2~mA overall beam current, so $14\times 10^{7}$ H$^{-}$ particles can only be achieved if 55\% or more bunches are removed on a microsecond-scale interval.

We provide several examples of PIP-II linac pulse structures that may be illustrative. The first example is PIP-II beam for Fermilab Booster operation. The micropulse structure is two $14\times 10^{7}$ H- bunches, separated by 6.2ns, then another 16.2ns until the next pair of bunchest, repeating every 22.4ns (to match to Booster RF buckets). The macropulse structure is the bunch pattern repeats for 0.6ms (averaging 2mA in that 0.6ms) and then no beam until the next pulse 50ms later (to match the Booster cycle time).

A second example is PIP-II beam for proposed mu2e-II experiment. In this case it is a sequence of ten $14\times 10^{7}$ proton pulses, then 1.693us until the next set of ten, repeating for 50ms (or until the next Booster pulse). The mu2e-II current averages only 0.13mA (leaving opportunities for other experiments to receive beam in the the 1.693us intervals). The proton beam (rather than H$^{-}$ beam) would be delivered for mu2e-II by using a single-pass beam-stripping foil (or series of foils) in the beamline between the end of the PIP-II linac and the mu2e-II target.

A third example is the capabilities for unchopped steady PIP-II pulses. Without bunch-by-bunch chopping, each bunch must be $8\times 10^{7}$ H$^{-}$ particles to fall under the 2~mA current limit. If a truly uniform beam is desired, it may be possible to increase the ns bunch-length (i.e. ``de-bunch'') in the beamline to experiment.

\begin{figure}[htp]
\begin{centering}
         \includegraphics[width=0.6\textwidth]{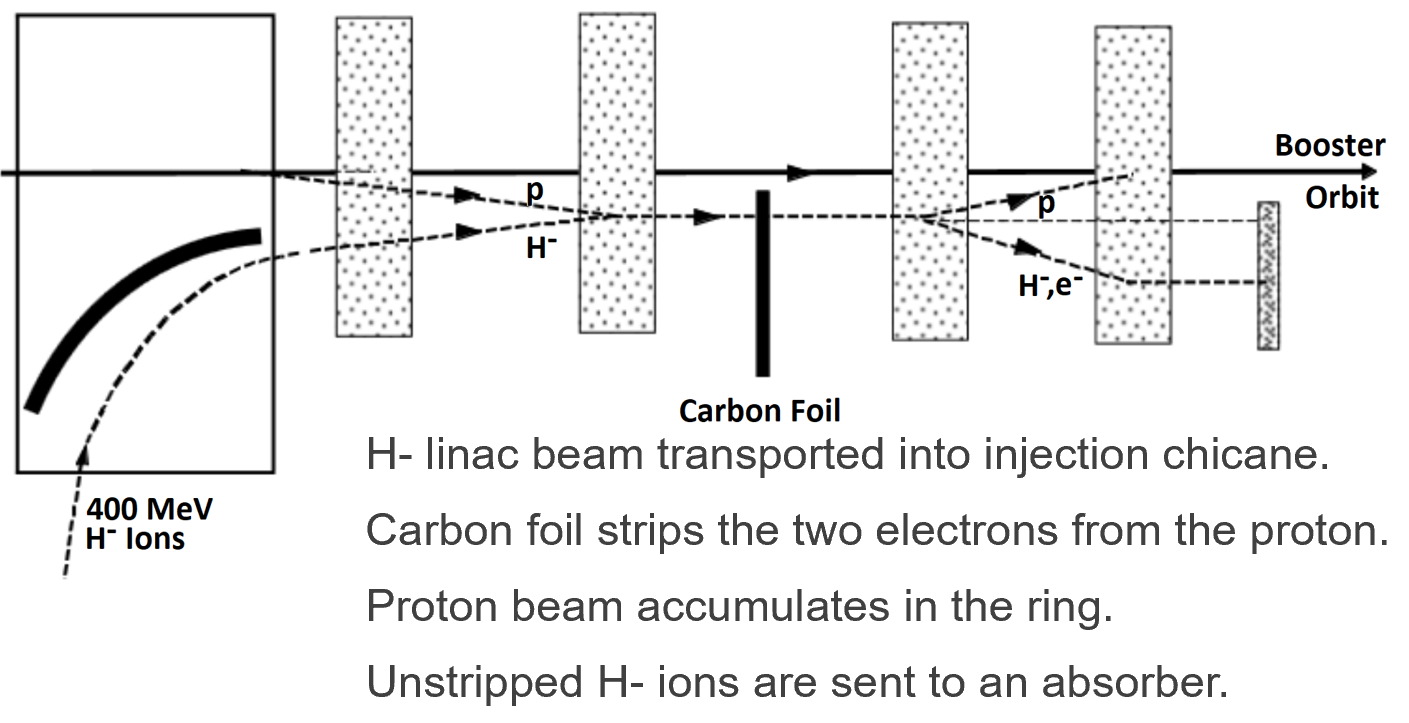}
        \caption{Foil-stripping injection of H$^{-}$ beam into a proton AR. Adapted from \cite{BoosterRookie}} 
        \label{fig:HminusInj}
\end{centering}
\end{figure}

Many particle experiments, accelerated-based dark-sector searches among them~\cite{PIP2BD}, require beams with substantially higher current and lower duty factor than is generally provided by linac beam sources. A well-established technique for adapting low-current high-duty $H^{-}$ beams to high-current low-duty proton beams is H$^{-}$ foil stripping injection (sometimes termed ``charge exchange'' stripping) into an accumulator ring (AR). Fig.~\ref{fig:HminusInj} shows a detail of the AR injection region. In H$^{-}$ foil stripping injection, the trajectories of H$^{-}$ particles from the linac coincide with the trajectories of proton particles circulating in the AR on a (usually Carbon) foil, and the phase-space density of circulating protons to be enhanced when the electrons are removed by the foil. In this way, the beam current can be enhanced tens to thousands of times over before encountering a ``space-charge limit'' in the number of particles stored in the ring or a ``foil heating limit'' in the density of particles interacting with the foil.

\subsection{ACE Scenarios and Capabilities}

\begin{figure}[htp]
\begin{centering}
         \includegraphics[width=0.6\textwidth]{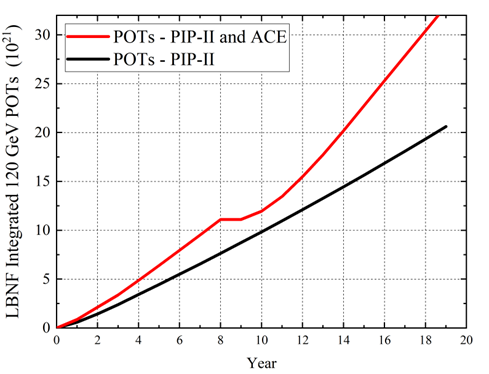}
        \caption{Timeline assumes that main Injector cycle time is decreased at Year 2 and Booster is replaced with a new 8~GeV machine at Year 8.} 
        \label{fig:ACE_POT}
\end{centering}
\end{figure}

The Fermilab Accelerator Complex Evolution (ACE) is a series of accelerator upgrade scenarios to achieve 2.4~MW beam power to the DUNE/LBNF program, improve accelerator reliability, and set the stage for next generation Intensity Frontier experimental program at Fermilab. To that end, engagement with the HEP community, including the accelerator-based dark sector search community, is critical for determining which accelerator upgrade scenario ACE should take. Fig.~\ref{fig:ACE_POT} shows the improvement in the integrated protons on target for the DUNE/LBNF program provided by the ACE upgrade.

The first phase of ACE is a plan to improve the Main Injector ramp from $\sim$1.2s to $\sim$0.7~s, which will result in a $\sim70\%$ improvement to 120~GeV DUNE/LBNF beam power (at the expense of pulsed to the 8~GeV beam program). This consists of upgrading Main Injector RF systems, magnet power supplies and service buildings as well as a series of infrastructure and controls upgrades to improve reliability of old systems. Ideally, this upgrade would occur as soon as possible to be ready for the end of the mu2e run, around 2033.

However, the largest and most critical system to replace is the Fermilab Booster that drives the faster ramping Main Injector (nearly 60 years old when PIP-II comes online). Consequently, the second aspect of ACE is to replace the Fermilab Booster with a new 8~GeV proton machine that will ensure the reliability of the DUNE/LBNF program and open up HEP opportunities at 2~GeV, 8~GeV, and 120~GeV. With DOE funding at a similar rate to the PIP-II upgrade, the Booster replacements phase of the ACE upgrade could be completed in around 2038 ($\pm$ 2 years).

\begin{figure}
\centering
\begin{subfigure}{0.45\textwidth}
         \includegraphics[width=\textwidth]{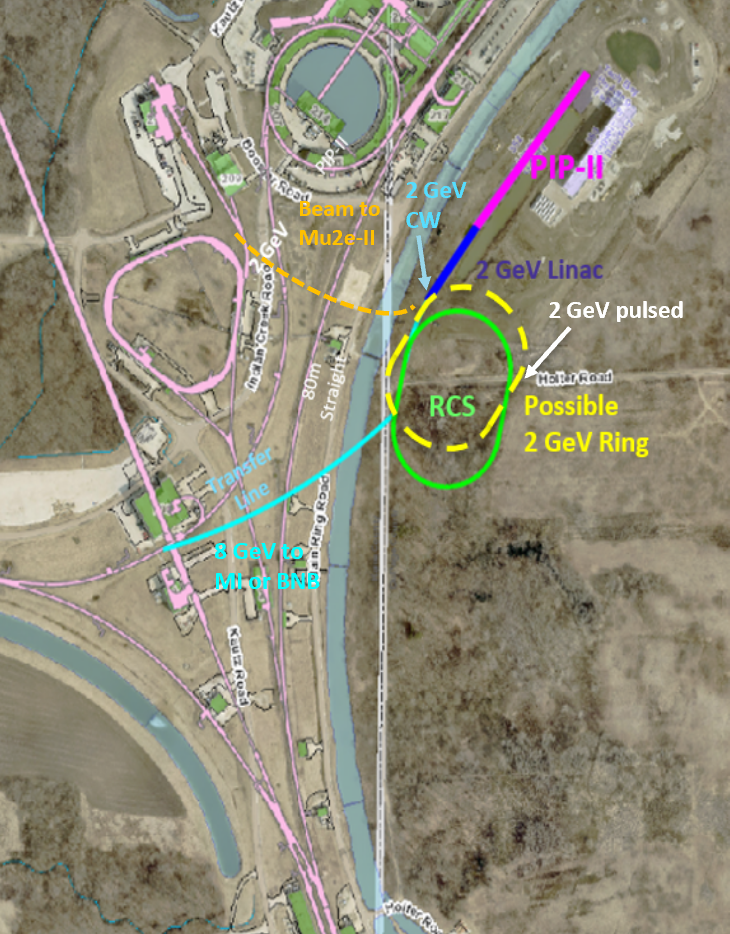}
         \caption{Potential siting of ACE RCS configuration.  The Tevatron field can accommodate other ARs or beamlines to operate alongside the 2-GeV Linac and 8-GeV RCS.}
         \label{fig:ACE_RCS}
\end{subfigure}
\hfill
\begin{subfigure}{0.45\textwidth}
         \includegraphics[width=\textwidth]{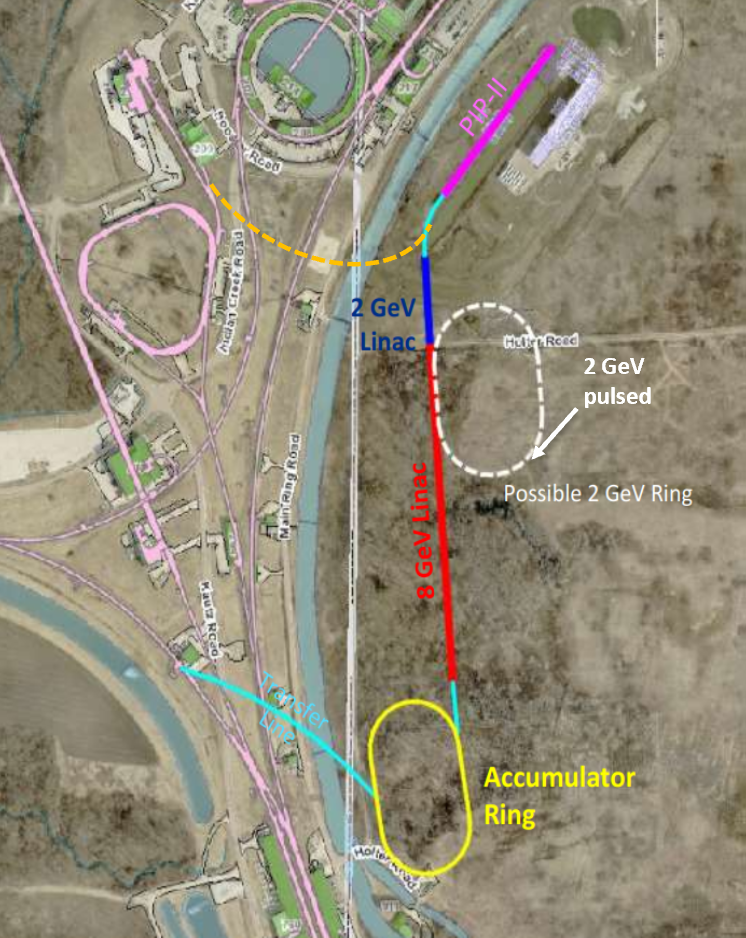}
         \caption{Potential siting of ACE Linac configuration. The Tevatron field can accommodate other ARs or beamlines to operate alongside the 8-GeV Linac and 8-GeV AR.}
         \label{fig:ACE_Linac}
\end{subfigure}
\end{figure}

Detailed scenarios were developed for six configurations of ACE booster replacements. The first three, termed ``RCS configurations'', call for a 2~GeV extension of the PIP-II linac and a new high-power 2-8~GeV rapid-cycling synchrotron (RCS). The remaining three, termed ``Linac configurations'' call for an 8~GeV extension of the PIP-II linac and a new 8~GeV AR. In either case, the PIP-II linac is brought out to (at least) 2~GeV, and an abundance of pulsed proton power is available at 8~GeV. Fig.~\ref{fig:ACE_RCS} and Fig.~\ref{fig:ACE_Linac} show the potential siting of the six configurations.

The linac beams available at 0.8~GeV, 2~GeV, and 8~GeV differ significantly across the six ACE configurations, as shown in Table~\ref{tab:ACE_Linacs} below. In the three 5~mA ACE scenarios (RCS v3, Linac v3, Linac v3), the PIP-II linac would no longer operate in CW-mode and the impact on any potential users would have to be evaluated. On the other hand, two ACE RCS configurations extend the CW capability out to 2~GeV.

\begin{table}[htp]
\centering
\begin{tabular}{| l || l | l | l |}
\hline
Linac Beam at & ~ & ~ & ~ \\
~ & 0.0-0.8~GeV & 0.8-2~GeV & 2-8~GeV \\
\hline
PIP-II & 2mA, CW & - & - \\
\hline
ACE~RCS~v1 & 2mA, CW & 2mA, CW & - \\
ACE~RCS~v2 & 2mA, CW & 2mA, CW & - \\
ACE~RCS~v3 & 5mA, 2ms, 20Hz & 5mA, 2ms, 20Hz & - \\ 
\hline
ACE~Linac~v1 & 2.7mA, CW & 2.7mA, 2ms, 20Hz & 2.7, 1.5ms, 10Hz \\
ACE~Linac~v2 & 5mA, 2ms, 20Hz & 5mA, 2ms, 20Hz & 5mA, 2ms, 10Hz \\
ACE~Linac~v3 & 5mA, 2ms, 20Hz & 5mA, 2ms, 20Hz & 5mA, 2ms, 20Hz \\
\hline
\end{tabular}
\caption{Linac capabilities of three ACE RCS configurations and three ACE Linac configurations considered as upgrades of PIP-II linac capability.}
~\label{tab:ACE_Linacs}
\end{table}

One of the three RCS configurations (ACE RCS v2) requires a 2~GeV AR to facilitate injection (although its circumference must match that of the RCS). In the other configurations, the GeV-scale AR rings are optional and would be designed to the needs of the GeV-scale experimental program. 

Table~\ref{tab:ACE_Pulse Power} gives the pulsed power potentially available at 0.8-2~GeV and 8-GeV. The 0.8-2~GeV requires a separate accumulator ring, with power and pulse structure limited by the performance that AR. At higher energies, greater AR performance is generally possible. All the ACE configurations enhanced 8~GeV beam power, although the range of beam powers span nearly an order of magnitude with 1200~kW beam power achieved with the ACE Linac~v3 configuration.

\begin{table}[htp]
\centering
\begin{tabular}{| l || r | r |}
\hline
Pulsed Power at & ~ & ~ \\
~ & 0.8-2.0~GeV$^{\ast}$ & 8~GeV \\
\hline
PIP-II & up to 2000~kW & 80~kW \\
\hline
ACE~RCS~v1 & up to 4000~kW & 160~kW \\
ACE~RCS~v2 & up to 2000~kW & 720~kW \\
ACE~RCS~v3 & 400~kW & 720~kW \\
\hline
ACE~Linac~v1 & up to 2000~kW  & 160~kW \\
ACE~Linac~v2 & 400~kW & 570~kW \\
ACE~Linac~v3 & 400~kW & 1200~kW \\
\hline
\end{tabular}
\caption{Potentially achievable pulsed beam power for three ACE RCS configurations and three ACE Linac configurations considered as upgrades of PIP-II upgrade. $^{\ast}$The 0.8-2~GeV pulsed power is only available if an AR is constructed, and with power and pulse structure limited by the performance of that AR. Only the ACE~RCS~v2 configuration requires a 2-GeV AR, the others consider an option. The 8~GeV beam power given above is what is available after providing pulses to the RR/MI for a 2.4~MW DUNE/LBNF program (about 160~kW).}
~\label{tab:ACE_Pulse Power}
\end{table}

The ACE upgrade is not intended to support a 120~GeV fast-extraction program to run concurrently with 2.4~MW DUNE/LBNF and the LBNF beamline cannot support beam powers in excess of 2.4~MW. However, ACE should be considered a prerequisite for achieving 3-5~MW at 120~GeV (i.e. multiple megawatt-scale programs) in the Main Injector in a future upgrade, beyond the scope of this workshop.

\subsection{Accumulator Rings in PIP-II and ACE Era} \label{sec:accrings}
\subsubsection{PIP-II Accumulator Ring (PAR)}
A PIP-II proton accumulator ring (PAR) is proposed to be built simultaneously with PIP-II. It could effectively serve several purposes: a) meeting the needs of delivering high intensity bunched beams for possible future rare processes experimental program by accumulating long low current PIP-II pulse of $H^-$ particles into few short high intensity  0.8-1 GeV proton bunches with $O$(100 kW) average beam power, b) improvement of the Booster operation for neutrino program with $>2$ MW out of the Main Injector due to elimination of very challenging $H^-$ charge-exchange injection system from the Booster – that can be more easily done in PAR. As the result, a single turn 2$\mu$s injection from PAR to the Booster would be greatly superior to the 0.55ms long pulse injection directly from the PIP-II linac and, therefore, will significantly reduce the beam losses in the Booster and allow higher intensity operation and potentially better reliability; c) staging for a potential future 1 GeV upgrade of the Booster injector energy.  

PAR is a low cost accumulator ring to be located adjacent to the PIP-II Booster transfer line (BTL) near the planned Booster injection point.  The $H^-$ charge-exchange injection into PAR will feature a number of modern design improvements over H$^-$ charge-exchange injection system into the Booster, including the extraction unstripped H- to an external absorber rather than increasing activation in the ring and better control of large-angle Coulomb-scattering losses off the injection foil.  This compact ring is designed to fit alongside the BTL to allow easy transfer of beam to and from the BTL as well as to a rare processes physics program to be located in the middle of the PAR. The PAR will accumulate the PIP-II beam with one of two planned RF systems.  The bunched beam will then be extracted to either the Booster in a single turn injection or to a beam dump experiment. The PAR extraction rate for the Booster beam cycles would remain at 20 Hz with the required accumulation time.  The rate for the other PAR users would be limited by the injection and pulsed extraction systems.  The baseline design is 100 Hz operation with 100 kW to 200 kW beam power. 

\begin{figure}[htp]
\begin{centering}
\includegraphics [width=\textwidth]{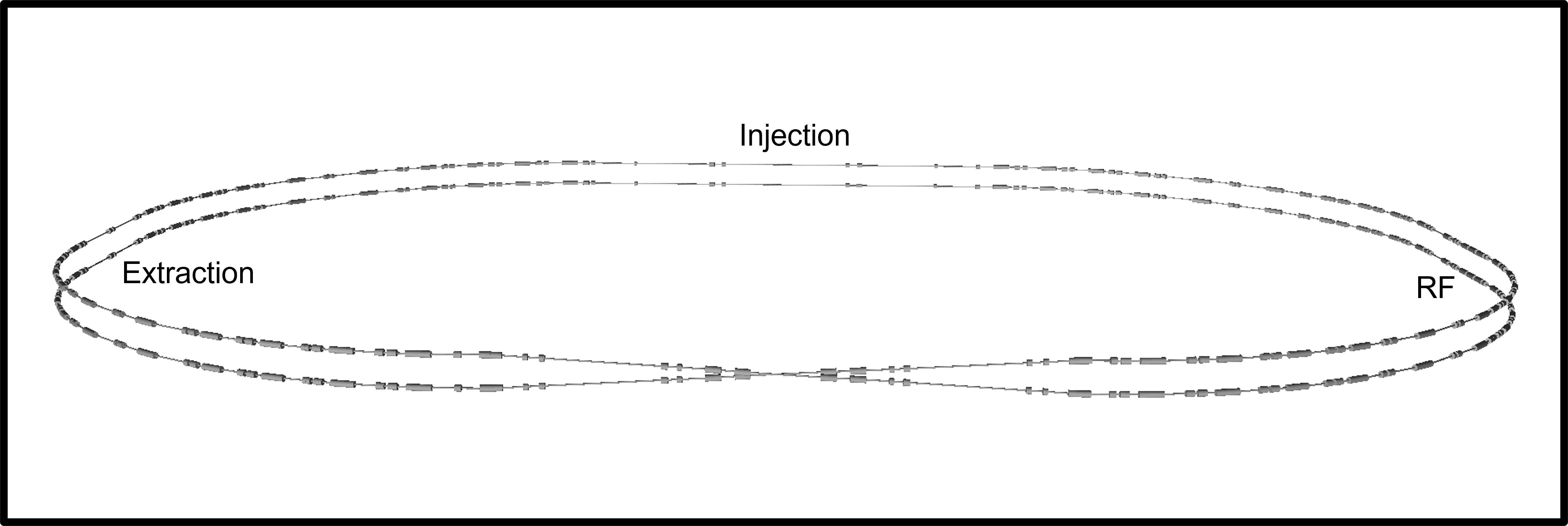}
 \caption{Schematic view of the PAR. A folded figure-8 design has the necessary injection region, two extraction regions, and two RF systems, all in an approximately 244 m circumference tunnel.} 
  \label{PAR}
\end{centering}
\end{figure}

The footprint of PAR is constrained by the Booster in the west, the Booster transfer line (BTL) in the south and PIP-II in the east. Working with the PIP-II civil engineers and Proton Source personnel, a suitable location for PAR was found that met all the constraints of the existing infrastructure and PIP-II plans. The limited spacing for PAR requires that the tunnel be smaller than the present Booster circumference. However, a desire to keep the PAR as a Booster loader required that we have a nearly identical harmonic number.  To accommodate the Booster injection needs, a folded accelerator and lattice design was developed. More details on the PAR lattice will be presented in the following subsection.
The PAR will cross the present Main Ring tunnel twice, which is unavoidable.  The crossing will be similar to the BTL but the rings will not be at the same elevation.

The placement of PAR will enable the commissioning of PIP-II to the BTL and Booster independently from the commissioning of PAR. The independence is achieved by having a fast switching magnet near the last BTL dipoles. The pulser magnet will either kick beam into the transfer line to PAR or remain on the BTL to Booster trajectory.  This method of beam transfer is only required for the commissioning period of the PAR.  Once commissioned, the pulsed magnet system is replaced with DC dipoles magnets and injection to the Booster will pass through PAR. One dipole bends beam to the PAR and another would bend the extracted PAR beam onto the BTL to be injected into the Booster before the BTL passes through beam pipe. 

The present civil construction plan for PAR has a service building located near the injection and extraction area to house the power supplies required for the pulsed systems.  The exact placement has yet to be determined but will be chosen to satisfy the needs of the planned dark sector (DS) physics program and other potential PAR users.  
\begin{figure}[htp]
\begin{centering}
\includegraphics [width=0.6\textwidth]{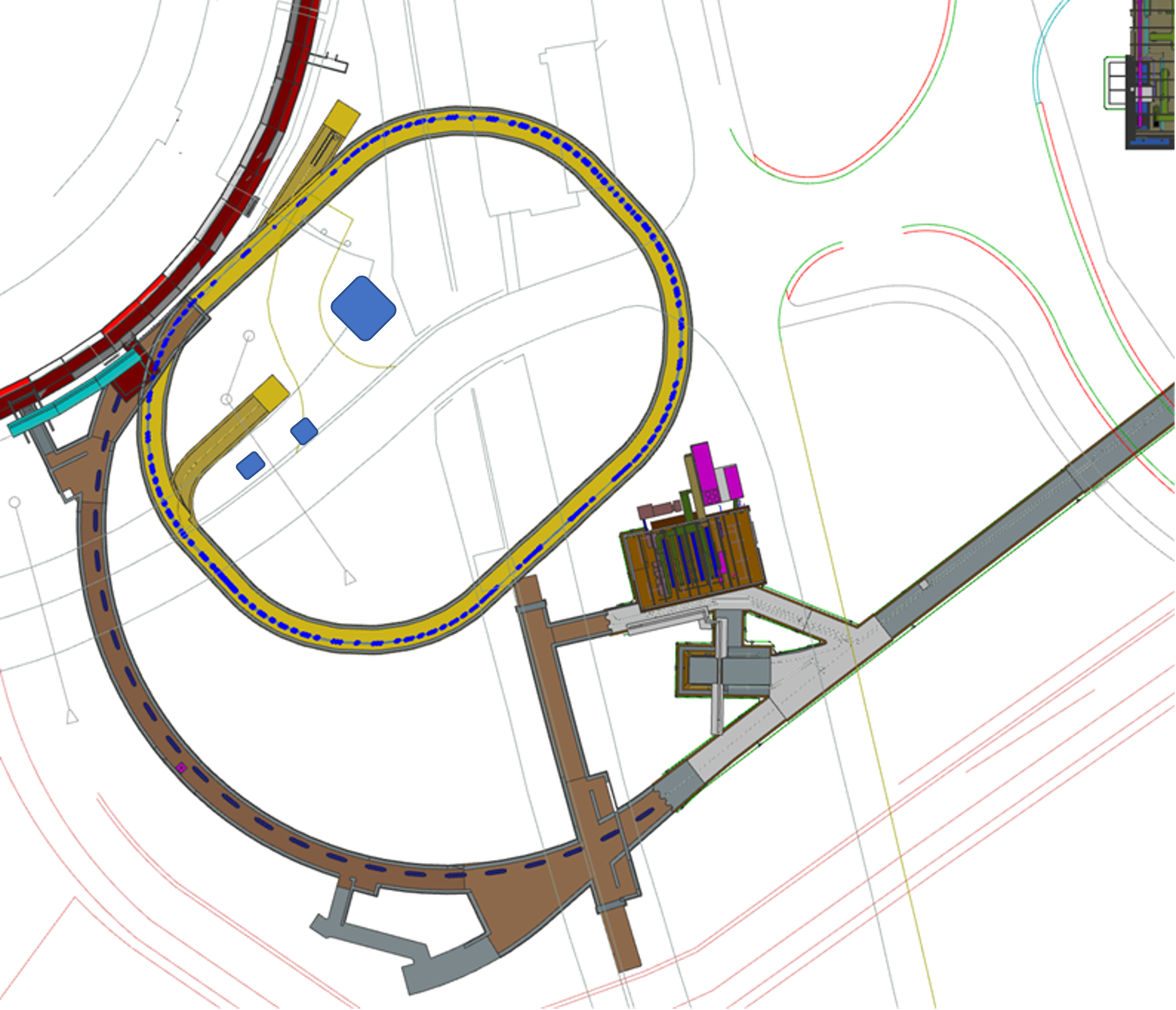}
 \caption{The location PAR is adjacent to Booster and Booster Transfer Line.} 
  \label{PAR_location}
\end{centering}
\end{figure}

\begin{figure}[htp]
\begin{centering}
\includegraphics[width=0.6\textwidth]{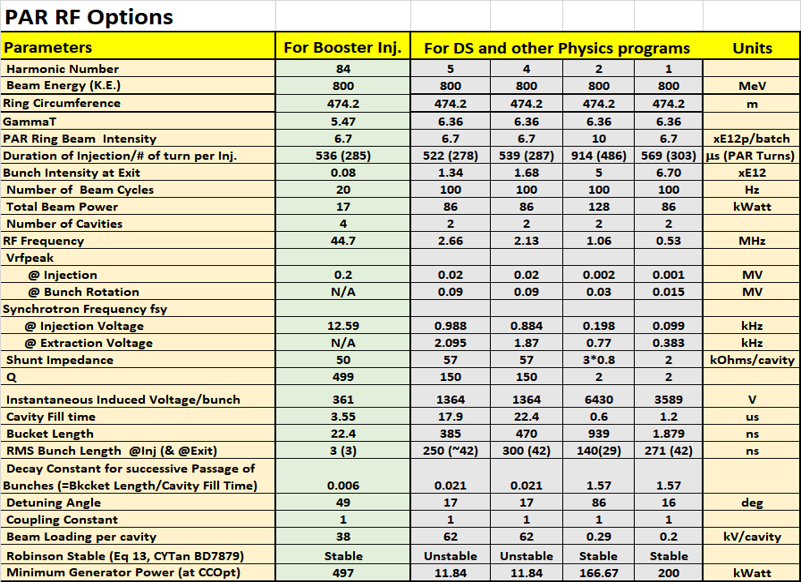}
 \caption{PAR option table with Booster and DS options.} 
  \label{PAR_RF_Table}
\end{centering}
\end{figure}

PAR has been designed around a list of key parameters which are presented in the PAR parameter table. As work continues on the PAR design, these parameters will be adjusted. However, it should be noted that they have not changed significantly over the past year.
The PAR beam pipe aperture is 28\% larger than the Booster dipole aperture.  This will allow for low loss accumulation and higher beam flux for the dump experiments.  This aperture is also the same as the new wide bore Booster RF cavities.  A larger aperture will make the RF cavity the aperture restriction which will require a new RF cavity design and testing program.  This aperture is well suited for the delivery of several hundred kW of beam power to a DS program.  Additionally, this aperture allows PAR to use existing FNAL magnets which are in storage - significantly reducing cost.
The PAR has looked at several designs to best match or optimize the dark sector physics reach.  The present idea is to use a ring harmonic number of 2.  This option looks to deliver the highest intensity and shortest bunch lengths.  The key aspects of the H=2 PAR can been seen in the PAR overview options figure.  The H=2 column is the base PAR dark sector design. The impact of the harmonic number for PAR can be seen in Fig.~\ref{PAR_RF_Table}, where a harmonic number of two is considered the baseline for DS operation. 

\begin{table}[htp]
\centering
\begin{tabular}{|| l || l | l ||}
\hline
Parameter & PAR Base & PAR Upgrade \\
\hline
Cycle Rate & 100 Hz & 100 Hz \\
Beam Intensity & $10\times 10^{12}$ in 2 bunches & $15\times 10^{12}$ in 2 bunches\\
Beam Power & 120 kW & 240 kW\\
Proton structure & 140 ns FBW per bunch & 140 ns FBW  per bunch \\
\hline
\end{tabular}
\caption{Key PAR specifications of both baseline and upgrade scenarios.}
\label{tab:PIU_MuC_Overview}
\end{table}

The baseline power that might be delivered to the DS dump experiment is estimated to be 125 kW.  This will require 100 Hz operation with 1E13 protons per cycle.  The intensity limit is based upon a conservative value for space charge at 800 MeV.  A future upgrade to the PIP-II linac to 1 GeV will allow for the PAR base intensity to be increased to 1.5E13 protons per pulse.  The higher energy and intensity will increase the delivered power to the DS program to 240 kW.  A DS physics program would be competitive at these power numbers and provide a new beam based DS program here at FNAL.  

\subsubsection{Compact PIP-II Accumulator Rings (CPAR)}

Three scenarios were developed for the PIP2-BD Snowmass paper~\cite{PIP2BD}, given by the table below~\ref{table:PIP2BD_rings}. PAR is the proposed AR described above, which every 20 Hz serves to facilitate injection into the Booster and at 100 Hz provides for a 100~kW beam dump physics program. For PAR, a detailed design has been completed and preliminary tracking performance verified. For RCS-SR and CPAR, the parameters can be re-optimized as the designs progress. The version of RCS-SR described here is for a 2-GeV AR that would have a circumference designed to facilitate injection into an ACE-RCS configuration and an transverse aperture appropriate for a MW-class beam program. The concept of Compact PIP-II Accumulator Ring (CPAR) is a more compact ring that are better optimized for the low-duty factor experiments. The version of CPAR described here is for a 100~m AR with the capability of rapid bunch-by-bunch extraction, a small aperture to save on initial AR costs, and a subsequent 1.2~GeV energy upgrade (perhaps as part of ACE).

\begin{table} [h]
\begin{center}
\begin{tabular}{|p{1.5cm}|p{1.5cm}|p{1.5cm}|p{1.5cm}|p{1.5cm}|}
\hline
Facility & Beam energy (GeV) & Repetition rate (Hz) & Pulse length (s) & Beam power (MW)\\
\hline
PAR & 0.8 & 100 & $2 \times 10^{-6}$ & 0.1 \\
CPAR & 1.2 & 100 & $2\times 10^{-8}$ & 0.09\\
RCS-SR & 2 & 120 & $2 \times 10^{-6}$ & 1.3 \\
\hline
\end{tabular} \caption{The parameters of three possible accumulator ring scenarios considered as low duty-factor upgrades to the PIP-II linac. Adapted from \cite{PIP2BD}.}\label{table:PIP2BD_rings}
\end{center}
\end{table}

Of the three-scenarios considered, CPAR delivers the best outcome for the PIP2-BD program because of its compact geometry and rapid short-pulse extraction rate~\cite{AMF,AMF_Workshop_23}. Further optimization is possible. Without significantly increasing project cost, AR performance can likely benefit from design work towards a compact accelerator lattice, rapid extraction kickers, and/or open-plane magnets. At increased project cost, the AR pulse intensity can also be straight-forwardly increased with a larger transverse aperture or a higher injection energy. An AR can be designed to accommodate a greater injection energy than it initially operates, allowing for the possibility of a staged power increase -for instance upgrading the energy from 0.8~GeV to 1.2~GeV allows for a factor of 3 increased in pulse power.

\begin{figure}[htp]
\begin{centering}
         \includegraphics[width=0.6\textwidth]{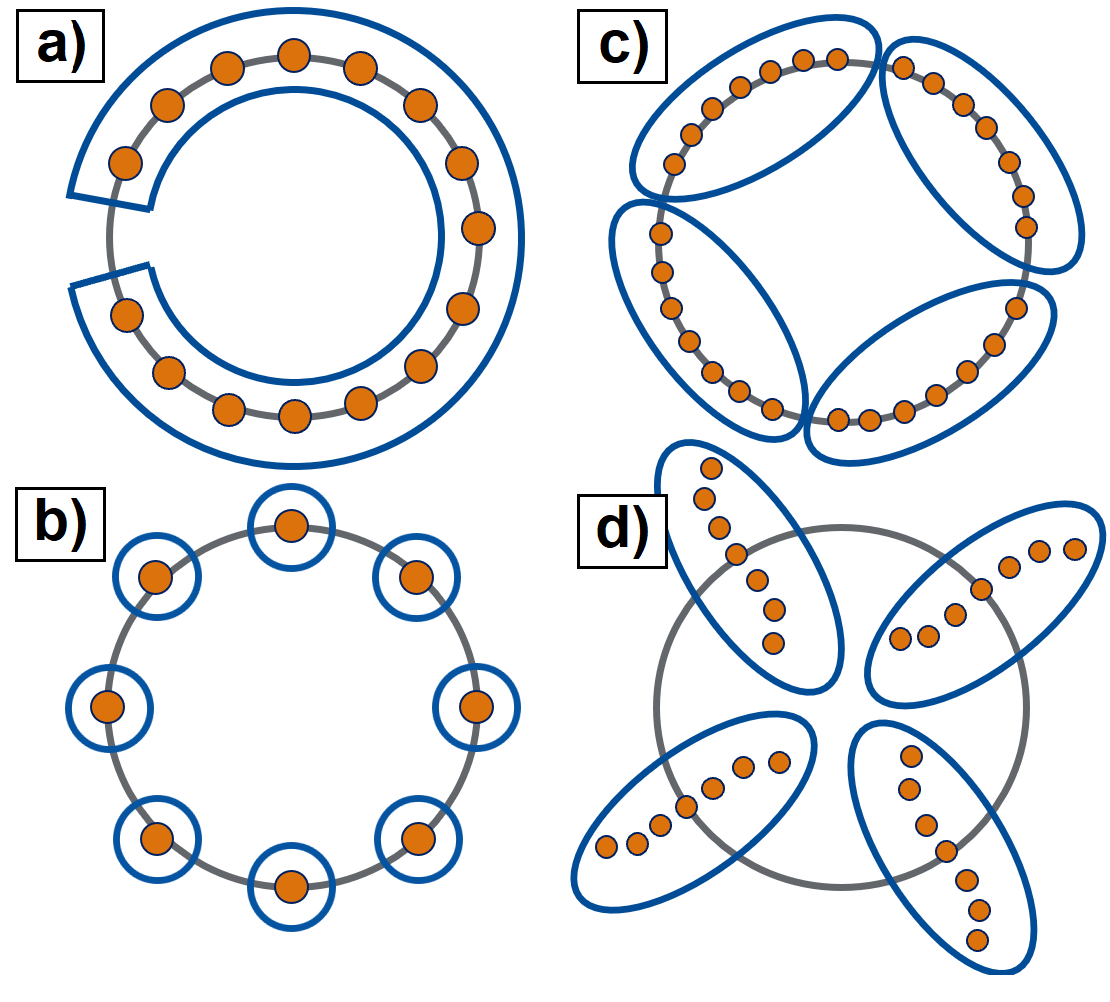}
        \caption{a) Direct Extraction mode. b) Bunch-by-Bunch Mode c-d) Bunch-Rotation mode, before/after rotation in the AR. Particles (in orange) that are extracted together are encircled in blue.} 
        \label{fig:CPAR_Modes}
\end{centering}
\end{figure}

In addition to the optimization of AR design parameters, the same AR can operate with multiple extraction modes depending on the needs of the experimental program (possibly switching between modes in the same run). Table~\ref{tab:CPAR_Modes} gives the pulse structure for three beam operating modes with a common CPAR design (at 0.8~GeV, more ambitious parameters than the \cite{PIP2BD} CPAR design). Fig.~\ref{fig:CPAR_Modes} illustrates the beam modes conceptually. The simplest is ``direct extraction'' mode, in which the ring is filled to maximum intensity and all particles in the ring are extracted simultaneously regardless of bunch structure (this maximizes pulse intensity). The next best-established is ``bunch-rotation'' in which the beam is initially gathered into a small number of long-bunches (4 in this case), an RF manipulation compresses those bunches (increasing the momentum spread), and then each of those long-bunches are extracted as individual pulses in quick succession. The last mode is ``bunch-by-bunch'' in which the beam is collected into short evenly space (in this case every other RF bucket) and a series of high rep. rate kickers extract each bunch as an individual pulse. The maximum kicker rep rate is a matter of further engineering work. 

\begin{table}[htp]
\centering
\begin{tabular}{| l || r | r | r | r | r |}
\hline
Operating & Pulse & Pulse & Extr. Rep. & Beam & Duty \\
Mode & Intensity & Length ($4\sigma$) & Rate & Power & Factor \\
\hline
Direct Extraction & $12\times 10^{12}$ & 540~ns & 100~Hz & 150~kW & $5.4\times 10^{-5}$ \\ 
Bunch Rotation & $2.2\times 10^{12}$ & $\sim$60~ns & 400~Hz & 110~kW & $1.6e\times 10^{-5}$\\
Bunch-by-Bunch & $0.8\times 10^{12}$ & $\sim$20~ns & 800~Hz & 82~kW & $1.2\times 10^{-5}$ \\
\hline
\end{tabular}
\caption{Three possible operating modes for the same CPAR ring design.}
\label{tab:CPAR_Modes}
\end{table}

Present design efforts are towards a 150~m 0.8~GeV ring with the capabilities described in Table~\ref{tab:CPAR_Modes} above. The major design features would also be compatible with a larger aperture version, with four times greater pulse intensity (for greater project cost) for all beam modes, Regardless of the aperture parameter, the AR would also be capable of a subsequent 1.2~GeV upgrade for twice the pulse intensity (or three times the pulse power) for all beam modes. These more ambitious parameters are consistent with the requirements for the proposed muon charged-lepton flavor violation (CLFV) called Advanced Muon Facility (AMF)~\cite{AMF}, but also a range of prospective beam powers for a beam dump physics facility.

\section{Theory Directions}~\label{sec:theory-directions}
\input{Contributions/Th}

\input{Contributions/mcp}
\section{Experimental Considerations}~\label{sec:exp-considerations}
 
\subsection{Opportunities for detectors with eV threshold.}~\label{subsec:ev-det}

\input{Contributions/eV_opportunities}

\subsection{Opportunities for detectors with keV threshold}~\label{subsec:kev-det}
\input{Contributions/keVopp}

\subsection{Opportunities for detectors with MeV threshold}~\label{subsec:mev-det}
\input{Contributions/MeV-Physics_Experimental_Considerations}

\section{Detector Technologies and Potential Experiments }\label{sec:det-tec}

\subsection{Charged-Coupled Devices} {\bf{\textcolor{red}{}}}
\input{Contributions/CCDs}

\subsection{Cryogenic Microcalorimenters}

\begin{figure}[thb]
\begin{center}
\includegraphics[width=2.4in]{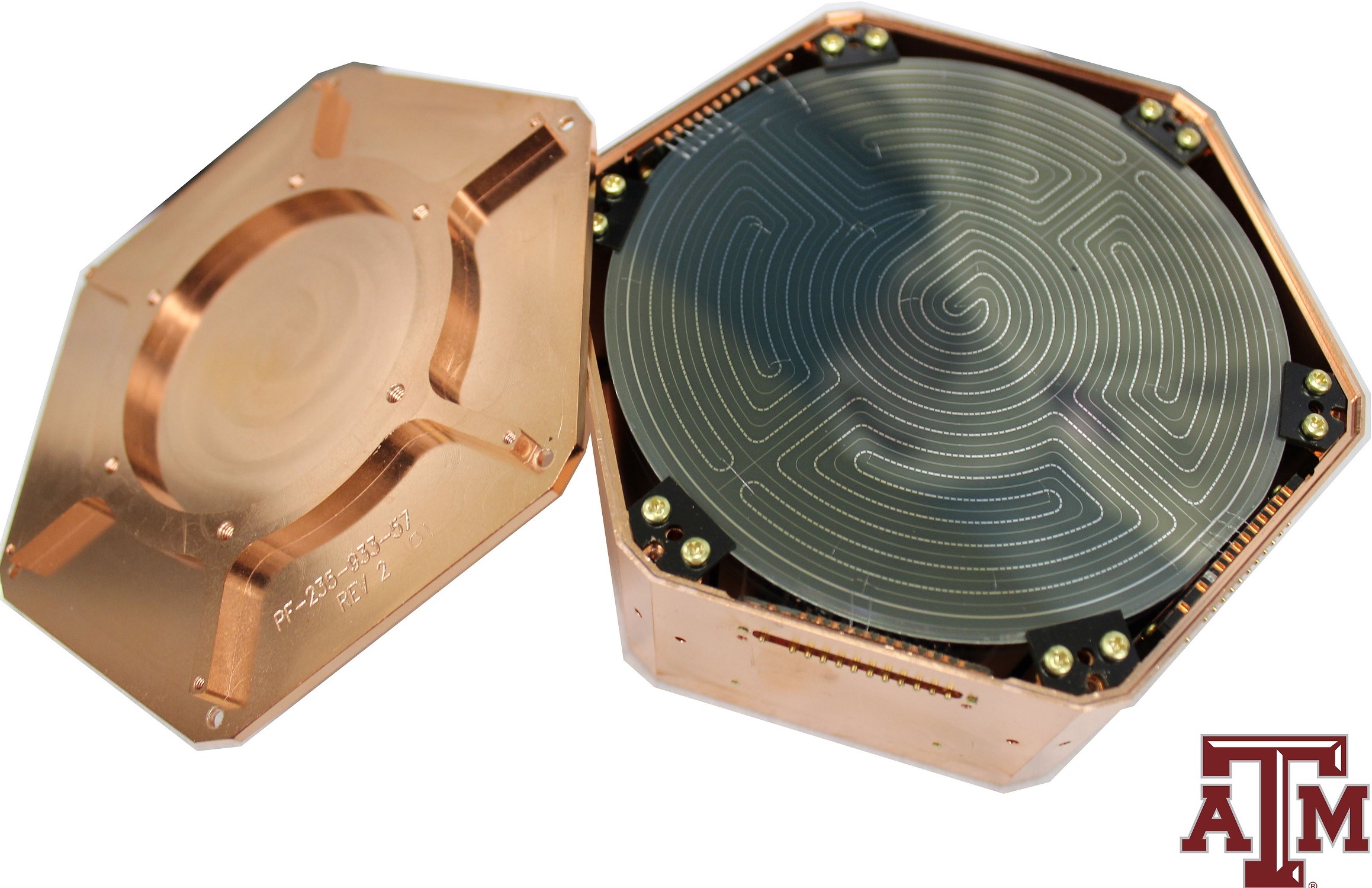}
\includegraphics[width=3.4in]{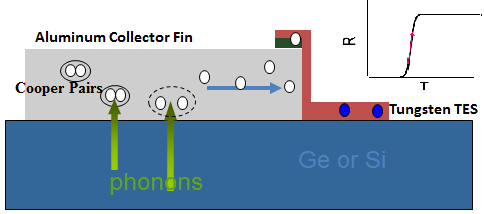}
\smallskip
\caption{{\bf(a)} Interleaved Z-senstive Ionization and Phonon (iZIP) detector developed for SuperCDMS and MINER, Data taken with an implanted $^{210}$Pb source, demonstrates excellent surface event rejection, \small {\bf(b)} Phonons, from interaction in crystal, break cooper pairs in superconducting aluminum fins, which are then collected by tungsten Transition Edge Sensors with high $\Delta R/\Delta T$ (inset).}
\label{fig:SuperCDMS_MINER}
\end{center}
\end{figure}

The baseline detector technology used by the SuperCDMS and the MINER experiment is a low temperature  ($\sim$ few mK) phonon-mediated semiconductor detector (kg-scale germanium/silicon) with photolithographically patterned Quasi-particle assisted Electro-thermal feedback Transition Edge Sensors (QET), as shown in Fig.~\ref{fig:SuperCDMS_MINER}. At the heart of the QETs are superconducting Transition Edge Sensors (TES) that are voltage biased to a stable quiescent point in the middle of the transition spontaneously held by the electro-thermal Feedback (ETF) mechanism ~\cite{CDMS-II:2009ktb}. The majority of the QET surface is made of large area aluminum fins that are connected to the much smaller area TES, whose function is to transform the phonon energy to quasi-particles. The quasi-particles diffuse and get trapped in the tungsten TES due to the smaller band gap (compared to aluminum), leading to a sharp temperature and thus sharp resistance increase. The primary function of the QET is to collect the energy and concentrate to the TES similar to the concave mirror of a reflective telescope. 

For the SuperCDMS SNOLAB Dark Matter search experiment, we use two different but complementary techniques. The iZIP detector simultaneously measures ionization and phonons.
In ER events, the particle loses energy through interaction with the valence electrons that leads to efficient ionization through sparse deposition of energy over a larger volume. In NR events, the particle loses energy through interaction with the nucleus, which leads to a dense deposition of energy that is not favorable to efficient ionization and produce approximately one-third (Lindahard) electron-hole pairs, compared to ER events. Thus, the ratio of ionization to phonon energy, called Ionization Yield , provides discrimination between ER and NR events. However, this discrimination stops at a keV-scale due to the limited ionization signal produced and the limiting charge readout noise. 

For the very low mass DM causing a nuclear recoil, the ionization signal may become smaller than the Signal-to-Noise Ratio (SNR) offered by standard ionization readout ruining the discrimination. To measure very low ionization, we use a High Voltage (HV) detector technology where the limited ionization signal is amplified to a much larger phonon signal using the NTL effect, albeit at the expense of background  discrimination.

The measurement principle is based on the fact that carriers drifting through crystals under an applied electric field generate additional phonons whose total energy is proportional to the number of carriers generated by the recoil as well as the applied bias voltage: $E_{Luke}=V_{bias}E/\epsilon$, 
 where $E$ is the energy of the interaction and $\epsilon$ is the average energy necessary to produce electron and hole pairs. Since the total signal is proportional to the bias potential, in the absence of any leakage current, the SNR improves proportionally to the bias and can be improved down to single electron-hole sensitivity.

A combination of the iZIP and the HV detectors, with well-demonstrated low-threshold performance, will provide excellent sensitivity to Dark Sector searches at PIP-II. 

\subsection{Scintillating Bubble Chamber} 
\input{Contributions/sbc}

\subsection{LArTPC ideas}~\label{subsec:LArTPC}
\input{Contributions/LArTPC}

\subsubsection{Low-threshold LArTPCs} 
\input{Contributions/LArTPCs_Low-Threshold}

\subsection{PIP2-BD}
\input{Contributions/pip2bd}

\subsection{DAMSA}~\label{subsec:damsa}
\input{Contributions/damsa}

\subsection{Coherent CAPTAIN-Mills}

The Coherent CAPTAIN-Mills (CCM) is a newly operating experiment to search for dark sector physics at the MeV mass scale. Highly motivated new vector portal dark sector models predict the existence of sub-GeV dark matter that can be tested at accelerator-based beam-dump experiments like CCM. They are sensitive to production of such dark matter allowing for probing early universe relic density limits. As well, CCM can probe for Axion Like Particles (ALP's) parameter space un-tested by previous experiments and cosmological constraints, and dark sector coupling to meson decay motivated by the enduring MiniBooNE anomaly.  Many of the CCM limits were shown in the corresponding limits for PIP2-BD, and CCM is therefore a theoretical and experimental proving ground for these long term physics goals.  CCM operates at the LANSCE Lujan Center which is a 100-kW neutron and stopped pion source that delivers an 800-MeV proton beam onto a tungsten target at 20 Hz with a short pulse width of 290 ns. The CCM detector is placed 23\,m from the target behind extensive shielding (5\,m steel and 3\,m concrete) and is a state of the art 10-ton liquid argon scintillation light detector. It is instrumented with 200 fast 8" Hamamatsu 5912-mod PMT's that can reconstruct scattering events from as low as 20\,keV thresholds up to hundreds of MeV.  The fast pulsing of the beam and nano-second time response of the detector is crucial for isolating relativistic dark sector events in time with the beam separating them from the prolific, but slower neutron backgrounds. A preliminary six week engineering run in 2019 has produced first physics results demonstrating the experiment works and significant dark sector searches are achievable with a stopped pion source and fast LAr detector~\cite{CCM:2021leg,CCM:2021yzc,CCM:2021jmk}.  With the lessons learned from the engineering run~\cite{CCM:2021leg} an improved CCM200 detector was built and began running in 2021.  The physics run will continue till 2025 to reach the predicted sensitivities.  Over the next year planned upgrades of the Lujan-PSR to a 120\,nsec short pulse mode and detector Cerenkov light reconstruction will reduce backgrounds and enhance signal sensitivity beyond current predictions.

\section{Conclusions}\label{sec:conc}
\input{Contributions/conclusions}

\addcontentsline{toc}{section}{References}
  \bibliographystyle{unsrt}
\bibliography{References}


\end{document}

%% file: authors.tex
\author[1]{A.~A.~Aguilar-Arevalo}
\author[2]{J.~L.~Barrow}
\author[3]{C.~Bhat}
\author[4]{J.~Bogenschuetz}
\author[5,6]{C.~Bonifazi}
\author[3]{A.~Bross}
\author[1]{B.~Cervantes}
\author[1]{J.~D'Olivo}
\author[7]{A.~De~Roeck}
\author[8]{B.~Dutta}
\author[9]{M.~Eads}
\author[3]{J.~Eldred}
\author[3]{J.~Estrada}
\author[3]{A.~Fava}
\author[10]{C.~Fernandes~Vilela}
\author[3]{G.~Fernandez~Moroni}
\author[3]{B.~Flaugher}
\author[3]{S.~Gardiner}
\author[4]{G.~Gurung}
\author[11]{P.~Gutierrez}
\author[4]{W.~Y.~Jang}
\author[8]{K.~J.~Kelly}
\author[8]{D.~Kim}
\author[3]{T.~Kobilarcik}
\author[2]{Z.~Liu}
\author[2]{K.~F.~Lyu}
\author[3]{P.~Machado}
\author[8]{R.~Mahapatra}
\author[11]{M.~Marjanovic}
\author[12]{A.~Mastbaum}
\author[3]{V.~Pandey}
\author[3]{W.~Pellico}
\author[13]{S.~Perez}
\author[14]{J.~Reichenbacher}
\author[13,15]{D.~Rodrigues}
\author[16]{A.~Sousa}
\author[3,9]{B.~Simons}
\author[17]{D.~Snowden-Ifft}
\author[3]{C.-~Y.~Tan}
\author[3]{M.~Toups}
\author[3]{N.~Tran}
\author[18]{Y.-T.~Tsai}
\author[19]{R.~G.~Van~de~Water}
\author[20]{R.~Vilar}
\author[21]{S.~Westerdale}
\author[4]{J.~Yu}
\author[3]{J.~Zettlemoyer}
\author[3]{R.~Zwaska}

\affil[1]{Instituto de Ciencias Nucleares, Universidad Nacional Autónoma de México, Mexico City 04510, Mexico}
\affil[2]{Department of Physics, University of Minnesota, Minneapolis, MN, 55455, USA}
\affil[3]{Fermi National Accelerator Laboratory, Batavia, IL, 60510, USA}
\affil[4]{Department of Physics, University of Texas, Arlington, TX, 76019, USA}
\affil[5]{International Center of Advanced Studies and Instituto de Ciencias Fisicas, ECyT-UNSAM and CONICET, Campus Miguelete – San Martin, Buenos Aires, Argentina}
\affil[6]{Universidade Federal do Rio de Janeiro, Instituto de Fisica, Rio de Janeiro, RJ, Brazil}
\affil[7]{CERN, European Organization for Nuclear Research, Geneva, Switzerland}
\affil[8]{Mitchell Institute for Fundamental Physics and Astronomy, Department of Physics and Astronomy, Texas A\&M University, College Station, TX, 77845, USA}
\affil[9]{Department of Physics, Northern Illinois University, DeKalb, IL, 60115, USA}
\affil[10]{Laboratorio de Instrumentacao e Fisica Experimental de Particulas (LIP), Av. Elias Garcia, 14-1, P-1000 Lisbon, Portugal}
\affil[11]{Homer L. Dodge Department of Physics and Astronomy, University of Oklahoma, Norman, OK, 73019, USA}
\affil[12]{Department of Physics, Rutgers University, Piscataway, NJ, 08854, USA}
\affil[13]{Department of Physics, South Dakota School of Mines and Technology, Rapid City, SD, 57701, USA}
\affil[14]{Universidad de Buenos Aires, Facultad de Ciencias Exactas y Naturales, Departamento de Fisica, Buenos Aires, Argentina}
\affil[15]{CONICET - Universidad de Buenos Aires, Instituto de Fisica de Buenos Aires (IFIBA), Buenos Aires, Argentina}
\affil[16]{Department of Physics, University of Cincinnati, Cincinnati, OH 45221, USA}
\affil[17]{Department of Physics, Occidental College, Los Angeles, CA, 90041, USA}
\affil[18]{SLAC National Accelerator Laboratory, Menlo Park, CA, 94025, USA}
\affil[19]{Los Alamos National Laboratory, Los Alamos, New Mexico, 87545, USA}
\affil[20]{Instituto de Fisica de Cantabria (IFCA), CSIC - Universidad de Cantabria, Santander, Spain}
\affil[21]{Department of Physics, University of California, Riverside, CA 92521, USA}

%% file: Contributions/TH.tex
The existence of new physics beyond the Standard Model(SM) is certainly well-motivated, since many puzzles, e.g., origins of neutrino masses and mixing, dark matter (DM), baryon abundance, fermion mass hierarchies, strong CP, etc. are yet to be resolved. The solutions to the existing riddles can emerge from new physics emerging from sub-GeV scale which can be investigated at  the ongoing proton accelerator based neutrino facilities, FASER, electron beam dump based experiments. Among, proton beam based neutrino facilities, CCM and  COHERENT with $\sim$~1~GeV proton beam, the J-PARC with $\sim$~3~GeV proton beam, and the Fermilab SBN program with 8 GeV BNB and 120 GeV NuMI beams  are providing opportunities to establish crucial new physics extensions of the SM addressing its several shortcomings.

The 1 GeV beam based experiments, e.g., CCM and COHERENT are sensitive to $\sim$ O(keV) to $\sim$ O(100) MeV energy depositions which are very interesting to search for a variety of new physics scenarios. Both COHERENT~\cite{COHERENT:2017ipa} and CCM\cite{CCM:2021leg} have already shown the feasibility of O(keV) searches and CCM has recently established the feasibility of searches associated with O(MeV) deposition~\cite{CCM:2021jmk}. The proposed PIP2-BD will also be sensitive to these regions using a 100ton LAr detector

The O(keV) region sensitivity is important to search for  
coherent scattering based nuclear recoil in these experiments which can investigate light dark matter models~\cite{CCM:2021jmk, COHERENT:2019kwz},
NSI~\cite{Dutta:2019eml,Giunti:2019xpr,Dutta:2020che,Han:2019zkz,DeRomeri:2022twg}, sterile neutrino parameter space  (via $\nu_{e,\mu}$ disappearances)~\cite{Blanco:2019vyp}. The SM background for new physics searches will emerge from the neutrino coherent elastic neutrino-nucleus scattering (CE$\nu$NS) processes which, however,  can be reduced by the timing and energy cuts on the prompt and delay neutrinos emerging from charged pion and muon decays respectively~\cite{Dutta:2019nbn,Dutta:2020vop}.

The O(MeV) region is important in investigating various new physics scenarios, e.g., ALP via decays, scattering, and absorption, dark photon, light dark matter via nucleus inelastic scattering, Milli-Charged particles, heavy neutral leptons, MiniBooNE anomaly, $N-\bar{N}$ oscillations, light dark matter via electron elastic scattering, etc. The SM background, in this case, emerges from neutrino-nucleus inelastic scattering and neutrino electron scattering which are small. In fact, the small neutrino backgrounds for the MeV regions make this stopped pion experiment very interesting for new physics searches compared to the neutrino facilities where higher energy beams are used which produce neutrinos from pions in flight. Further, the proximity of the detectors at  ($\sim 20$ m) and high-intensity proton beams ($\sim 10^{23}$ POT)  allow PIP2-BD to investigate complementary regions of new physics parameter spaces.

\subsection{Production of new physics particles} 
\begin{figure}[h]
\centering
    \includegraphics[width=0.49\textwidth]{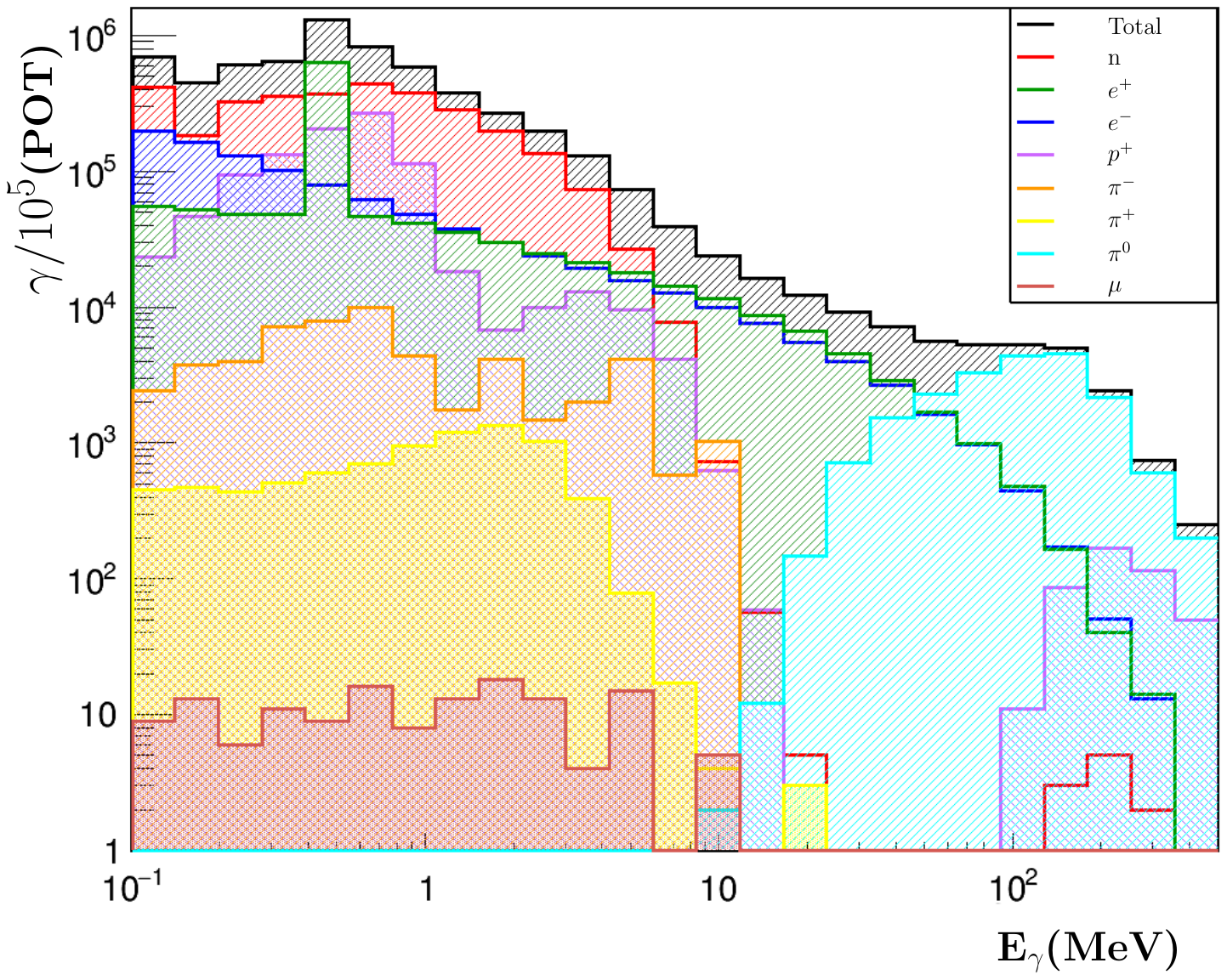}
    \includegraphics[width=0.49\textwidth]{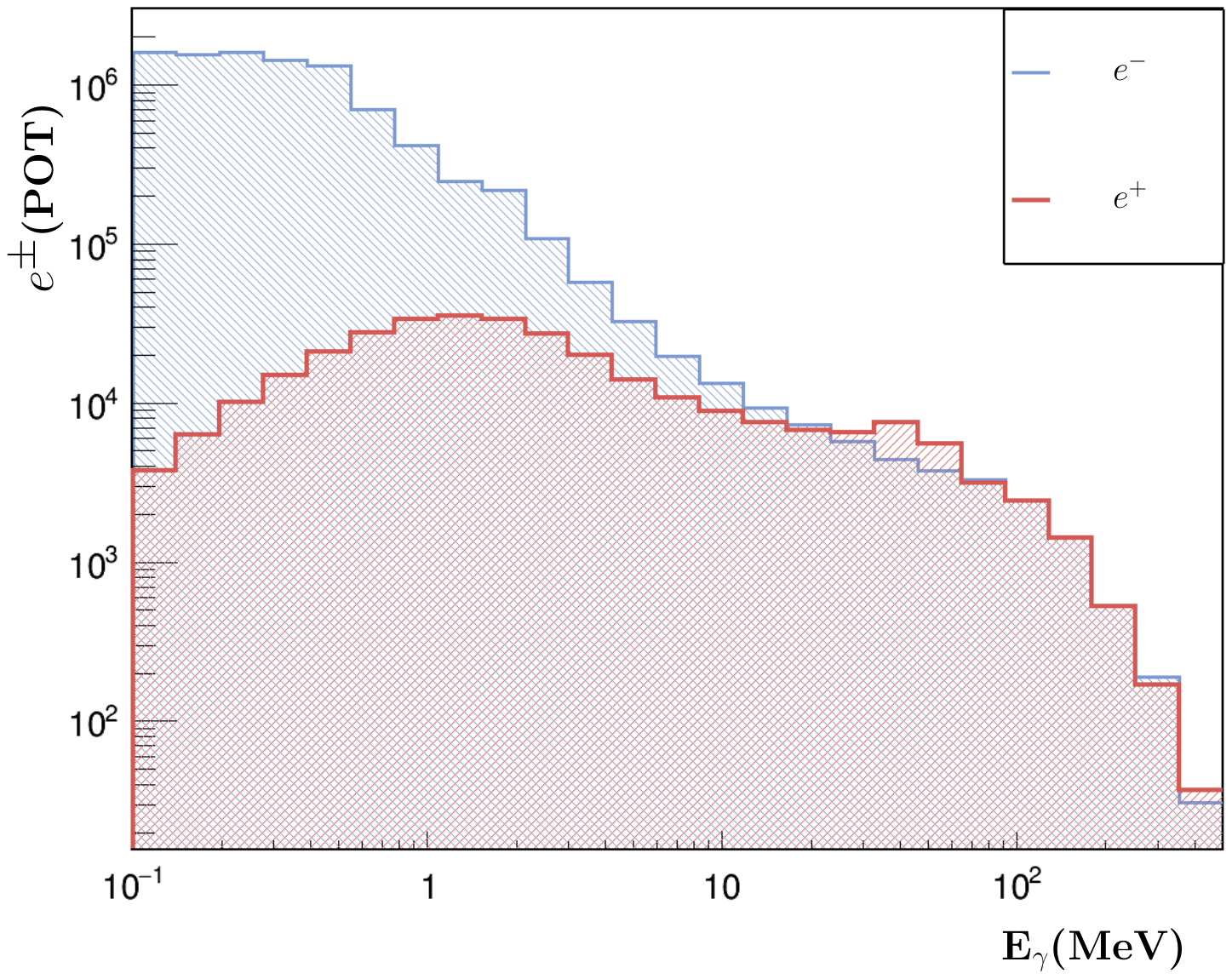}
    \caption{MC energy spectra for photons (top plot) and $e^\pm$ (bottom plot) at the Lujan source, simulated with \texttt{GEANT4 10.7} using the \texttt{QGSP\_BIC\_HP} library~\cite{GEANT4:2002zbu} by generating $10^5$ protons incident on a tungsten target.
    The different photoproduction sources are shown as non-stacked histograms in the top plot, with the total rate shown in black~\cite{Dutta:2020vop}.}
    \label{fig:photonspec}
\end{figure}

\begin{figure}[h]
\centering
    \includegraphics[width=0.49\textwidth]{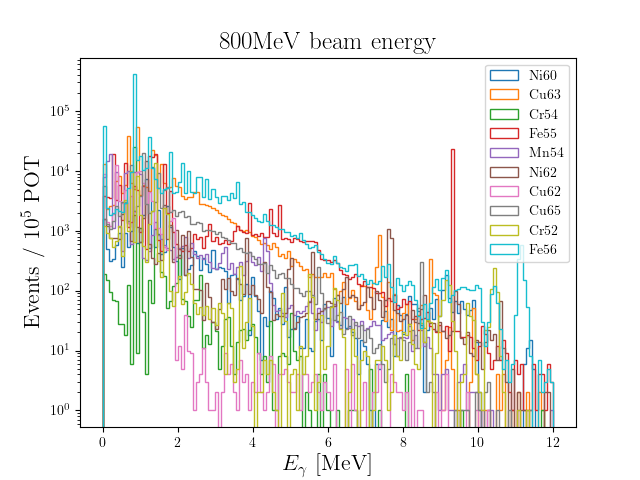}
    \caption{MC energy spectra for photons for $10^5$ protons incident on a carbon target.}
    \label{fig:photonspec2}
\end{figure}

When a high-intensity proton beam $\sim 1$ GeV strikes a target,  high-intensity flux of $\gamma$, $e^\pm$, $\pi^{\pm,0}$, $\eta$, de-excitation $\gamma$s are produced. In fig.~\ref{fig:photonspec}, we show the photon and $e^\pm$ flux for a 800 MeV beam hitting a Tungsten target. The typical numbers of $\pi^+,\,\pi^0$ are about $\sim$ 0.08/POT and $\pi^-$ is mostly absorbed. In general, $\pi^0$ is not produced at rest and the resulting $\pi^0$ flux is slightly in the forward direction~\cite{Dutta:2020vop}. We also show the de-excitation photons in Fig.\ref{fig:photonspec2}. These lines emerge when the proton beam hits a lighter target, e.g., carbon, Be etc. We use carbon as a target material for the figure.
\subsection{Neutrinos as Signal and as Background}

Protons with energies $>$ 300 MeV produce large numbers of positively charged pions, these pions can lose energy in dense material, stop and decay after coming to rest.  Clean stopped-pion neutrinos can be produced with proton energies of the order of 1 GeV or less~\cite{Alonso:2010fs}~(that limits the decay-in-flight component), and with a dense target to allow the pions to stop and decay at rest. The dominant neutrino production from stopped pions is from the weak-interaction two-body \textit{prompt} decay $ \pi^{+} \rightarrow \mu^+ + \nu_\mu$  ($\tau \sim$ 26 ns) followed by a three-body \textit{delayed} decay of muons $\mu^{+} \rightarrow e^+ + \nu_e + \bar{\nu}_\mu$ ($\tau \sim$ 2.2 $\mu$s) producing a well known spectrum shape. The prompt $\nu_\mu$ is monoenergetic (29.8 MeV) while $\nu_e$s and $\bar{\nu}_\mu$s energies spread out up to $m_\mu/2$. The stopped pion decay spectrum is relatively well known,  radiative corrections to pions and muons decaying at rest account for a tiny fraction of additional uncertainty~\cite{Tomalak:2021lif}. \\

These tens of MeV neutrinos can scatter off the target nucleus in the detector either via (i) neutral current CE$\nu$NS producing keV scale energy nuclear recoil signature, or (ii) charged current (only $\nu_e$s) or neutral current (all favors) inelastic scattering producing MeV scale energy signatures. These neutrinos can be treated as a signal for various SM and BSM studies, and they form the background to various BSM physics signals as described further in Ref.~\cite{Pandey:2023arh}. The pulsed time structure of these neutrinos gives a strong handle on suppressing the background.

\subsubsection{Elastic Scattering and keV Scale Physics}

In the CE$\nu$NS case, the neutrino scatters off an entire nucleus, exchanging a $Z^0$ boson and transferring some of its momenta to the nucleus as a whole, the scattered nucleus remains in its ground state.  For a few tens of MeV energy neutrinos and scattering off medium-sized nuclei, a dominant fraction of interactions are expected to be of coherent type. The differential cross-section of the process is written as:
\begin{equation}\label{Eq:cevns_xs}
\frac{\mathrm{d}\sigma}{ \mathrm{d}T} = \frac{G^{2}_{F}}{\pi} M_{A} \left(1-\frac{T}{E_{\nu}}-\frac{M_A T}{2 E^2_\nu}\right)~\frac{Q^2_{W}}{4}~F_{W}^2(Q^2), 
\end{equation}
where $G_F$ is the Fermi coupling constant, and $Q_W$ is the weak nuclear charge given as $Q^{2}_{W} = [g_p^V Z+g_n^V N]^2 = [(1-4\sin^2\theta_\text{W}) Z-N]^2$. N (Z) are neutron (proton) numbers, $\theta_W$ is the weak mixing angle, $E_\nu$ is the neutrino energy, $T$ is the nuclear recoil energy, and $M_A$ is the nuclear target mass. $F_{W}^2(Q^2)$ is the weak form factor of the nucleus. The weak-interaction charge of the proton is small compared to that of the neutron, so the CE$\nu$NS rate primarily depends upon $N^2$, the square of the number of neutrons. \\

The uncertainty on the CE$\nu$NS cross section is dominated by the weak form factor of the nucleus and is estimated to be a few percent level~\cite{VanDessel:2020epd}, the radiative corrections form a tiny fraction of the total uncertainty~\cite{Tomalak:2020zfh}. Since the uncertainties on SM predicted CE$\nu$NS cross-section is relatively small, CE$\nu$NS cross-section measurements allow testing of SM weak physics (e.g. weak nuclear form factor of a nucleus and weak mixing angle) or probing many new physics signals (e.g. non-standard interactions, sterile neutrinos, light dark matter) where CE$\nu$NS form the primary irreducible background. Any deviation from the SM predicted event rate either with a change in the total event rate or with a change in the shape of the recoil spectrum, could indicate new contributions to the interaction cross-section. The experimental signature of CE$\nu$NS is low-energy nuclear recoils $T$ of $\mathcal{O}$(keV). The pulsed time structure of these neutrinos gives a strong handle on suppressing the background for BSM physics signals. 

\subsubsection{Inelastic Scattering and MeV Scale Physics}

In the inelastic NC or CC scattering, where a single $W^+$ (CC) or $Z^0$ (NC) boson is exchanged between neutrino and target nucleus, the neutrino excites the target nucleus to a low-lying nuclear state, followed by nuclear de-excitation products such as gamma rays or ejected nucleon. The interaction cross sections for these processes do not have the $N^2$ enhancement therefore they are typically  an order of magnitude smaller than that of CE$\nu$NS process. The observable final-state particles of these inelastic scattering have typical energies of the same order as the incident neutrino energies, $\mathcal{O}$(MeV). Inelastic neutrino-nucleus cross-sections in these tens of MeV regime are quite poorly understood. Theoretical understanding of these processes is also relatively poor, due to the strong dependence of the interaction rates on the specific initial- and final-state nuclear wavefunctions, and experimental measurements are sparse~\cite{VanDessel:2020epd, Pandey:2014tza, Pandey:2016jju}. The neutrinos from the galactic core-collapse supernova carry similar energies as the well-understood stopped-pion neutrinos, therefore measuring inelastic neutrino interaction off nuclei (in particular off argon) can provide a unique opportunity to enable future neutrino experiments' (e.g. DUNE) capability to detect supernova neutrinos~\cite{DUNE:2023rtr}, please see Sec.~\ref{subsec:LArTPC}. \\

In addition to probing standard model nuclear structure physics and enabling supernova detection capabilities, inelastic neutrino-nucleus scattering also forms the irreducible background to some BSM signals including ALP and inelastic DM-nucleus scattering that produce similar MeV scale signature. Alternate to widely explored elastic scattering keV scale signals of e.g. dark photon models, the inelastic scattering of the same DM candidates has a smaller cross section but the MeV scale gamma rays signature can have higher signal-to-background ratios~\cite{Dutta:2023fij, Dutta:2022tav} (discussed later). The decay of nuclei excited by the inelastic scattering of DM is an unexploited channel that has a significantly lower background compared to similar searches using the elastic scattering channel. These models can be explored at decay at rest facilities using detectors with MeV scale signal sensitivity. The detectors with MeV energy threshold are discussed in Sec.~\ref{subsec:mev-det}. 
\subsection{New physics model sensitivities at PIP2-BD: A few examples}\label{subsec:newphys} We now discuss new physics model sensitivities at PIP2-BD. Examples include ALPs, dark photon and light dark matter. All these examples are final states involving O(MeV) electromagnetic energy.  This is certainly not an exhaustive list and analyses such as those involving HNLs, $N-\bar{N}$ oscillations, dark pion, mirror neutron models, explanations of MiniBooNE excess events will be included in the future.   
\subsubsection{Axion-like particles via decays and scattering} High intensity gamma sources at  a neutrino experiment can allow us explore axions via Primakoff production processes first pointed out in~\cite{Dent:2019ueq}. Generic models of ALPs with couplings to photons and electrons can be investigated at PIP2-BD. These interactions can be parameterized:
\begin{equation}\label{ALPlagrangian}
\mathcal{L}_{\rm ALP} ~\supset~ -\frac{g_{a\gamma}}{4}\,a\,F_{\mu\nu}\tilde{F}^{\mu\nu}\,-\,g_{ae}\,a\,\bar e \,i \gamma_5\, e\,,
\end{equation}
$\tilde{F}^{\mu\nu}$) $F_{\mu\nu}$ is the electromagnetic (dual-)field strength tensor. We will adopt a simplified model approach by considering two limiting cases: in the first, we set $g_{a e}=0$, so that the ALP phenomenology is completely determined by its electromagnetic interactions parameterized by $g_{a\gamma}$; and in the second case, we assume that $g_{ae}$ is sufficiently large to dominate ALP production and detection. 

\begin{figure}[H]
    \centering
    \includegraphics[scale=0.6]{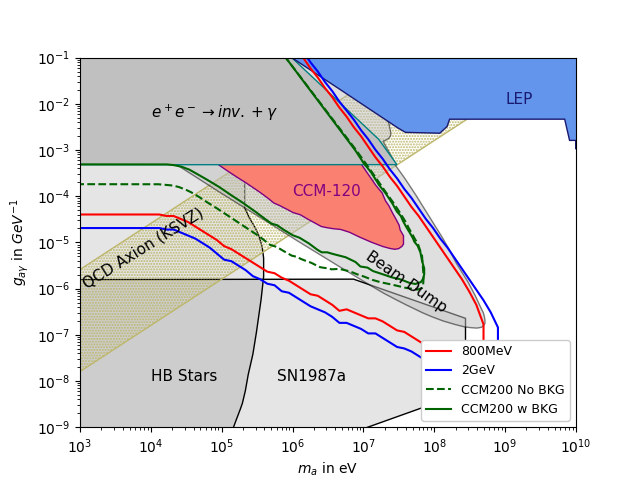}
    \caption{[Preliminary] ALP  search sensitivity using $g_{a\gamma\gamma}$ coupling~\cite{alppip}}
    \label{Fig:pip2axiongagamma}
\end{figure}

In Fig.~\ref{Fig:pip2axiongagamma}, we show the PIP2 sensitivity for the $g_{a\gamma\gamma}$ vs $m_a$ parameter space. Here the ALPs are produced by the Primakoff process and detected by the inverse Primakoff and decay to two-photon final states.

\begin{figure}[H]
    \centering
    \includegraphics[scale=0.6]{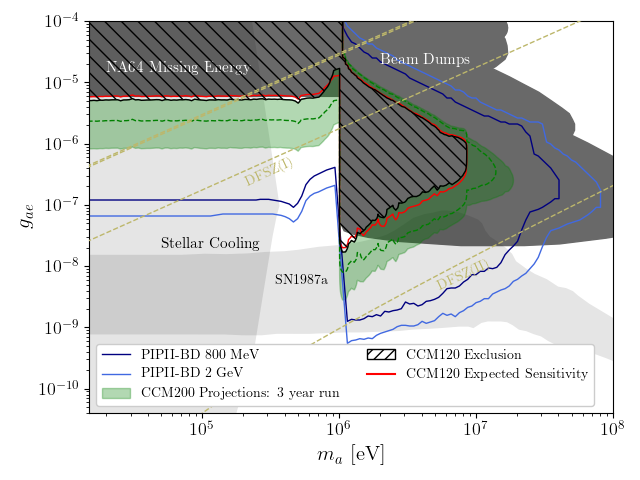}
    \caption{[Preliminary] ALP  search sensitivity using $g_{ae}$ coupling~\cite{alppip}}
    \label{Fig:pip2axiongae}
\end{figure}

In Fig.~\ref{Fig:pip2axiongae}, we show the PIP2 sensitivity for the $g_{ae}$ vs $m_a$ parameter space. Here the ALPs are produced by the Compton, associated, resonance production process and detected by the inverse Compton and decay to $e^+e^-$ final states.


\subsubsection{Axion-like particles via transition lines}
Another ALP production mechanism is to exploit the coupling of ALPs to nuclei. The decay rate ratio  $\Gamma_a / \Gamma_\gamma$ for nuclear decay $N^* \rightarrow N + a/\gamma$. Since ALPs are pseudoscalars, ALP $a$ is associated with MJ transitions (magnetic multipole transitions with angular momentum J).
The coupling $g_{aNN}$ is given by~\cite{Waites:2022tov}
\begin{equation}
\mathcal{L}_{aN} = ia \bar{\psi}_N\gamma_5({g^0_{aNN}}+g^1_{aNN}\tau_3)\psi_N 
\end{equation}
where $\psi_N= \left( \begin{array}{c}
     p \\
     n 
\end{array} \right)$. The branching ratio for the transitions to ALPs is~\cite{Avignone:1988bv}
\begin{eqnarray}
\label{eq:br}
\left( \frac{\Gamma_a}{\Gamma_\gamma}\right)_{\text{MJ}} &=& \frac{1}{\pi \alpha} \frac{1}{1+\delta^2} \frac{J}{J+1} \left( \frac{|\vec{p}_a|}{|\vec{p}_\gamma|} \right)^{2J+1} \nonumber 
\\
&\times& \left( \frac{ g_{aNN}^0 \beta + g_{aNN}^1}{(\mu_0-1/2)\beta + \mu_1 - \eta} \right)^2,
\end{eqnarray}
where $\beta$ and $\eta$ are nuclear structure factors, which have default values $\beta=1$, $\eta=0.5$ in the absence of nuclear data to support their calculation.

\begin{figure}[H]
    \centering
    \includegraphics[scale=0.75]{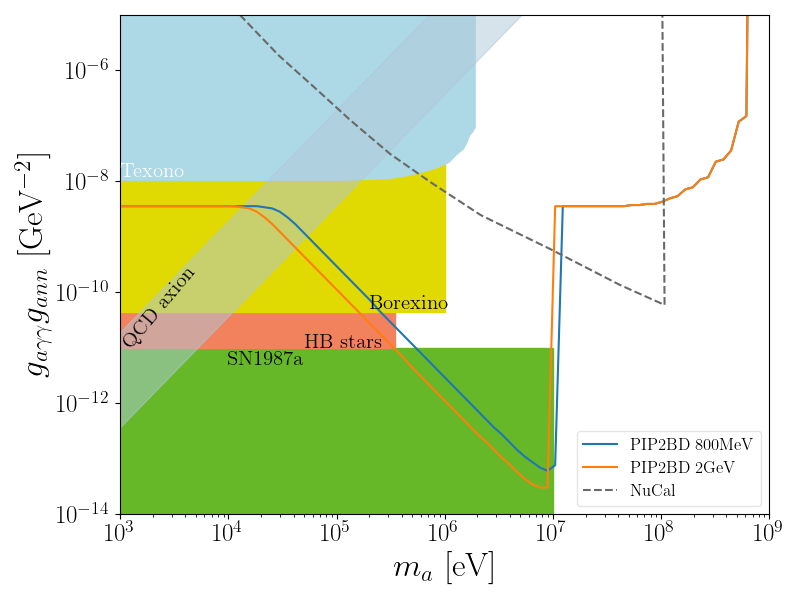}
    \caption{[Preliminary] ALP  search sensitivity using $g_{aNN}\times g_{a\gamma\gamma}$ coupling~\cite{absp}}
    \label{Fig:pip2axiongann}
\end{figure}

The GEANT4 simulation of PIP2-BD target Be provides several transition lines. Similar transition lines are utilized in the context of IsoDAR~\cite{Waites:2022tov}. Lighter nuclear targets with lower energy beams exhibit these kinds of excited lines. In Fig.\ref{Fig:pip2axiongann}, we show $g_{aNN}\times g_{a\gamma\gamma}$ as a function of $m_a$. PIP2-BD can probe larger regions of parameter space compared to any other experiment using Ni60, N15 lines. For the plot, we use the axions produced from the transition lines and detection lines via two photon decays (green and orange lines) while axions produced via Primakoff and detection via nuclear excitation lines (by absorption) at the target (red and blue lines). 

\subsubsection{Dark photon} 
\begin{figure}[H]
    \centering
    \includegraphics[scale=0.5]{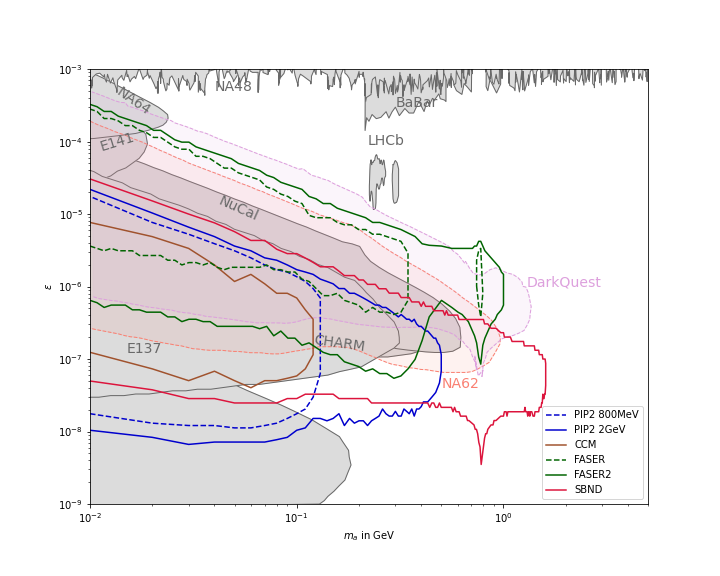}
    \caption{[Preliminary] Dark photon parameter space to be probed at PIP2-BD~\cite{darkphoton}}
    \label{Fig:pip2darkphoton}
\end{figure}

Dark photons can be searched at PIP2BD, using decays, scattering and nuclear absorptions. We can use  the following Lagrangian
\begin{equation}L\supset e_f\epsilon^f A^{\prime}_\mu\bar{f}\gamma^\mu f\end{equation} where $e_f=e Q_f$. If we use $A^\prime\rightarrow e^+ e^-$, we show the parameter space that can be probed at PIP2-BD~\cite{darkphoton} in Fig.~\ref{Fig:pip2darkphoton}. We find that the complementary regions of parameter space can be covered at PIP2-BD compared to other ongoing neutrino experiments and FASER. The complementary feature emerges due to the  near location ($\sim$ 20 m) of the detectors.  

\subsubsection{Light dark matter via inelastic nuclear scattering}\label{subsubsec:inelastics}
Light DM emerging from the decays of a vector mediator, for example, a dark photon, has been proposed in numerous studies as a viable DM candidate~\cite{deNiverville:2015mwa,Batell:2014mga,Dutta:2019nbn}  and the existing searches have looked for the elastic scattering signature of DM in the detectors of pion decay-at-rest experiments, e.g., COHERENT, CCM. The interaction Lagrangian for fermionic, $\chi$, and scalar, $\phi$, DM coupled to the SM via the dark photon is expressed as
\begin{eqnarray}
    \mathcal{L}_f &\supset& g_D A'_\mu \bar \chi \gamma^\mu \chi + e \epsilon Q_q A'_\mu \bar q \gamma^\mu q \\
    \mathcal{L}_s &\supset& \lvert D_\mu \phi \rvert^2 + e \epsilon Q_q A'_\mu \bar q \gamma^\mu q
    \nonumber
    \label{eq:lag}
\end{eqnarray}
where $g_D$ is the dark coupling constant, $\epsilon$ is the mixing parameter, $Q_q$ is quark's electric charge. The dark photon  can be produced through pion capture, pion/eta decay, and photons emerging from cascades. Among these production processes, the pion decay provides the dominant contributions.

Recently, we executed a similar DM search strategy via the inelastic channel~\cite{Dutta:2022tav,Dutta:2023fij}, which makes use of the photon spectrum produced through the decay of excited nuclear states ($N^\ast\rightarrow N\gamma$). Though this channel has smaller rates compared to the elastic channel it has a significantly reduced background containing an irreducible component coming from neutrino inelastic scattering. Due to the larger energies deposited during inelastic scattering the sensitivity in this channel is not limited by the detector threshold.

To a good approximation, the inelastic cross-section to a given final state $J_f$ is:
\begin{align}
    \frac{d\sigma^{DM}_{inel}}{d\cos\theta} &= \frac{2e^2\epsilon^2 g_D^2 {E'}_\chi {p'}_\chi}{(2m_N E_r + m_{A'}^2 -\Delta E^2)^2} \frac{1}{2\pi} \frac{4\pi}{2J+1} \\
    &\times \sum\limits_{s_i,s_f} \vec{l} \cdot \vec{l}^* \frac{g_A^2}{12\pi}
    |\langle J_f|| \sum_{i=1}^A \frac{1}{2}\hat{\sigma_i} \hat{\tau_0}|| J_i\rangle|^2 \nonumber
\end{align}
where $\Delta E$, $m_N$, and $J$ are the excitation energy, nuclear mass and spin, respectively. 
The DM currents, $\vec{l}$, depend on the DM spin under consideration and we consider both fermionic and scalar DM:
\begin{eqnarray}
    \hspace*{-4mm}\sum\limits_{s_i,s_f} \left(\vec{l} \cdot \vec{l}^*\right)_f &=& 3- {1\over 4E_\chi {E'}_\chi} \left[ 2 \left(p_\chi^2 + {p'}_\chi^2-2m_NE_r \right) + 3m_\chi^2 \right] \nonumber \\
     \hspace*{-4mm}\sum\limits_{s_i,s_f} \left(\vec{l} \cdot \vec{l}^*\right)_s &=& {1\over 2E_\phi {E'}_\phi} \left(p_\phi^2 + {p'}_\phi^2-2m_NE_r \right)
    \nonumber
\end{eqnarray}

\begin{figure}[H]
    \centering
    \includegraphics[scale=0.5]{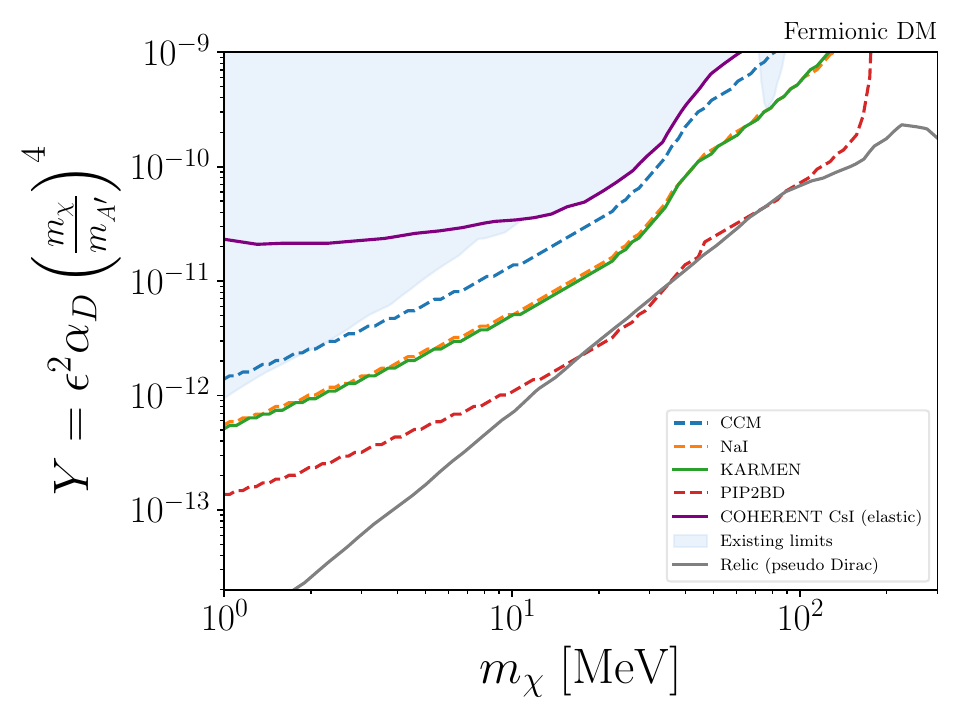}
     \includegraphics[scale=0.5]{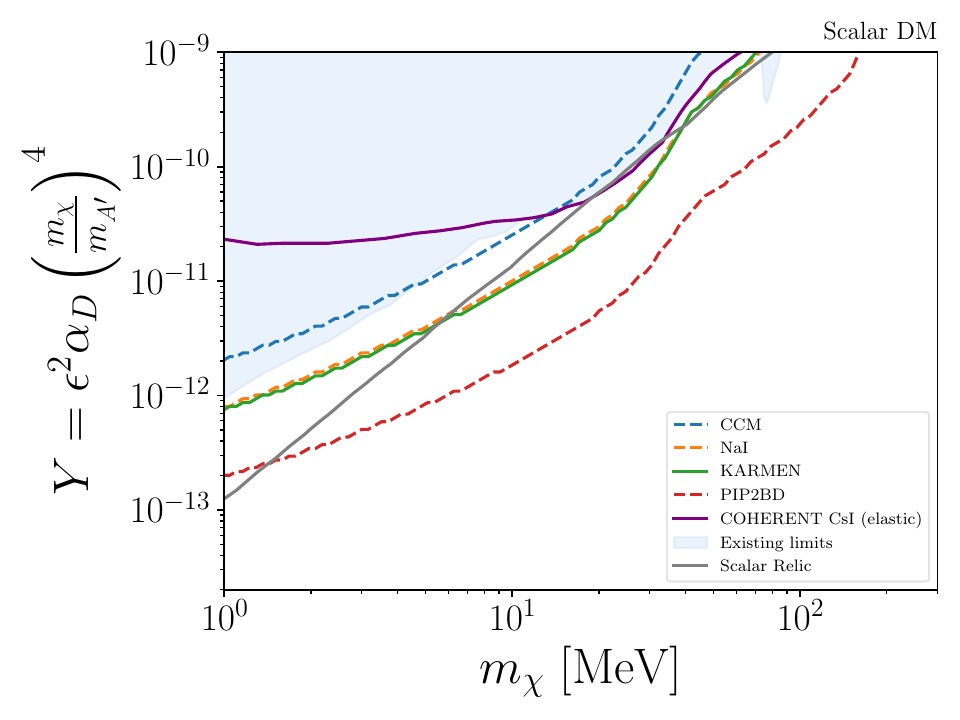}
    \caption{Light dark matter via inelastic nuclear scattering channel PIP2-BD where we use $m_{A'}/m_{\chi}=3$, and $g_D=0.5$~\cite{Dutta:2023fij}}
    \label{Fig:pip2darkmatterinelastic}
\end{figure}

In Fig.~\ref{Fig:pip2darkmatterinelastic}, we show the current limits from KARMEN (existing data~\cite{Maschuw:1998qh}), ongoing CCM, and PIP2-BD expected sensitivity reach.  The excitation of nuclei via neutral current $\nu$ scattering was observed by the KARMEN experiment using the $^{12}$C($\nu$, $\nu'$)$^{12}$C$^*(1^+,1; 15.1$ MeV) reaction at the ISIS neutron source~
\cite{KARMEN:1998xmo,KARMEN:1991vkr,Maschuw:1998qh}. We use this measurement to constrain the parameter space~\cite{Dutta:2023fij}.

\subsubsection{Dark sector models to explain the MiniBooNE excess}
Neutrino-based new physics explanations have been popular solutions to the MiniBooNE electron-like excess events~\cite{MiniBooNE:2008yuf,MiniBooNE:2018esg,MiniBooNE:2020pnu}. The dark sector-based light mediator-motivated solutions have constraints from the MiniBooNE dump results when the light mediators emerge from the neutral meson decays~\cite{MiniBooNEDM:2018cxm,Jordan:2018qiy}. However, the helicity unsuppressed charged meson 3 body decays can provide a solution to the excess since due to the focussing horns MiniBooNE detector observes much higher flux of light mediators from the charged mesons compared to the neutral mesons decays~\cite{Dutta:2021cip}.

Recently, we explained the MiniBooNE excess using (pseudo)scalar and vector mediators emerging from the charged pion decays which subsequently produce an inverse Primakoff-like-scattering at the detector. For example,
\begin{equation}
\begin{tikzpicture}
        \begin{feynman}
            \vertex (a) {\(\pi^\pm\)};
            \vertex [right=1.2cm of a, blob] (b) {\(\, \, \, \, \, \)};
            \vertex [right=2.0cm of b] (f2) {\(X(a,\,\phi, A')\)};
            \vertex [below left=0.3cm of b] (xl);
            \vertex [above right=1.2cm of b] (f1) {\(\nu\)};
            \vertex [below right=1.2cm of b] (f3) {\(\ell\)};
            
            \diagram* {
                (a) -- [scalar] (b) -- [fermion] (f1),
                (b) -- [anti fermion] (f3),
                (b) -- (f2)
            };
        \end{feynman}
    \end{tikzpicture}
    \begin{tikzpicture}
           \begin{feynman}
                    \vertex (o3) {\(\)};
                    \vertex [above=1.0cm of o3] (p) {\(+\)};
           \end{feynman}
    \end{tikzpicture}
    \begin{tikzpicture}
    \begin{feynman}
         \vertex (o1);
         \vertex [above left=0.7cm of o1] (f1) {\(X\)};
         \vertex [above right=0.7cm of o1] (i1){\(\gamma\)};
         \vertex [below=0.7cm of o1] (o2);
         \vertex [right=0.7cm of o2] (f2) {\(N\)};
         \vertex [left=0.7cm of o2] (i2) {\(N\)};

         \diagram* {
           (i1) -- [boson] (o1) -- (f1),
           (o1) -- [edge label={\(Y\)}] (o2),
           (i2) -- [fermion] (o2),
           (o2) -- [ fermion] (f2),
         };
        \end{feynman}
       \end{tikzpicture}
      \nonumber
\end{equation}
This inverse Primakoff scattering helps to explain the energy and angular spectra of the electron-like excess events~\cite{Dutta:2021cip}. This explanation can purely emerge from a dark sector that does not have any neutrino interactions.

\begin{figure}[H]
    \centering
    \includegraphics[scale=0.4]{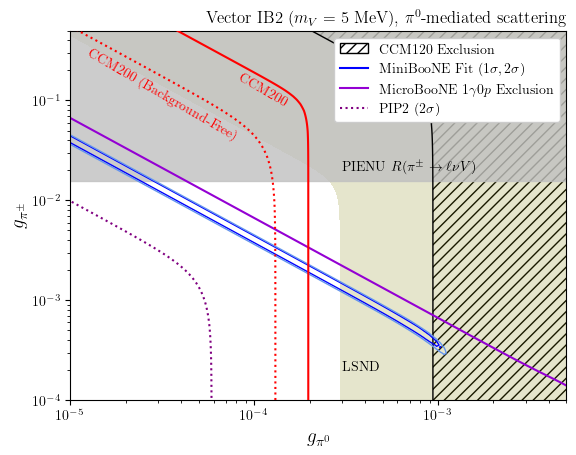}
    \caption{[Preliminary]Dark sector solution parameter space sensitivity at PIP2-BD~\cite{dscmd}}
    \label{Fig:pip2dscmd}
\end{figure}

The existence of high-intensity sources for charged and neutral pions at the stopped pion experiments can probe these dark sector models in a complementary way. In Fig.\ref{Fig:pip2dscmd}, we show the MiniBooNE excess fit in the parameter space of dark sector models where the couplings of light mediators to charged and neutral pions are assumed to be model independent. We find that the ongoing CCM and the future PIP2-BD will cover a large region of the excess parameter space via $\pi^0$ and $\pi^\pm$ production processes where the signal is $1\gamma +0$p. For this example, we used a light vector mediator which can be produced from $\pi^+,\, \pi^0$ decays (via$\pi^\pm\rightarrow l\nu V$ and $\pi^0\rightarrow\gamma v$) which then produces a photon via scattering at the target by exchanging $\pi^0$ with the nucleus(via $\pi^0-\gamma-V$ and $\pi^0-N-N$ interactions).  We also show the MicroBooNe sensitivity region in the same parameter space. 

The dark sector models also give rise to $e^+e^-$ signals via inelastic dark matter where $\chi_1$ emerging from the decays of light mediator can upscatter to $\chi_2$ which subsequently decays into  $\chi_2\rightarrow\chi_1 e^+e^-$~\cite{Dutta:2021cip}.  The 2 GeV proton beam energy at PIP-II will help us to probe the parameter space of this signal since the $\chi_2$ upscattered production as required for the explanation of the excess is mostly inaccessible at 1 GeV proton beam energy.

\subsection{Millicharged particles} 

The physics origin and the nature of charge quantization are unknown~\cite{Dirac:1931kp}, and the search for millicharged particles (MCPs) could help reveal the underlying physics. 
Further, fractionally charged particles exist in the SM: quarks and antiquarks have an electric charge $\pm 1/3$ or $\pm 2/3$ that of the electron. 
Millicharged particle (mCP) models are extensions of the SM where a new particle is introduced with a very small electric charge. This can be achieved by a new fermion charged under the Standard Model hypercharge with a small charge, e.g.,  
\begin{equation}
    \mathcal{L}_{\rm mCP}=i\Bar{\chi}(\slashed\partial-i\varepsilon^\prime g^\prime \slashed B+ M_{mCP})\chi \,,
\end{equation}

\noindent where $\chi$ is the particle, $\slashed B$ is the SM electroweak vector boson and $\varepsilon$ is the millicharge.

More generally, we can focus on the millicharged particle couples to photons (e.g., the hypercharged particle after electroweak symmetry breaking),
\begin{equation}
    \mathcal{L}_{\rm mCP}=i\Bar{\chi}(\slashed\partial-i\varepsilon e \slashed A+ M_{mCP})\chi \,,
\end{equation}
For instance, MCPs may emerge in models where a dark photon has minor kinetic interaction with the Standard Model (SM) photon. This interaction leads to an extremely minuscule charge for particles in the dark sector~\cite{Holdom:1985ag,Holdom:1986eq}. In such a scenario, another vector boson, referred to as $\slashed B'$, needs to be taken into account ~\cite{Chang:2018rso,Antypas:2022asj}. It's also worth noting that mCPs have been proposed as potential dark matter (DM) candidates ~\cite{Pospelov:2007mp} and can help explain various experimental anomalies~\cite{Gninenko:2006fi,Liu:2019knx,Agrawal:2021dbo}. 
Because of these reasons, experiments frequently target mCPs in the MeV to GeV mass range.

\begin{figure}[t]
    \centering
    \includegraphics[width=0.4\textwidth]{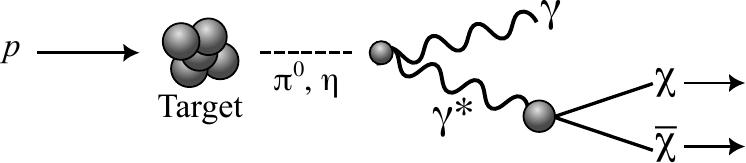}
    \caption{Schematic of millicharged particle production in the NuMI beam via a meson decay and a virtual photon~\cite{SENSEI:2023gie}.}
    \label{fig:schematic_decay}
\end{figure}

In numerous neutrino experiments, accelerator-based neutrino beams originate from a high-intensity proton beam colliding with a fixed target. Assuming mCPs exist, they could be produced in line with the neutrino beam through photon-mediated decays of scalar mesons, vector mesons, and direct Drell-Yan processes during each collision (Fig.~\ref{fig:schematic_decay}).

The high-intensity PIP-II beam offers a unique opportunity to explore millicharged particles within the MeV domain.

\subsection{Heavy Neutral Leptons}
The neutrino oscillation is clear evidence of physics beyond the Standard Model (SM) as the neutrinos are exactly
massless within the SM. It indicates the existence of new particles and/or new interactions in the neutrino sector,
and the Heavy Neutral Lepton (HNL) is a particularly motivated candidate for such a new physics
(see, \emph{e.g.}, Refs.~\cite{Boyarsky:2009ix,Drewes:2013gca,Dasgupta:2021ies,Abdullahi:2022jlv,Batell:2022xau} 
and references therein).
If the HNL is lighter than the pion and the muon, the HNL can be produced from the decay of pions and muons,
and hence the PIP2-BD experiment can be an HNL factory due to its large statistics of stopped pions and muons.
In the following, we estimate the future sensitivity of the PIP2-BD experiment on the HNL that mixes dominantly
with either muon neutrinos or electron neutrinos.

The HNL interacts with SM particles through the mixing with neutrinos as
\begin{align}
	\mathcal{L}
	&= \bar{N}\left(i\slashed{\partial} -m_N\right)N
	- \frac{g}{\sqrt{2}}U_{lN}^*\bar{l} \slashed{W}^- N
	- \frac{g}{2\cos\theta_W} U_{lN}^*\bar{\nu}_l \slashed{Z} N
	+ (\mathrm{h.c.}),
\end{align}
where $N$ is the HNL with its mass $m_N$, $g$ is the SU(2) gauge coupling, $l$ is the SM charged lepton
(we focus on $l = e$ or $\mu$), $\nu_l$ is the SM neutrino, $\theta_W$ is the weak mixing angle,
and $U_{lN}$ is the mixing angle between the HNL and the SM neutrino.
To be specific, we take $N$ the Dirac fermion.
We focus on the mass region $m_N < m_{\pi}$ with $m_\pi$ the (charged) pion mass. 
In this case, the main HNL production mode is the decay of stopped muons and pions. 
In the parameter space of our interest, the mean decay length of the HNL is significantly longer than the laboratory scale. 
Therefore, after being produced at the target, a small portion of the HNL
travels downward and decays into an electron-positron pair plus a neutrino inside the detector.
The total event number of such a decay inside the detector is estimated as
\begin{align}
	N^{(i)}_{ee} &= \sum_{i = \mu, \pi} N_i \times \epsilon_\mathrm{det}
	\times \frac{1}{\Gamma_i}\int dE_N
	\frac{d\Gamma(i \to f N )}{dE_N}
	\times
	\frac{L_\mathrm{det}}{\gamma\beta c\tau_{N\to ee\nu}},
	\label{eq:Nee_total}
\end{align}
where the superscript $i$ indicates whether the HNL is from the muon or pion decay,
$N_i$ is the total number of stopped muons and pions, $\epsilon_\mathrm{det}$ is the angular coverage
of the detector, $\Gamma_i$ is the total decay width of the muon and pion,
$d\Gamma(i\to fN)/dE_N$ is the differential decay rate of $i$ with $E_N$ the HNL energy.
Here $f = e+\nu$ for $i = \mu$ and $f = \mu$ or $e$ for
$i = \pi$ (note that the chirality flip can be supplied by the HNL mass in the electron mixing case).
The mean decay length is given by a product of the HNL (partial) decay length at rest $c\tau_{N \to ee \nu}$
and the relativistic factor $\gamma \beta = \sqrt{E_N^2-m_N^2}/m_N$,
and $L_\mathrm{det}$ is the detector length.

\begin{figure}[t]
	\centering
 	\includegraphics[width=0.495\linewidth]{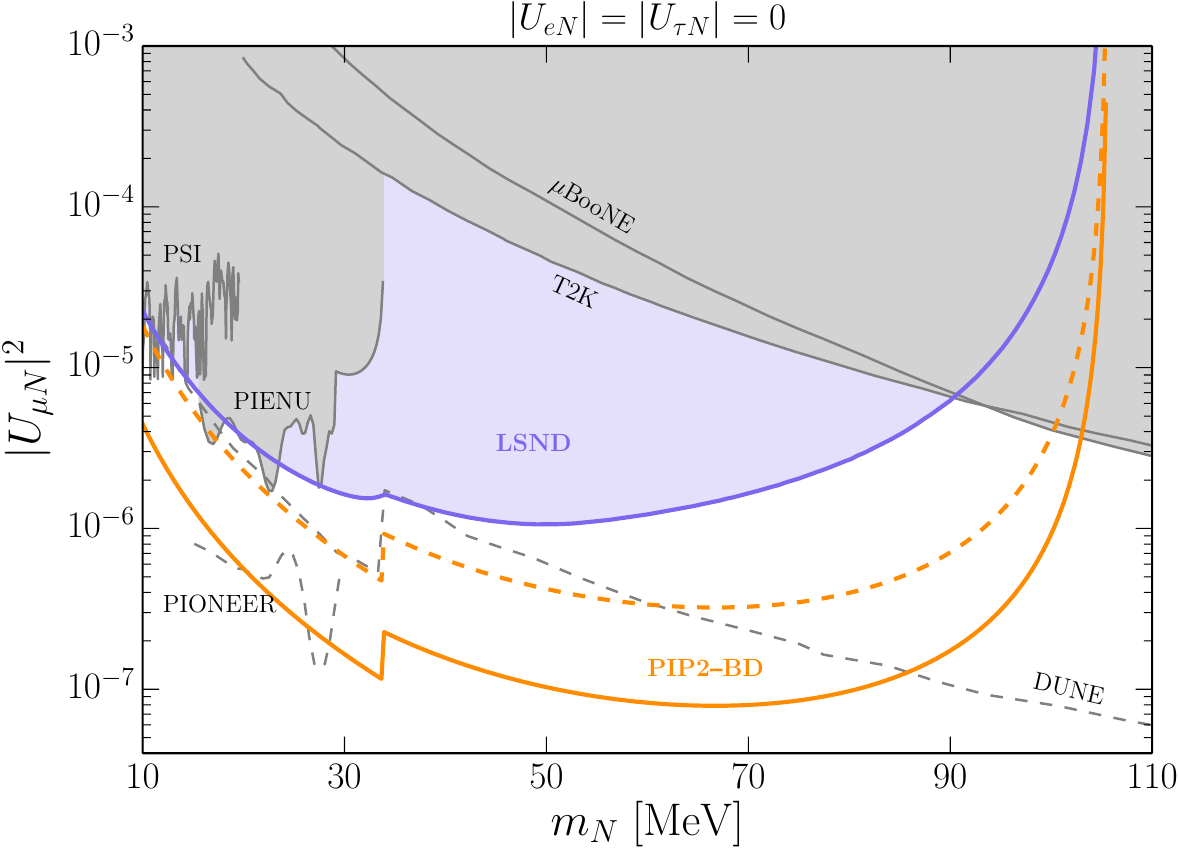}
	\includegraphics[width=0.495\linewidth]{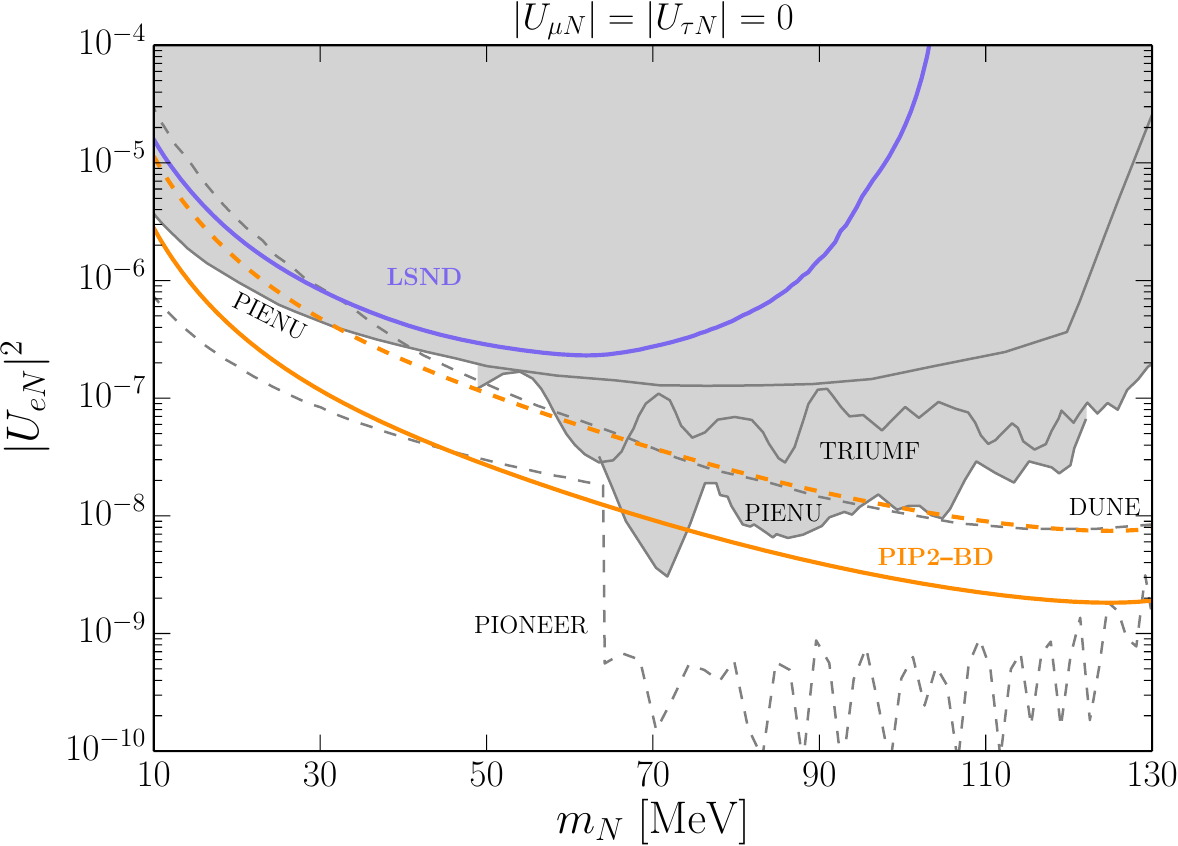}
	\caption{
	The sensitivity of PIP2-BD on $\vert U_{l N}\vert^2$ is shown in orange color, 
	where the solid (dashed) orange line corresponds to 3 (50) events of the HNL decay inside the detector~\cite{Ema:2023buz}.
	The existing limit from the LSND experiment~\cite{LSND:2001akn,Ema:2023buz} is shown in blue, which
	demonstrates the potential of stopped muons and pions as the HNL source.
	The previous constraints (all at 90\% C.L.) from 
	PSI~\cite{Daum:1987bg}, TRIUMF~\cite{Britton:1992xv},
	PIENU~\cite{PIENU:2017wbj,PIENU:2019usb,Bryman:2019bjg},
	T2K~\cite{T2K:2019jwa,Arguelles:2021dqn}, 
	and $\mu$BooNE~\cite{MicroBooNE:2021usw,Kelly:2021xbv},
	as well as the expected future sensitivities of DUNE~\cite{Berryman:2019dme} 
	and PIONEER~\cite{PIONEER:2022yag,PIONEER:2022alm}, are also shown in gray color
	(the data adapted from \texttt{Heavy-Neutrino-Limits}~\cite{Fernandez-Martinez:2023phj}).
	\emph{Left}: the muon mixing case. \emph{Right}: the electron mixing case.
 }
	\label{fig:mixing_parameter}
\end{figure}

In the following estimate, we take $1.2\times 10^{23}$ as the total number of proton-on-target and
0.1 as the formation rate of stopped $\pi^+$ (and hence $\mu^+$) per proton,
which results in $N_\pi = N_\mu = 1.2\times 10^{22}$.
We assume that the active volume of the detector is cylindrical in shape, with 4.5\,m in height and 4.5\,m in diameter, 
located 18\,m away from the HNL production point, which fixes the other parameters as
$\epsilon_\mathrm{det} \simeq 3.9\times 10^{-3}$ and $L_\mathrm{det} = 4.5\,\mathrm{m}$.
Without any dedicated study on backgrounds at this moment, 
we may draw the lines that correspond to 3 and 50 events of the HNL decay inside the detector, 
assuming 75\,\% of event acceptance.
In Fig.~\ref{fig:mixing_parameter}, we show the future sensitivity of PIP2-BD, together with the expected sensitivities of
DUNE~\cite{Berryman:2019dme} and PIONEER~\cite{PIONEER:2022yag,PIONEER:2022alm}.
The solid orange line corresponds to 3~events, while the dashed orange line corresponds to 50~events.
The figure shows that PIP2-BD has the potential to explore the new parameter region, 
in particular in the muon mixing case.

%% file: Contributions/eV_opportunities.tex
Significant developments were made on low threshold detector technologies in recent years. This effort is mostly driven by low mass direct dark matter searches and includes phonon sensitive calorimeters \cite{SuperCDMS:2020hcc, Petricca:2017zdp},  semiconductor detectors \cite{SENSEI:2020dpa}, and scintillating bubble chambers using noble liquid detectors \cite{hawleyherrera2023sbcsnolab}.

In many cases the number of dark sector signal events in a beam dump experiment expected for a detector has been shown to scale linearly with the detector threshold. This is the case for the kinetic mixing models discussed in \cite{Dutta:2023fij} and also for the millicharged particles model discussed in \cite{mcpRoni2019}. The power of low threshold detectors in beam dump dark sector experiments has been recently demonstrated by the world leading result in millicharged particles(mCP) established by the SENSEI experiment with a 9 g-day exposure in the MINOS beam at Fermilab~\cite{barak2023sensei}.

The high intensity of the PIP-II proton beam, together with the novel low threshold technologies present a unique new opportunity for the development of a low threshold dark sector program at Fermilab. As a demonstration of this we have estimated the flux of mCPs for a detector in a PIP-II beam dump, compared with the flux for SENSEI at MINOS, shown in Fig.\ref{Fig:milli_flux}.

The main challenge for taking advantage of this opportunity consists of managing the backgrounds at low energies. This includes the environmental radiation background along with the background produced by the beam. 

Extensive studies of environmental background have been performed at the shallow underground facility at Stanford during the early stages of the superCDMS program \cite{platt2018} with a overburden of 17 meters water equivalent. In those studies it was demonstrated that with a 10 cm lead shield, the event rate in a keV threshold germanium detector can be reduced to 100 events/kg/day/keV (dru). When an additional veto is introduced to reject the events produced by cosmic muons it achieves a significant reduction on the background rate in the higher energy (MeV) regime. Extensive studies were also completed for neutron background at that site. It was observed that cosmic muons hitting the 10 cm lead shield used to suppress gammas generate a significant flux of additional neutrons that are produced by spallation. In the same study it was demonstrated that a neutron moderator shield of 30 cm was needed to suppress these neutrons. For low threshold detectors  at the PIP-II beam dump facility we will assume a shallow site with a overburden of approximately 20 m.w.e. with environmental background rates similar to those reported at the Stanford shallow underground facility. The beam timing structure could provide additional suppression of the environmental background.

A realistic simulation is needed of the low energy background associated with the beam to perform more detailed studies.

\begin{figure}[t]
    \centering
    \includegraphics[width=1.0\textwidth]{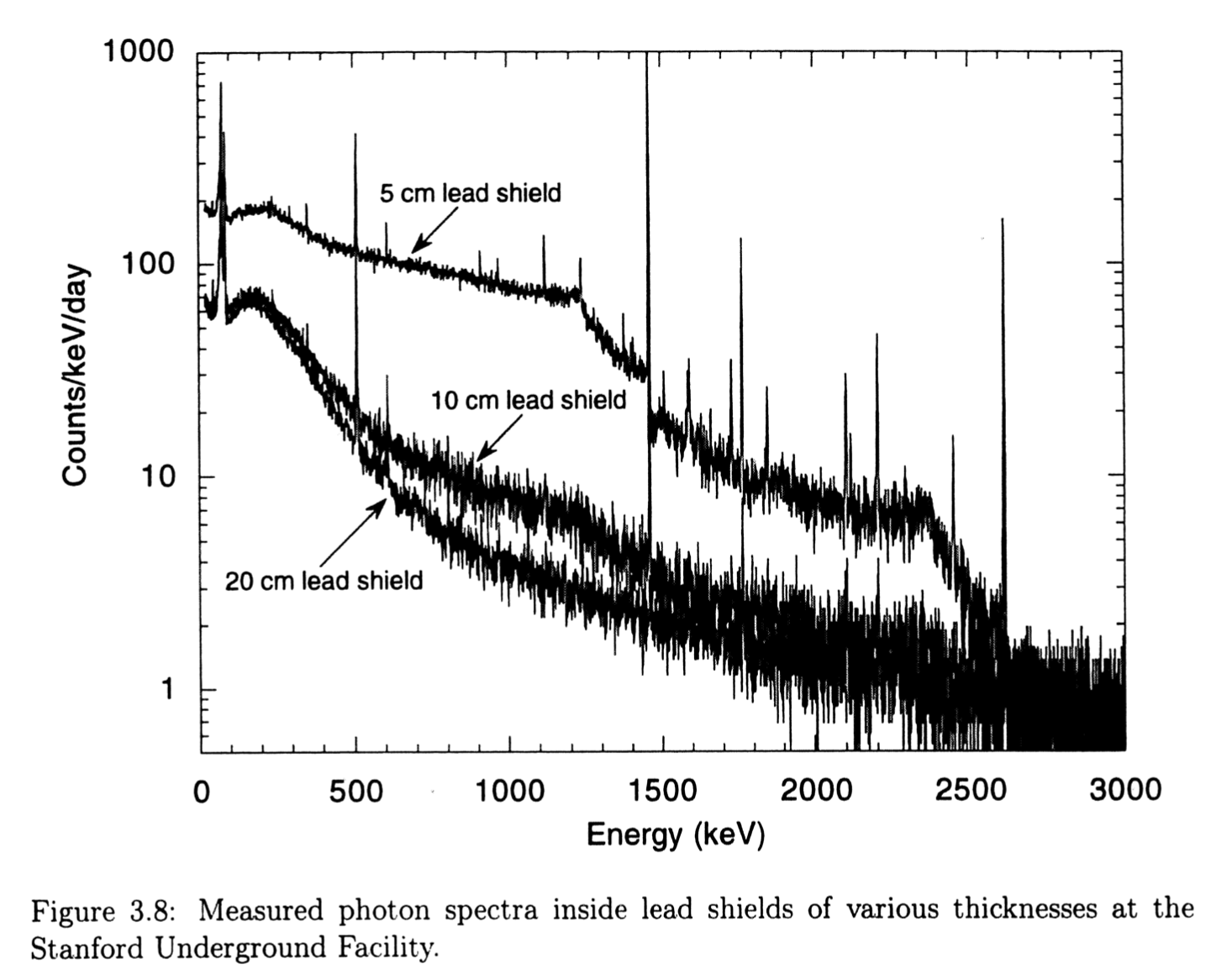}
    \caption{}
    \label{fig:shallow}
\end{figure}

\subsubsection{ALP production and detection}

Proton beam dumps produce not only a large quantity of hadrons such as pions, but also a large quantity of other particles, including photons, electrons, and positrons. Therefore, they also provide excellent opportunities to search for axion-like particles (ALPs). ALPs can be produced in the beam dump via Primakoff processes and propagate to the detector. At the same time, ALPs can also be produced by a Compton-like process. 

Sensor technologies with an eV-energy threshold can benefit from the increase in the number of low-energy ALPs produced at the beam dump as part of the electromagnetic shower. The inverse Compton scattering and the axioelectric channel enable new detection channels for low-energy threshold detectors. 

\subsubsection{DM production and detection}

 Recent results from the COHERENT and Coherent CAPTAIN-Mills experiments [40, 41]
have demonstrated how detectors capable of measuring coherent elastic neutrino-nucleus scattering (CEvNS) with very low energy thresholds can also
be used to set limits on vector portal and leptophobic DM at proton beam dumps. Low-energy threshold detectors enable access to the DM-nucleus scattering channel with energy depositions in the keV range. The conversion efficiency to a detectable signal (for example ionized charge) leaves a signature of less than few hundred of eVee for detection.


%% file: Contributions/keVopp.tex
The proton collisions with a proposed PIP-II beam dump produce both charged and neutral mesons depending on the proton energy. At $<2$~GeV, pion production is the dominant source of mesons. At energies $>2$~GeV which is achieved under some the of ACE scenarios that are described in Sec. \ref{sec:accrings}. Under the beam dump scenario, a large fraction of the mesons decay-at-rest, producing a clean and well-known neutrino spectrum and multiple flavors. The pion decays-at-rest produce a prompt monoenergetic $\nu_{\mu}$ at 30 MeV and a delayed $\nu_e$ and $\bar{\nu_{\mu}}$ with a Michel spectrum that has an endpoint at $\frac{m_{\mu}}{2}$. Having a handle on the meson production is important to reduce flux-related uncertainties.  

The discovery of coherent elastic neutrino-nucleus scattering (CEvNS) by the COHERENT collaboration~\cite{COHERENT:2017ipa, COHERENT:2020iec} opens the door for exciting physics explorations with a keV-scale threshold detector. At a pion decay-at-rest neutrino source, these interactions produce nuclear recoils of $\mathcal{O}$(10~keV). The ability to detect these signatures also opens up the possibilities of dark sector searches and sterile neutrinos where the detectable signature is similar to CEvNS as a low-energy nuclear recoil. It is important to understand 

The addition of an accumulator ring to the PIP-II linac allows optimal sensitivities to light dark matter models described in Sec.~\ref{subsec:newphys}. The bunched structure of the beam allows for a powerful reduction in steady state backgrounds. A beam bunch timing of $\mathcal{O}$(10~ns) as explored for the C-PAR concept allows in some light dark matter models to even separate the boosted dark matter signal from the prompt neutrino. The timing between the prompt and delayed neutrino signals still allows for understanding the scale of the neutrino backgrounds.

A stopped-pion neutrino source also allows for a definitive sterile neutrino search taking advantage of the monoenergetic $\nu_{\mu}$ signature such that the oscillation will vary only on the baseline $L$ assuming two identical detectors are placed into the experimental hall. Searches for an eV-scale sterile neutrino is a current focus of the Short Baseline Neutrino (SBN) Program~\cite{Machado:2019xpc} and one of the main possibilities to explain the anomalies seen in previous short-baseline neutrino experiments such as LSND and MiniBooNE.

%% file: Contributions/MeV-Physics_Experimental_Considerations.tex
Fixed target experiments at PIP-II beam energy scales enable exploration of MeV scale physics which are difficult to access at the collider experiment environments.
The high intensity proton beams in a sufficiently long target or thick dump at the PIP-II generate a large number of mesons and photons that could couple to dark sector mediators and produce dark sector particles (DSP).
This section presents several experimental considerations must be reflected into designing an experiment in order to take full advantage of the capabilities of the facility.
\subsubsection{DSP Signature Categories}~\label{subsubsec:mev-dsp-signatures}
The DSP direct observations from PIP-II beam dump require high beam flux, large-mass high density detector for scattering signatures, large volume low-density detector for decay signatures.
They would also benefit from low energy threshold for expanding search kinematic phase space and underground location or an overburden of 20WME for low cosmic ray background.
Finally, the MeV level neutrinos from the low energy pions in the target or dump could provide an opportunity for BSM signatures based on oscillatory behaviors.

\subsubsection{Signal and Background Considerations} ~\label{subsubsec:mev-sig-bck}
Most signatures for dark sector particle (DSP) discovery include leptons ($e^{\pm}$, $\mu^{\pm}$) and photons in the final states, as described in section~\ref{sec:theory-directions}.
The signals with two EM particle final states from DSPs such as the Axion-like particles (ALPs) or dark photon decays have clear advantage over that of single EM particle final state.
The impact of the $\nu-N$ interaction backgrounds is less for more EM particle final states.
In addition, thanks to the low proton beam energy, the uncertainties in $\nu-N$interaction modeling effect is expected to be small compared to the higher energy beam case, such as in DUNE.

While the number of neutrinos produced from the beam in the target/dump for PIP-II energy level, the beam related neutrons (BRNs) become primary backgrounds  especially for the experiments to be stationed at a shorter distance to the beam source, such as DAMSA described in section~\ref{subsec:damsa}
Therefore, it is essential to factor in the selection of detector technologies and the experimental environments

\paragraph{Neutron Induced Backgrounds and Reduction Considerations}

Neutrons constitute a critical background directly for CEvNS searches via nuclear recoils and more indirectly for the rare event ALP search in DAMSA 
via neutrons capturing on nuclei in the target and surrounding detector materials, including shielding materials. Neutron capture gamma-rays of energies of around $10\,$MeV and even above can be produced \cite{Reichenbacher:2023ncb}, that could be detected as false signals in rare event ALP searches due to the relative poor energy resolution of an economical Ecal detector at these visible energies. 
It is therefore paramount to not only shield neutrons appropriately, but also to a) select the shielding materials accordingly to ensure that capture gamma-ray energies are predominantly below a critical threshold and b) to shield subsequent capture gamma-rays, 
in addition to precise coincidence timing capabilities of the detector with respect to the proton beam on target, in order to reject beam-correlated neutron induced events. 

In addition to the dominant background from beam-correlated neutrons that can have kinetic energies of hundreds of MeV produced via spallation in the target, 
there are also ambient sources of neutron backgrounds.  Ambient neutron sources are primarily neutrons from the hadronic component of the cosmic ray with energies that can extend even beyond the GeV range \cite{Parvu:2021ezc}, but that could be significantly reduced in rate by a thick enough overburden of for example a few meters of dirt. 
Further, there are neutrons with energies in the regime of $10-100\,$MeV induced by the muonic component of cosmic rays via deep inelastic scattering (DIS) and muon capture reactions (muCap) in the detector, its shielding and overburden materials. All ambient cosmic induced neutrons are random in time and cannot entirely be shielded with a reasonable overburden, 
but they could be tagged and vetoed by an active outer muon and proton recoil veto counter system 
made up for example of hydrocarbon scintillators read out by fast PMTs \cite{Reichenbacher:1998ne,KARMEN:2002zcm}. 
Moreover, there is ambient neutron background with a flux of about $10^{-5}\,$neutrons/$cm^2$/sec stemming from the shielding and overburden itself \cite{Reichenbacher:2023ncb}, as they contain largely varying traces of $^{238}U$ and $^{232}Th$. 

Fission neutrons with a few MeV energies are produced by spontaneous fission of $^{238}U$. 
Neutrons with energies up to $10\,$MeV can be produced via ($\alpha$, n) fusion reactions. 
It is important to consider all sources of beam-induced and ambient neutron backgrounds and their subsequent neutron capture gamma-rays, 
and design the target absorber, the detector and its immediate shielding materials, 
as well as the thickness of the overburden accordingly. 
The target absorber should consider efficient neutron absorbers, such as boron, gadolinium or lithium added to polyethylene moderator in order to reduce beam-correlated neutron backgrounds. Further, deep minima in the neutron cross section on e.g. iron in the resonance energy regime of around $100\,$keV have to be considered and mitigated 
by adding layers of different neutron moderator materials \cite{Reichenbacher:2005nc,Reichenbacher:2023nca}. 
In order to reduce cosmic induced neutron backgrounds, a fast enough active muon and proton recoil veto counter system is desired in addition to sufficient passive shielding. Ambient neutron background from spontanous fission of $^{238}U$ and ($\alpha$, n) fusion reactions should be mitigated by pre-assayed shielding material selection, additional neutron moderator shielding towards the surrounding shielding materials, in addition to neutron capture gamma-ray shielding of the inner detectors.

\subsubsection{Beam, Target and Dump Considerations} ~\label{subsubsec:mev-bm-tgt-dmp}
The primary goal of the beam, target and dump selection is to ensure creating as many source particles for signal production as possible, while enabling minimizing the source of backgrounds.
Some of the considerations for the beam are (1) the optimal $E_{p}$, (2) beam timing structure such as continuous wave versus pulses, why and at what spacing, (3) optimal beam transverse size and why (4) can the accelerator meet these requirements at a reasonable cost and timescale?
Considerations for the target are (1) what would be the optimal target material and its dimensions, (2) would dump work better?
Finally, if a good combination of the beam, target and dump is accomplished, it becomes a facility.  Therefore, the natural question is what other physics can be explored.
Given the opportunities that PIP-II facility provides, it would be ideal to have a common study tool infrastructure for these can be performed.

\subsubsection{Detector Considerations}
The primary goal of the detector is to ensure capturing the signal at the high efficiency, while enabling backgrounds mitigation at the hardware and analysis level.
The characteristics of detector depends on the signal and the kinematic phase space of the DSP. 
The decay signatures requires large volume, low density detector.
What would be the most optimal dimensions of the detector?
Can we implement decay volume in the upstream of the detector?
What would be the thinnest wall thickness that can help minimizing background interactions, such as those from BRNs?

Some of capabilities of the detector to accomplish the above goals must be considered are (1) the position, momentum and energy resolutions for MeV range signals whose final state particle energies are of the order few 100s of MeV, (2) are the good timing capabilities needed? How good do they have to be and why? (2)low energy threshold capability is definitely advantageous for the expansion of kinematic phase space and (4) What would be the most optimal material and the dimensions of the detector to accomplish the goal?

\subsubsection{Detector Technologies}
For low $E_{p}$ beams, such as 800MeV PIP-II LINAC onto beam dump, BRNs rather than the beam generated neutrinos would become the primary background.
Taking into account the signal final states, the following detector capabilities are needed:
\begin{enumerate}
    \item Excellent EM particle identification -- LArTPC, fine granular total absorption calorimetry, magnetized detectors sandwiched with a few layers of silicon detector
    \item High position and angular resolution -- LArTPC
    \item Low energy threshold -- LArTPC, total absorption calorimetry
    \item Fine energy, momentum and invariant mass resolution -- fine granular total absorption calorimetry, LArTPC, gas TPC 
    \item High precision vertex pointing capability -- Some combination of precision si layers sandwiched in longitudinal layers of fine granular total absorption calorimeter
    \item Fast timing (ideally at sub-ns level) with a well separated beam pulses helps a great deal in beating down the BRN backgrounds -- Crystal or Scintillator based photon detectors
    \begin{itemize}
    \item The necessary timing depends heavily on the distance from the beam source to the detector
    \item For 10m distance, it takes ~30ns for the speed of light -- This parameter depends heavily on the mass of the signal particle
    \end{itemize}
\end{enumerate}

\subsubsection{Potential MeV Scale Detectors for PIP-II}
LArTPC detectors provide precision 3D image along with energy measurements.
Given that several versions of these detectors have been built and operating, LArTPC is an excellent candidate to meet the timeline for the completion of the PIP-II LINAC.
In particular, the $2\times 2$ prototype DUNE LArTPC near detector which is to be operating soon has a pixelated PCB read out through LArPIX cold electronics chip.
A new version of cold readout electronics called the Q-Pix chip which is equipped with self-triggering capability is under intense R\&D. 
Given these, a similar version of a pixelated LArTPC detector read out by a Q-Pix chip could be ready in time for the 2029 time scale when the PIP-II LINAC is to complete.

Another detector that can meet the detector considerations discussed in the previous sections is the DAMSA experimental concept which utilizes a fast timing fine granular total absorption EM calorimeter, described in detail in section~\ref{subsec:damsa}.
Such calorimeter can be made of the standard scintillation counters read out by the SiPM's and could be augmented by a small number of silicon layers that can provide precision vertex pointing capability.

%% file: Contributions/CCDs.tex
Charge-Coupled Devices (CCDs) are pixelated semiconductor sensors that are commonly made of silicon. Ionizing radiation interacting in the CCD substrate generates electron-hole pairs. A substrate voltage is applied to drift the generated charge carriers towards the CCD surface, minimizing charge recombination. Potential wells under biased metal electrode structures at the sensor's surface allow for charge collection. During readout, charge is transferred pixel by pixel along each parallel register (column of pixels) in the CCD towards the serial register (last row of pixels) by sequentially clocking the potentials applied to the surface electrodes. Then, charge is transferred along the serial register towards the CCD output stage where charge is read.

In conventional scientific CCDs, widely used in astronomy due to their low readout noise ($\sim$3.5~$e^-$), the common floating-diffusion output stage only allows for charge destructive readout. The new-generation skipper-CCDs implement a floating-gate output stage where multiple ($N_{skp}$) non-destructive measurements of the same charge packet are allowed. By averaging the $N_{skp}$ independent samples off-chip, the impact of the low-frequency noise is greatly reduced, and the readout noise decreases as $\sigma=\sigma_1/\sqrt{N_{skp}}$, where $\sigma_1$ corresponds to the readout noise of one sample. By increasing $N_{skp}$, sub-electron noise can be achieved, allowing to precisely count the number of electrons in each charge packet. However, as the sensor's readout time is proportional to $N_{skp}$, performing multiple measurements limits the sensor's time resolution. Performing smart readout~\cite{2020smartskipper} has been proposed to decrease the readout time of the skipper-CCDs while maintaining their electron-counting capability. Some other efforts related to CCD technology which aim to increase the readout rate are also being explored, such as CCDs with Single electron Sensitive ReadOut (SiSeRO) stages, Multi-Amplifier Sensing (MAS) CCDs and CMOS with skipper output stages~\cite{Sofotalk2023}.

The Skipper-CCD technology has been demonstrated to be highly competitive in searching for sub-GeV DM, leading to the current state-of-the-art limits in several DM-electron interactions for masses below $\sim$5~MeV~\cite{SENSEI:2020dpa, DAMICM2023}. Several active and planned experiments form part of the ongoing effort to search for sub-GeV DM with skipper-CCDs, aiming to increase their sensitivities through understanding and reducing their low-energy backgrounds and having larger detector masses. Within these, the most ambitious one is Oscura, planning to deploy a low-background $\sim$10-kg skipper-CCD detector at SNOLAB by 2028. New ideas on sensors packaging, cryogenics and electronics for Oscura have been developed during its R\&D stage. Derived from these ideas, we put together the largest skipper-CCD instrument ever built, in terms of active mass ($\sim$80 g) and number of channels (160). Details on this system, currently operating at FNAL with single-electron resolution, can be found in Ref.~\cite{Chierchie2023}. This massive instrument, or a similar one, can be used for beam-dump searches.

\subsubsection{Skipper-CCD detector at MINOS}
SENSEI running at MINOS has recently demonstrated to have the world-leading exclusion limit for millicharged particles (mCPs) between 30 and 380~MeV using this technology~\cite{barak2023sensei}. Motivated by these results and the development of the Oscura experiment, an engineering test conducted with an active mass of 1 kg of Skipper-CCD was proposed to investigate millicharged particles by utilizing the NuMI beam with the detector located in the MINOS underground area~\cite{perez2023early}.
Preliminary estimations indicate that this engineering run could become a pioneering endeavor in the exploration of low-mass mCPs, with the potential to provide world-leading limits.

Various strategies were assessed ranging from the most basic scenario, which only considers single tracks, to more advanced approaches that treat tracks with two or more hits as potential mCP signals. The former strategy proves the most effective in the low mass range. However, as the mCP mass increases to above 200 MeV, the background becomes more prominent and the resilience of tracks becomes a more competitive factor.

\subsubsection{Skipper-CCD detector at PIP-II}
\input{Contributions/mCP_eVthresh_SkipperCCDs}

%% file: Contributions/mCP_eVthresh_SkipperCCDs.tex
\begin{figure}[H]
    \centering
    \includegraphics[scale=0.5, angle=-90]{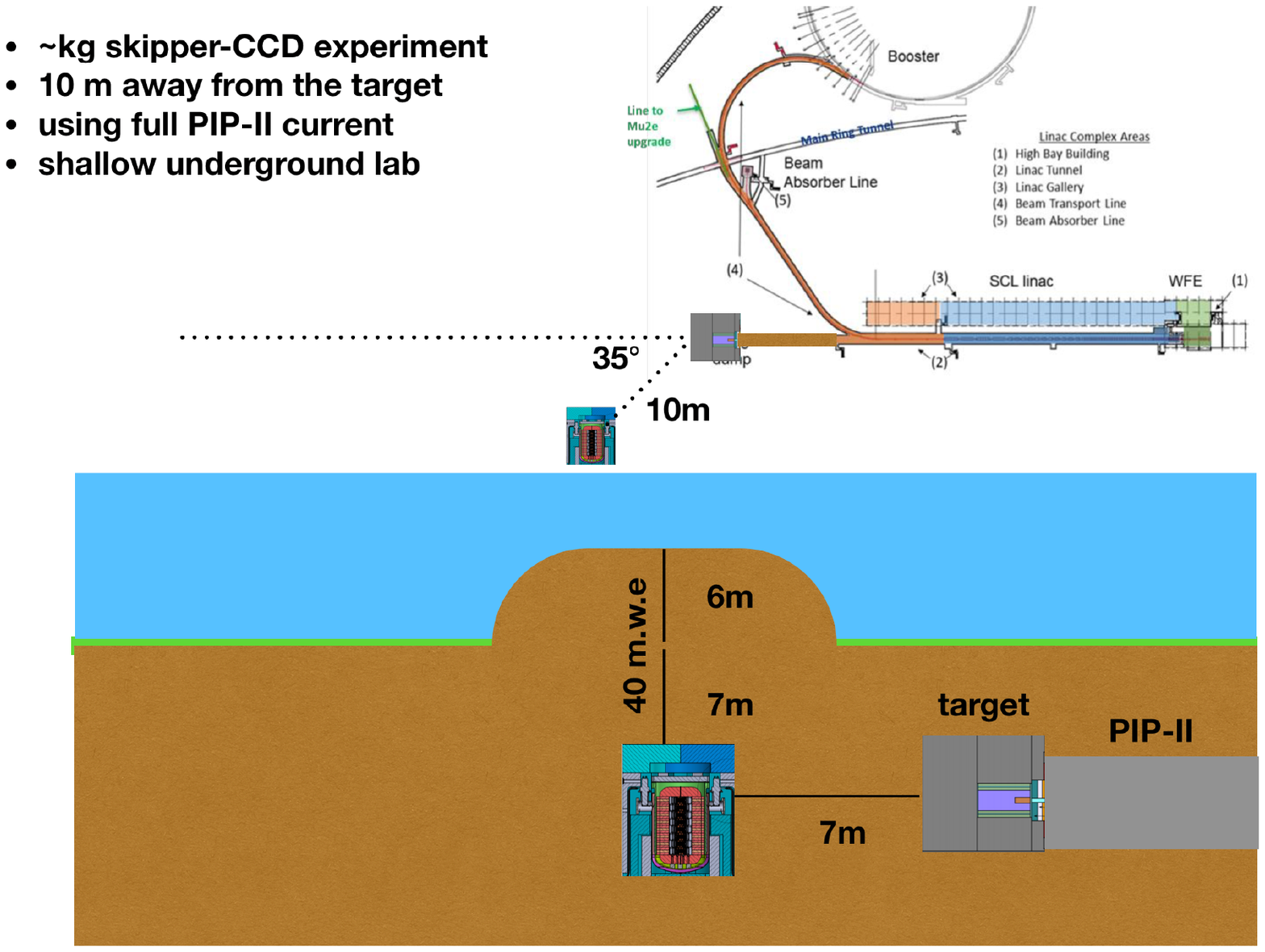}
    \caption{Concept for a kg scale skipper-CCD off-axis experiment in the PIP-II beam dump.}
    \label{Fig:pip2skipper}
\end{figure}

\begin{figure}[H]
    \centering
    \includegraphics[scale=0.35]{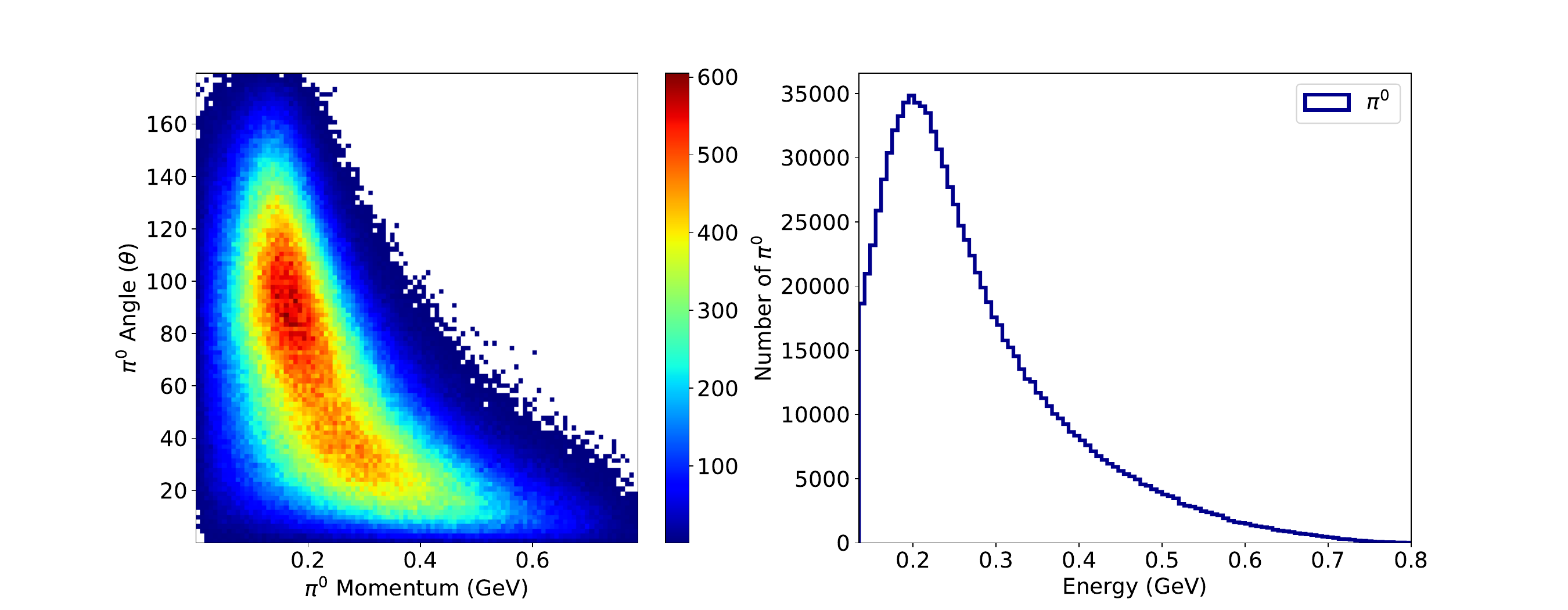}
    \caption{Left: Momentum-angle distributions for $\pi^0$ being produced by a $800$~MeV proton beam hitting a fixed Carbon target. The color scale is a relative measure of how many pions have the corresponding angle and momentum. Right: Energy spectra for neutral Pions coming out of the beam dump. }
    \label{Fig:pion_dist}
\end{figure}

We simulated $\pi^0$ mesons generated by protons striking a fixed target using the BdMNC code~\cite{deNiverville:2015mwa}. We analyzed the case of a $800$~MeV proton beam using the Burman and Smith parametrization of the expected $\pi^0$ distribution~\cite{}. This distribution depends on the number of protons and neutrons in the target. The simulations assumed a Carbon target, $Z=14$, and simulated $10^6 \pi^0$ being created at the beam dump. The two-dimensional momentum-angle distributions for this case is shown in Fig.~\ref{Fig:pion_dist}.

As can be noted from Fig.~\ref{Fig:pion_dist} (left), a lot of the pions being produced by the proton beam have an angular dispersion higher than 1 radian or 60 degrees with respect to the beam axis. The energy distribution of these pions is shown in Fig.~\ref{Fig:pion_dist} (right).

Neutral meson decay is the main production channel for millicharged particles and specifically in the case of the $\pi^0$, millicharged particles are produced from $\pi^0\rightarrow\chi\bar{\chi}\gamma$ which are analogous to Dalitz decays. For these three body decays the branching ratio can be related to that of $\pi^0\rightarrow\gamma\gamma$ using

\begin{equation}
    \text{Br}(\pi^0\rightarrow\chi\bar{\chi}\gamma)=2\varepsilon^2\alpha_{EM}\text{Br}(\pi^0\rightarrow\gamma\gamma)I^{(3)}\Big(\frac{m^2_{\chi}}{m^2_{\pi^0}}\Big)
    \label{branching_ratio}
\end{equation}

where $\varepsilon$ is the millicharge or the coupling of these particles to the SM photon, $\text{Br}(\pi^0\rightarrow\gamma\gamma)=0.988$, and $I^{(3)}(x)$ is a dimensionless function defined as

\begin{equation}
    I^{(3)}(x)=\frac{2}{3\pi}\int^1_{4x}dz\sqrt{1-\frac{4x}{z}}\frac{(1-z)^3}{z^2}(2x+z)
\end{equation}

\begin{figure}[H]
    \centering
    \includegraphics[scale=0.35]{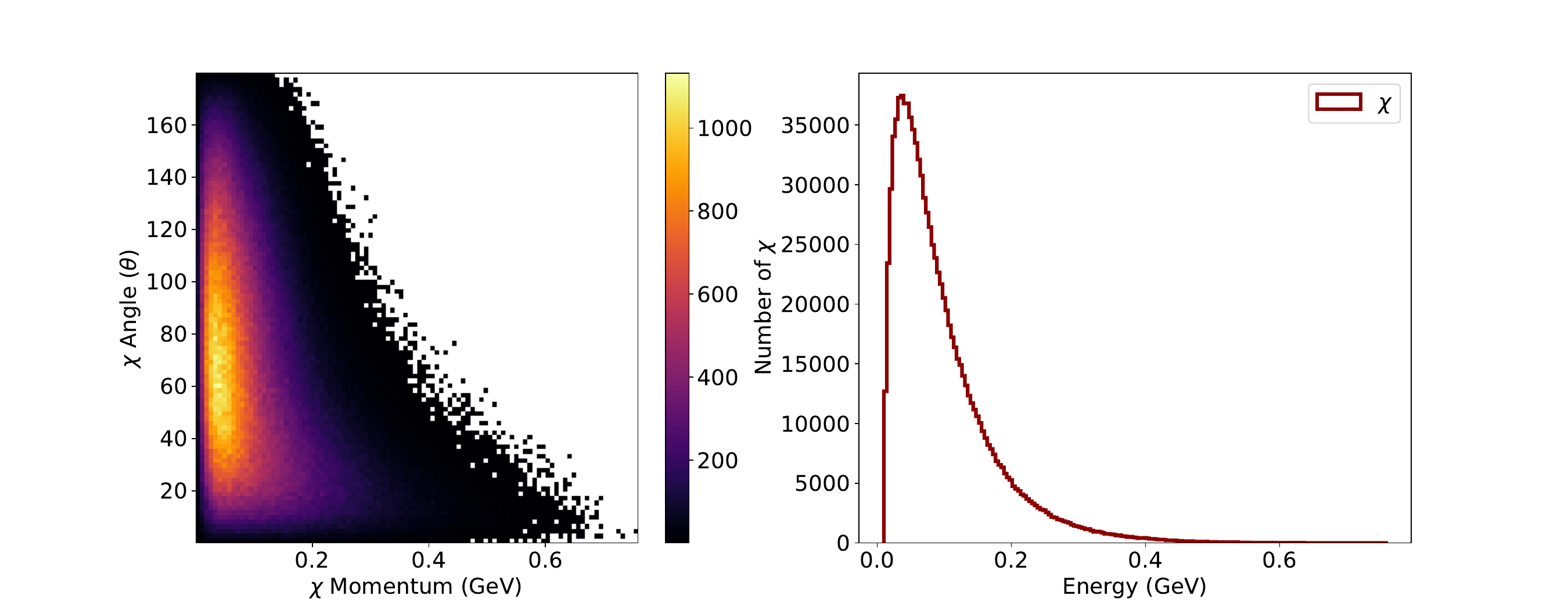}
    \caption{Left: Momentum-angle distributions for mCPs $\chi$ after pion decays, the angle is measured relative to the beam axis. The color scale is a relative measure of how many particles have the corresponding angle and momentum. Right: Energy spectra of produced millicharged particles from pion decays at a mass of 10MeV.}
    \label{Fig:milli_dist}
\end{figure}

In order to carry out $\pi^0$ decays into millicharged pairs we used the momentum of pions provided by BdMNC in combination with the python package \textit{phasespace}~\cite{Navarro2019} 
Decay kinematics can easily be taken into account for different $\chi$ masses using \textit{phasepace}. The two-dimensional momentum-angle distribution of mCPs coming from $\pi^0$ decays is shown in Fig.~\ref{Fig:milli_dist}.

\begin{figure}[H]
    \centering
    \includegraphics[scale=0.35]{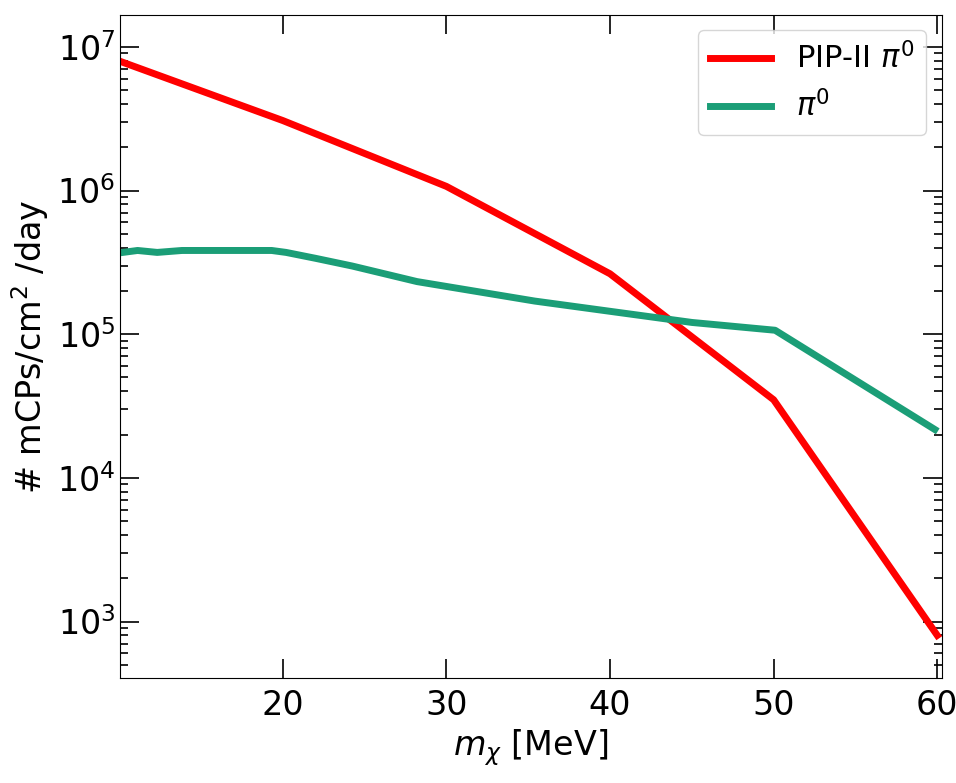}
    \caption{Millicharged flux arriving to a detector placed 10m away from a fixed Target at PIP - II, compared with the amount of mCPs arriving to the SENSEI detector placed in the NuMI beamline.}
    \label{Fig:milli_flux}
\end{figure}

As we can see from Fig.~\ref{Fig:milli_dist}, mCPs have a relatively large angular distribution. This means that in order to maximize the expected signal for a dark sector search experiment, the detector should be placed off axis. The optimized location for a detector hunting for these types of particles would be around 35 or 36 degrees. For a detector being placed at 35 degrees we expect around 9600 mCPs coming from the $10^6$ pions being generated from the target without taking into account the branching ratio. This means that we expect only 0.96\% from the particles being generated to reach the detector if none of them are deflected or their energy is affected by materials between the target and the detector. For this case, energy losses and deflections are important and need to be taken into account more carefully due to the fact that these mCPs have order MeV of energy, Fig.~\ref{Fig:milli_dist}. Following \cite{harnik2019millicharged}, the deflection angle is proportional to the millicharge $\varepsilon$ and inversely proportional to the energy of the mCPs. If the detector is placed near the target, then we expect a higher flux of mCPs to reach the detector but that also creates challenges when trying to implement tracking strategies to mitigate backgrounds.
To quantify the mCP flux in terms of the area of a cone subtended from 35 to 36 degrees we simply integrate over the expected flux. The total expected rate of mCPs is also affected by the millicharge being considered and the mass of the particle as described by Eq.~\ref{branching_ratio}. In Fig.~\ref{Fig:milli_flux}, we show the expected rate of mCPs for different masses for a detector placed 10 meters away from the target, 0.1 $\pi^0$ per POT and $10^{21}$ POT per day.

\begin{figure}[H]
    \centering
    \hspace*{-2.1cm}
    \includegraphics[scale=0.3]{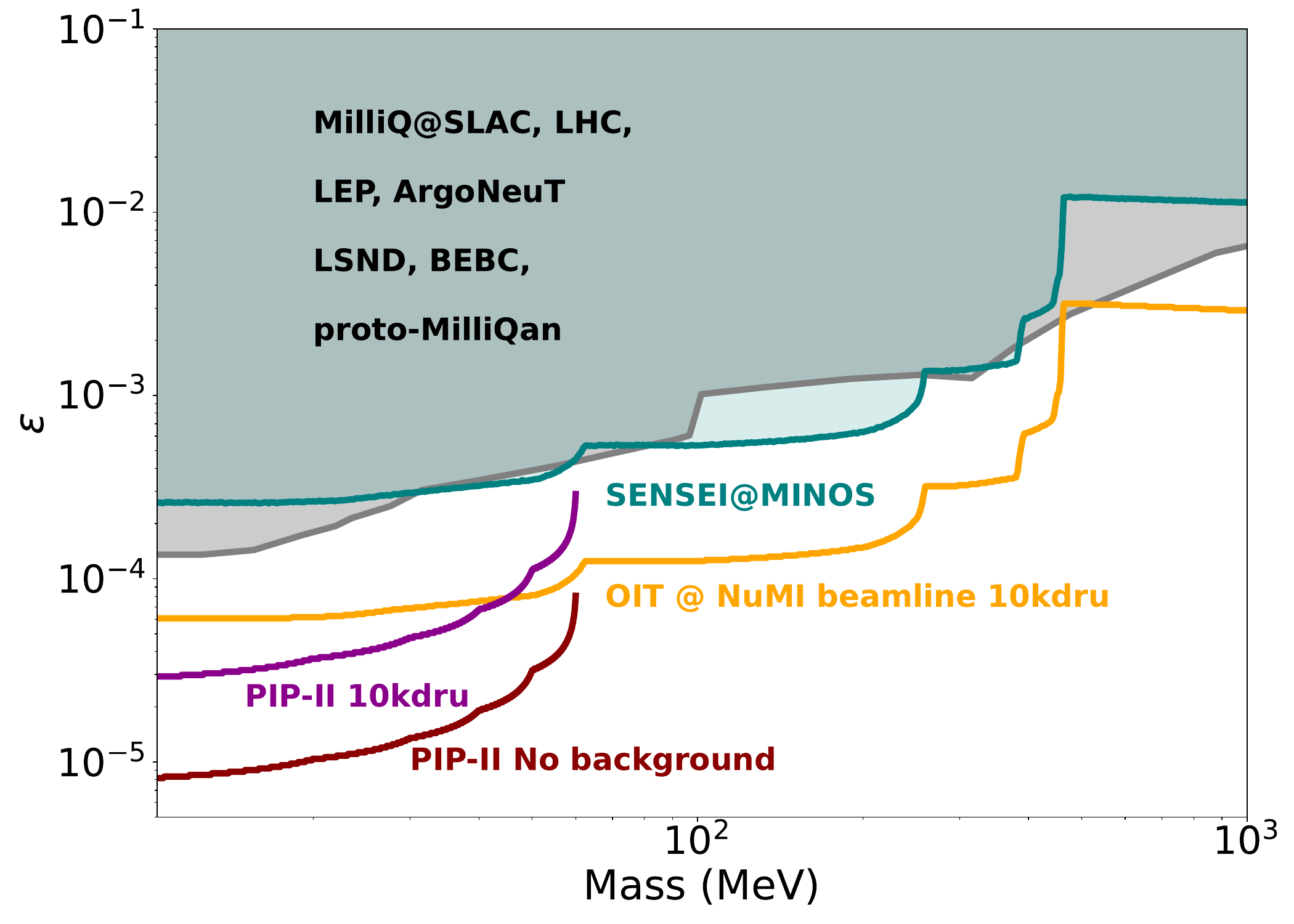}
    \caption{Projected constraints for millicharged and mass of mCPs with a 90\% C.L comparing with previous experiments, and the projections for the Oscura Integration Test installation at the NuMI beamline. Both projections are assuming 10~kdru, which is the level of background reached by surface experiments and a 1kg-year total exposure. In cyan the curve represents the 95\% C.L limit obtained by SENSEI using a Skipper-CCD in the NuMI beamline \cite{SENSEI:2023gie}.}
    \label{Fig:Limits}
\end{figure}

Skipper-CCDs are a demonstrated technology able to hunt for these particles with an unprecedented sensitivity. Using what we learned from previous searches, we can calculate the sensitivity to mCPs for a 1kg-year Skipper-CCD experiment placed near the PIP-II target. An assumption that most mCPs arrive at the detector with $\beta\approx1$, breaks down as the mCP mass reaches higher energies. The results are shown in comparison with previous mCP searches and the Oscura Integration Test projection (OIT) in Fig.~\ref{Fig:Limits}

\subsubsection{Synergy with LHC}

\begin{figure}[h]
\centering
\subfloat[PEP]{
\includegraphics[width=0.5\textwidth,height=5.3cm]{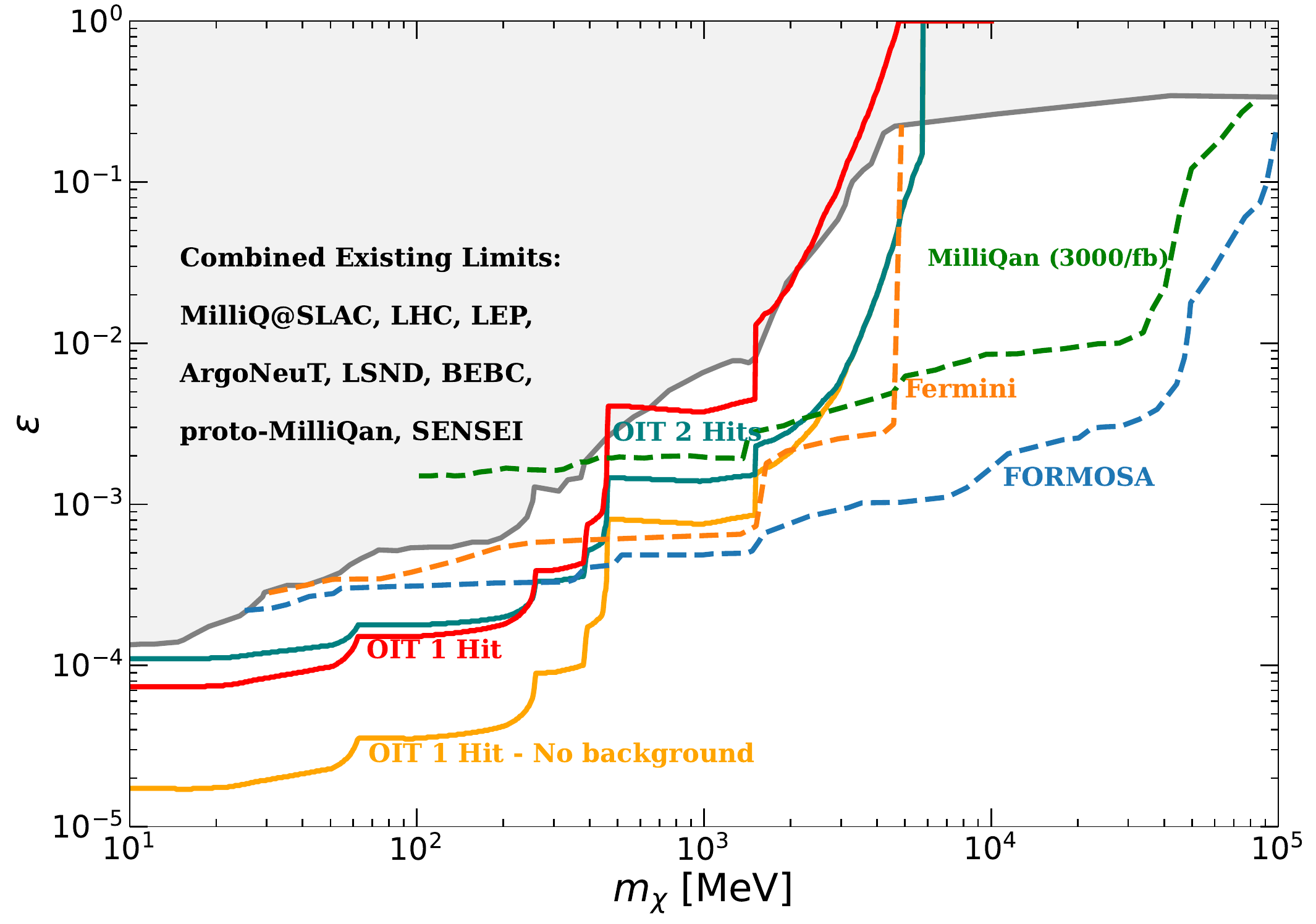}}
\subfloat[LHC]{
\includegraphics[width=0.5\textwidth,height=5.3cm]{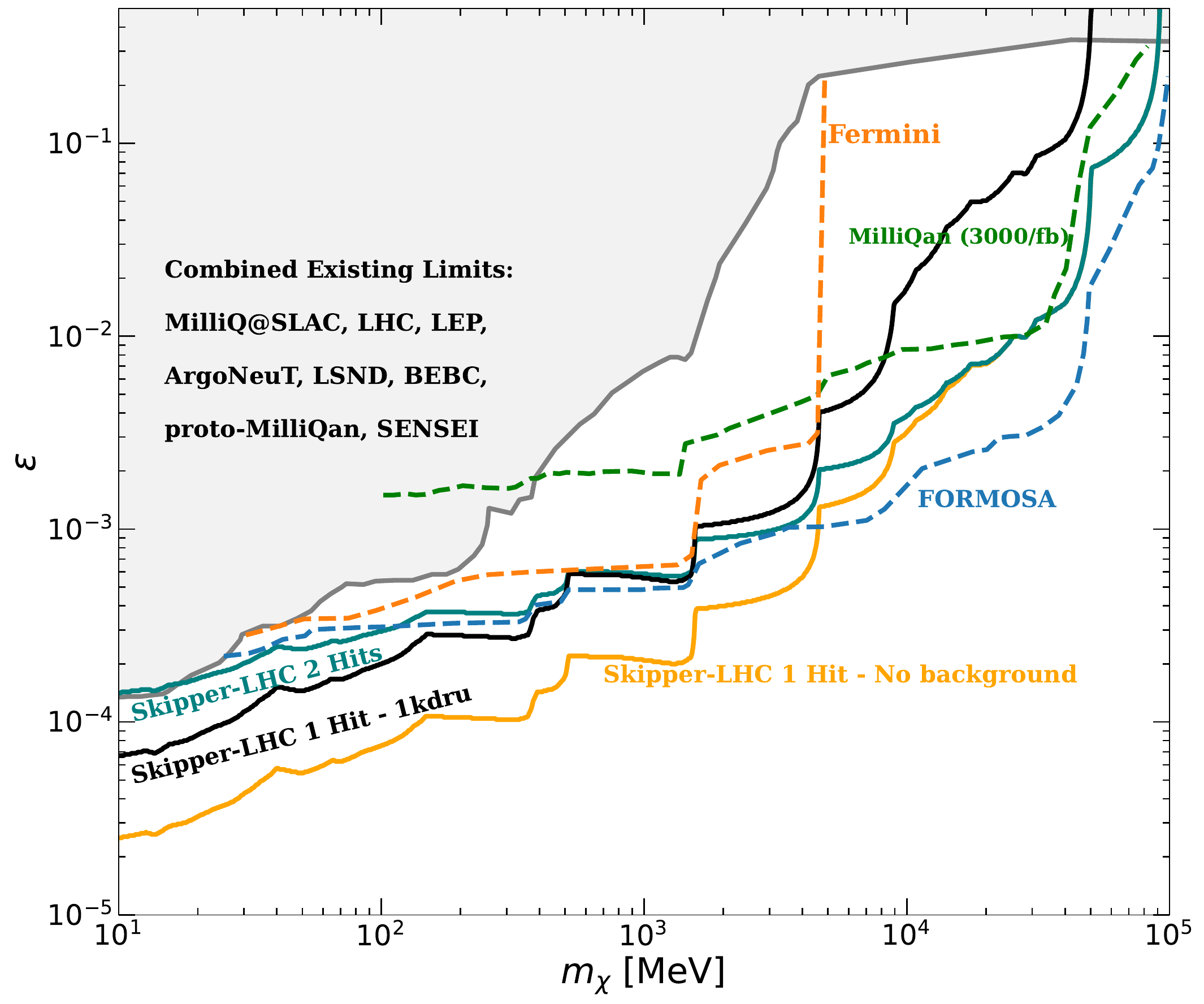}}
\caption{Projection for mCPs sensitivity placed in a 120 GeV proton beam for Skipper-CCDs over a range of detector masses on the left, and the projection using the expected luminosity delivered by HL-LHC of $5\cdot10^{34}/cm^2/seg$. The plot also shows the results from previous experiments and projections from MilliQan at LHC, FORMOSA at HL-LHC and FerMINI at NuMI/DUNE sites}
\label{Fig:Zhenliu}
\end{figure}

Skipper-CCDs have shown to be a very competitive technology for searching for mCPs using high-intensity proton beams as described in the previous section. The projections for Skipper-CCD detectors given enough detector mass could surpass the projects from FORMOSA~\cite{FORMOSA} or FerMINI~\cite{FerMINI}, shown in Figure ~\ref{Fig:Zhenliu}, until the sensitivity rapidly degrades for mCP masses of $\geq 0.4$~GeV. 

The 14~TeV proton beam at the HL-LHC could cover the high mass mCP search region. Currently, there is one experiment designed to search for mCPs running at LHC called milliQan ~\cite{milliqan}. A previous experiment, the "milliQan demonstrator” ~\cite{PhysRevD.102.032002}, was located at the CMS experimental site and aligned with the CMS interaction point(IP) 70 meters underground. The detector is shielded for most particles produced in the LHC collisions due to the 17~m of rock between the CMS IP and the demonstrator. About one percent of the full milliQan detector and 37.5~fb$^{-1}$ of data collected already provides competitive constraints on millicharged particles with a mass of 20–4700~MeV/c$^2$ and Q=e $\sim$ 0.01–0.3. This result also shows the range improvements achievable with more energetic beams. An upgrade to this experiment is currently running at the LHC and is planned for the HL-LHC with 3000~fb${^-1}$. 

Recently, another proposal called FORMOSA with similar technology than milliQan has been done. It will be located at a different place in the ring, at 480 meters downstream from the ATLAS interaction point in UJ12 for FORMOSA-I. Because FORMOSA looks at the forward detector instead of the central one, there is up to a factor of ~250 higher mCP rate compared to milliQan, but the backgrounds to worry about are different (cosmic vs beam induced).
There are also plans for a Forward Physics Facility (FPF)~\cite{FPF}, located several hundred meters from the ATLAS interaction point and shielded by concrete and rock. FPF will host a suite of experiments to probe standard model (SM) processes and search for physics beyond the standard model (BSM).
If we are able to install a few skipper CCD modules at this new facility FPF, we could improve the limit based on the projections at low-energy beams and shown on the right side of Fig.~\ref{Fig:Zhenliu}. The FPF is planned to be built during long shutdown 3 from 2026–28, with support services and experiments being installed starting in 2029. This provides an opportunity to propose and install the CCDs modules there by 2029. The schedule is very similar to the simple PIP-II which will allow to explore the full mass range at same time.
So far, there is a big gap for heavier ($\sim$GeV) low-charged particles, which requires a dedicated experiment to be covered. Adding the skipper CCDs to the already programmed experiments at LHC could help a lot to this. There will have to be a big effort in studying the backgrounds in that environment different from PIP-II, for instance the production of neutrinos and muons from the collisions or others. 
The right side of Fig.~\ref{Fig:Zhenliu} shows the prediction for 1kg of skippers CCDs located at FPF. This technology extend the sensitivity region further than any other experiment up to masses of almost 1GeV including 1000 dru of background and up to 5 GeV without any background. 

%% file: Contributions/sbc.tex
\begin{figure}
\begin{center}
    \includegraphics[height=3in]{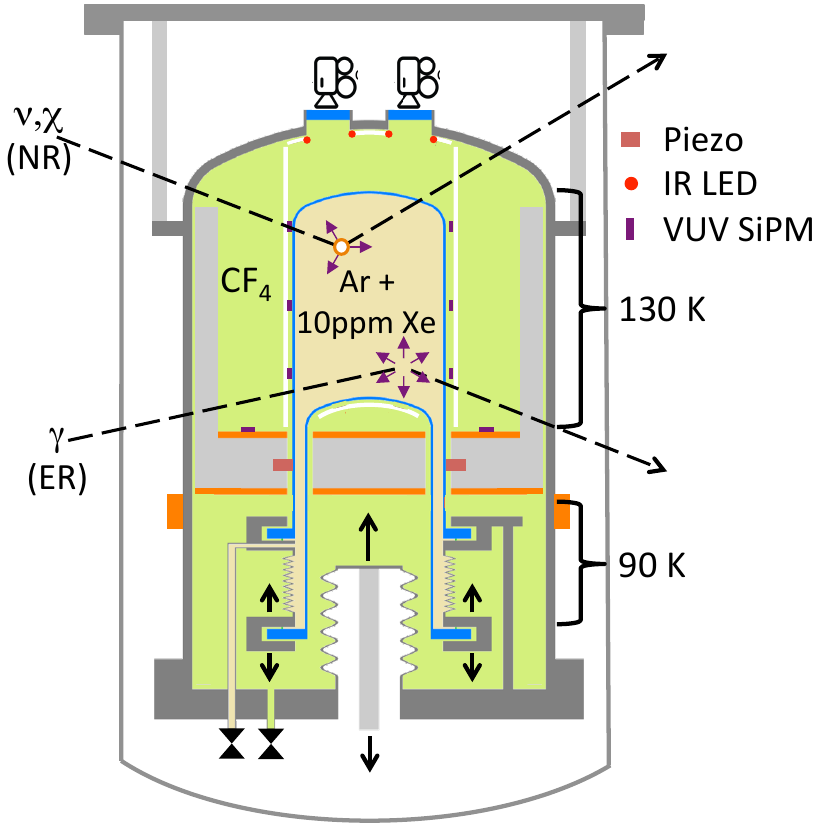}
    \includegraphics[height=3in]{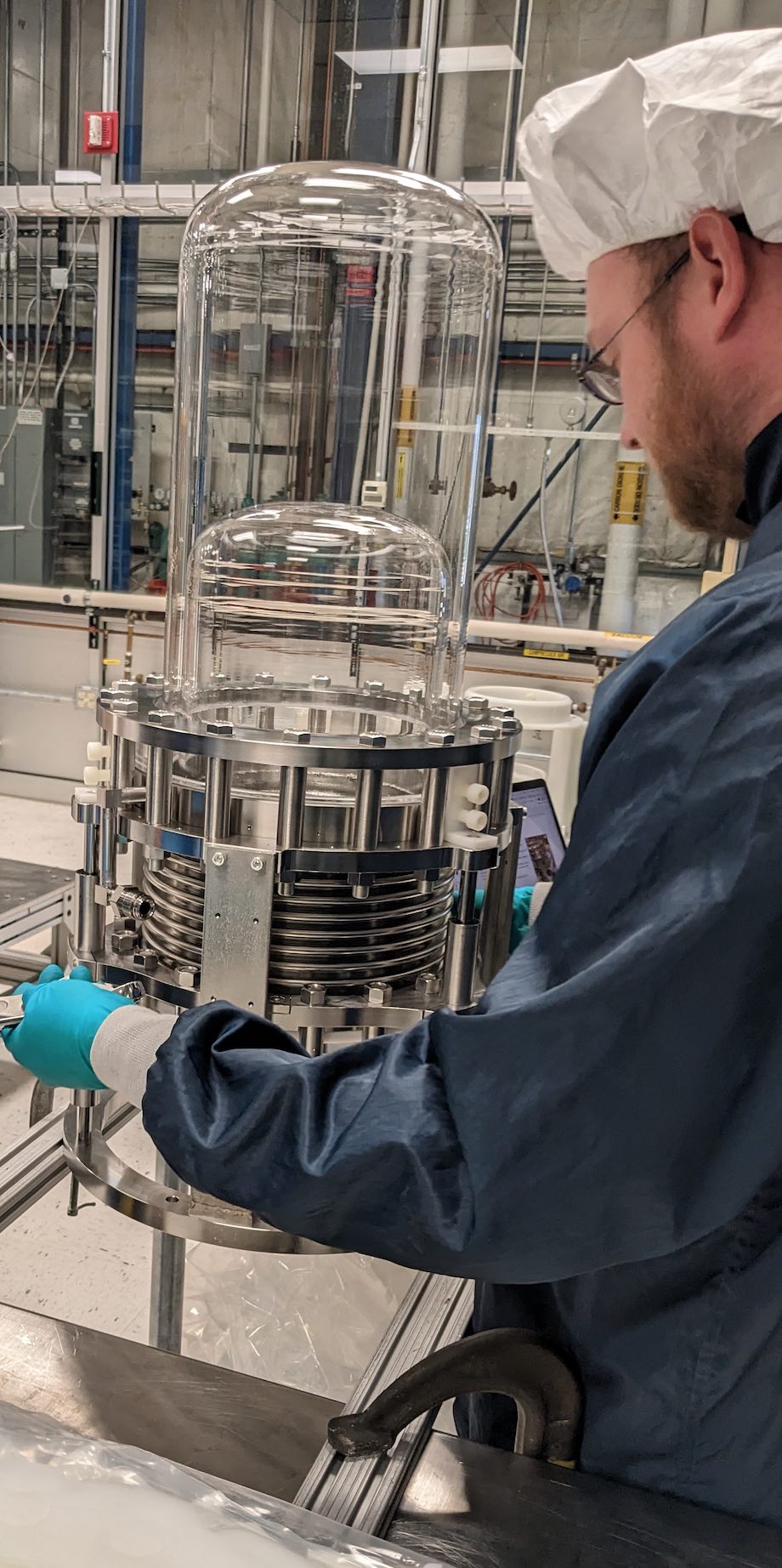}
    \includegraphics[height=3in]{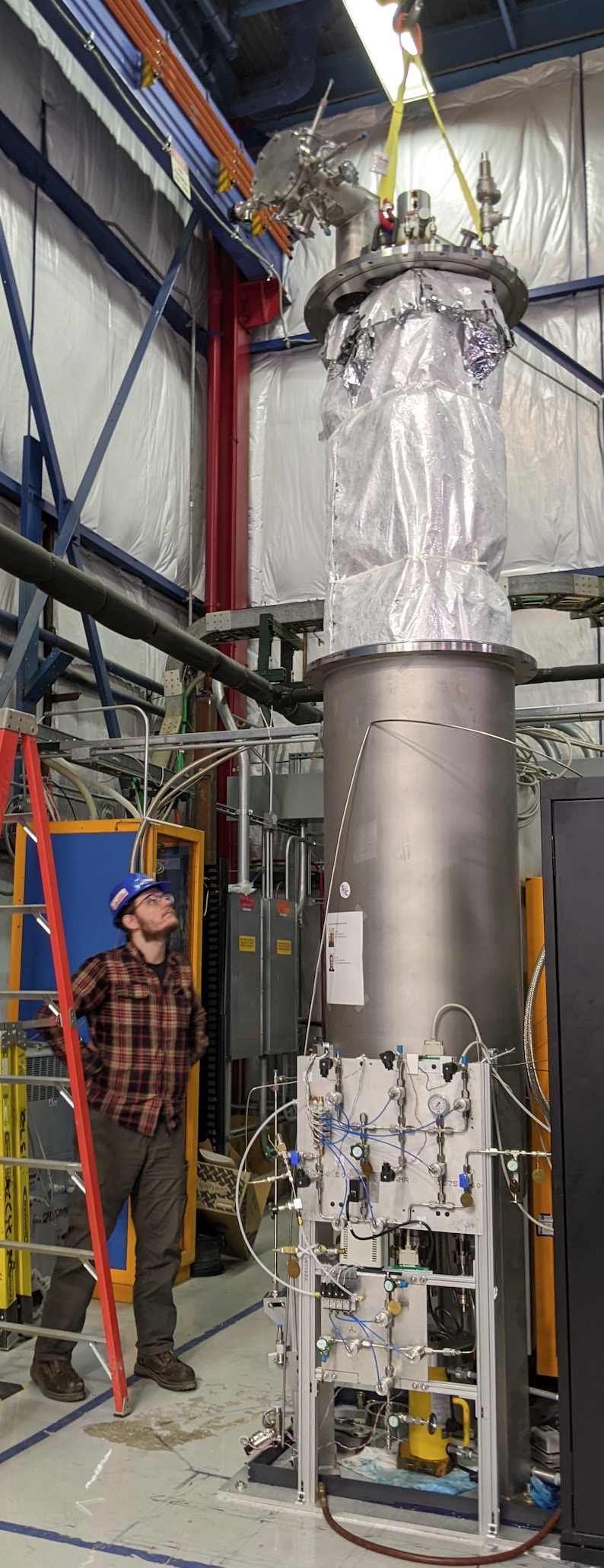}    
    \caption{\label{F:SBC} Schematic and images of the SBC-LAr10 bubble chamber \cite{Alfonso-Pita:2022akn}.  Nuclear recoils from neutrons, neutrinos or dark sector particles nucleate a single bubble in the bubble chamber with a coincident flash of scintillation light.  Electron recoils create scintillation light only.}
\end{center}
\end{figure}

Bubble chambers offer a powerful tool for detecting low-energy nuclear recoils, particularly when faced with large electron-recoil backgrounds:  the COUPP and PICO dark matter detection experiments have demonstrated $\mathcal{O}(10^{10})$ electron recoil rejection \cite{PICO:2019rsv}, operating in modes where every nuclear recoil above threshold creates a single bubble in the detector, but electron recoils create no bubbles at all.  Liquid-noble bubble chambers enhance this discrimination, lowering the minimum threshold for a nuclear-recoil-only search from $\mathcal{O}$(keV) to $\mathcal{O}$(100-eV), and at the same time add a scintillation signal can can be used for energy reconstruction at recoil energies above a few keV \cite{Baxter:2017ozv,BresslerPhD}.  The first physics-scale demonstration of the liquid-noble bubble chamber technique will be SBC-LAr10 (Fig.~\ref{F:SBC}), a 10-kg argon bubble chamber set to turn on in the MINOS tunnel at Fermilab in 2024 \cite{Alfonso-Pita:2022akn}, performing precision calibrations of sensitivity to sub-keV nuclear recoils with a target threshold of 100 eV.

The existing SBC physics program is focused on light (1--10 GeV) dark matter direct detection and reactor neutrino CE$\nu$NS measurements, \cite{Alfonso-Pita:2022akn,alfonso-pitaNewPhysicsSearches2022,cennstheorygroupatif-unamPhysicsReachLow2021}, but these detectors are also unique and powerful tools at a beam dump facility.  Their excellent background rejection (from the combination of ER-discrimination, mm-resolution 3-D position reconstruction, and scintillation-based energy reconstruction) enables a stopped-pion CE$\nu$NS measurement even in continuous-wave mode, while in pulsed mode the timing from the scintillation signal ($\sim$10-ns resolution) can be used to identify prompt relativistic particles.  This allows for example the separation of the prompt mono-energetic pion-decay neutrino from delayed muon-decay neutrinos.  The SBC chamber design also accommodates a variety of targets, allowing measurements of neutrino interactions not just in noble liquids (Ar, Xe) but also in other inert liquids such as N$_2$ and CF$_4$ --- targets that give unique sensitivity to axial-vector coupling, but that scintillate too weakly to show the CE$\nu$NS signal in any channel besides bubble nucleation.  Finally, bubble chambers are scalable.  The design of the 10-kg pathfinder scales readily to a 1-ton target.

Operating a bubble chamber near a beam dump will require a shallow underground site, as the bubble chamber is intrinsically a low-rate detector, capable of detecting only $\mathcal{O}(10^3)$ nuclear recoils per day per device.  An overburden of several meters water equivalent to eliminate the hadronic portion of cosmic ray showers is both necessary and sufficient for bubble chamber operation.

%% file: Contributions/LArTPC.tex
A strong physics motivation to deploy a liquid-argon time-projection chamber (LArTPC)
at the PIP-II Beam Dump facility is to provide important supporting measurements
for DUNE.
One of DUNE's primary physics goals is detection of neutrinos produced by
core-collapse supernova explosions.
These neutrinos, with energy of a few to a few tens of MeV,
carry 99\% of the gravitational potential energy of the supernova,
and are uniquely suited to characterize the features of the supernova and
its explosion mechanism.
They also allow measurements for studying particle physics, such as
neutrino mass hierarchy and non-standard interactions from neutrino mixing
in extreme conditions which cannot be obtained in terrestrial laboratories.
In addition, measurements of supernova neutrino fluxes can be used to constrain
a swath of parameter space of physics beyond the Standard Model (BSM).
Given that detection of all flavors of supernova neutrinos is desired,
DUNE, based on the LArTPC technology, is uniquely sensitive to electron
neutrinos by the charged-current (CC) interaction,
and will provide complementary information to other massive neutrino
detectors, which are mainly sensitive to $\overline{\nu}_e$.

The detected neutrino fluxes are convolved with neutrino cross sections and
detection resolution, while the original neutrino fluxes are of interests in
astrophysics and particle physics.
To disentangle the neutrino fluxes, it is important to understand the neutrino
cross sections and detector resolution.
However, the $\nue$--Ar CC cross sections at the energy regime of supernove neutrinos
have never been measured,
and the uncertainties owing to the large variation from different theoretical models
are therefore relevant.

The recent released paper from the DUNE collaboration suggests that with a
20\% precision of total $\nue$--Ar CC cross section,
it is achievable to obtain a decent measurement of supernova neutrino flux
parameters in DUNE.
An ongoing study confirms this statement, and further indicates that measurements
of the $\nue$--Ar CC cross section with a well-characterized artificial neutrino
source and a detector functionally equivalent to the DUNE far detector
can significantly improve the supernova neutrino flux measurements in DUNE.
While the statistical uncertainties are not considered in both the studies,
they strongly motivate precise measurements of the $\nue$--Ar CC cross
sections in a few tens MeV, which aligns with the energy range of neutrinos
produced from pions decay at rest.

LArTPCs detect charged particles, which can be produced from neutrino interactions
or decays from an electrically neutral particle.
The charged particles ionize the argon atoms, producing ionization electrons
and scintillation light.
The light is collected in $\sim$10~ns by the light detection system,
determining the event time, while the ionization electrons drift, along with
the high electric field, typically 500V/cm, towards the anode in milliseconds
and are eventually collected by the charge detection system with millimeter
granularity at the anode.
With the constant drift velocity, the drift time represents the position of
the event along the drift coordinate (time projection), and therefore
3-dimensional event kinematics can be obtained with a few 1-dimensional
or one 2-dimensional instrumentation.
Further, the millimeter tracking capabilities enable BSM opportunities,
for example, search for axion-like paricles decaying into two photons or $e^+e^-$.
However, the trade-off is the millisecond-long events, which will include
a number of background,
such as cosmic rays, beam-related neutrons, and other environmental and
radiological sources.
A pulsed beam, well-considered shielding and veto systems, are hence extremely
important,
and will determine the optimal detector dimension and the sensitivity
of the measurements.

LArTPCs have been used in measuring the GeV neutrino interactions and in
searching for keV recoils from dark matter.
Detection of MeV-scale particles is not fully explored, and a number of
R\&D programs are underway.
Examples include the field structure R\&D for modular LArTPCs and a grid activated
multi-scale pixel (GAMPix) charge detection system at SLAC.
Specifically, the GAMPix design aims to implement 500-µm pixels triggered by
coarser, millimeter scale wire grid,
and can reach a noise level of 50~$e^-$, thereby efficiently collecting sub-MeV
scale charge deposition.
Precise calorimetry and electron drift distance can also be obtained by
combining signals on the wires and on the pixels.
It is originally proposed for MeV-scale $\gamma$-ray detection in space,
but also serves as a candidate of the charge detection system in LArTPCs
measuring MeV-scale neutrinos and searching for BSM particles.

For detection of keV activities, as an alternative to dual phase detectors, R\&D efforts are ongoing to investigate the feasibility of single phase LAr-TPC with a keV detection threshold to simplify the detector design and facilitate its scalability to very large masses. This is done by attempting the production of secondary ionization and/or scintillation directly in the liquid phase in close proximity of the anode, through electric fields with a local intensity of the order of hundreds of kV/cm. If successful, this would provide an extremely promising detector technology for application in the PIP-II beam dump complex.

%% file: Contributions/LArTPCs_Low-Threshold.tex
Dual-phase LArTPCs using underground argon (low in \ce{^39Ar}~\cite{darksidecollaborationDarkSide50532dayDark2018a}) have been successfully used by the DarkSide-50 dark matter direct detection experiment to search for dark matter with nuclear couplings ranging from \SI{10}{\TeV\per\square c} down to \SI{1.2}{\GeV\per\square c}---\SI{40}{\MeV\per\square c} accounting for the Migdal effect~\cite{darkside-50collaborationSearchLowmassDark2023,darksidecollaborationSearchDarkMatterNucleon2023}. DarkSide-50 also constrained electron-scattering dark matter with \SIrange{16}{1000}{\MeV\per\square c} masses and dark photons and axion-like particles between \SI{30}{\eV\per\square c} and \SI{20}{\keV\per\square c}~\cite{darksidecollaborationSearchDarkMatter2023}.
This sensitivity was achieved using both scintillation (S1) and ionization (S2) signals for energy depositions above $\SI{\sim3}{\keV}$ electron recoil (\SI{\sim12}{\keV} nuclear recoil), and the S2-only channel for lower energies.
The S2-only channel achieved sub-keV thresholds due to the amplification of ionization electrons in the gas pocket, allowing them to be detected with near-perfect efficiency. 
DarkSide-50 calibrated its ionization response to electron recoils down to \SI{\sim180}{\eV} and to nuclear recoils down to \SI{\sim400}{\eV}~\cite{darksidecollaborationCalibrationLiquidArgon2021}. 
Single-electron signals are estimated to correspond to \SI{\sim20}{\eV} electronic recoils (set by the energy required to produce an electron-ion pair) and \SI{\sim 140}{\eV} nuclear recoils, though sensitivity at the lowest energies is limited by spurious electron noise~\cite{GlobalArgonDarkMatterCollaboration:2022czd}. 
The DarkSide-LowMass experiment is now being planned as a tonne-scale LArTPC optimized for a low-threshold S2-only search, detailed in Ref.~\cite{GlobalArgonDarkMatterCollaboration:2022czd}.
Depending on its spurious electron background rate, \SIrange{2}{4}{e^-} thresholds may be achievable, corresponding to \SIrange{20}{60}{\eV} electron equivalent, or \SIrange{140}{600}{\eV} nuclear recoil.

A similar detector at a beam dump facility may see sub-keV S2-only signals. Spurious electrons typically follow primary ionization events by \SIrange{5}{50}{\milli\second} timescales; if they can be mitigated with beam timing cuts, lower energies may be within reach.
A dual-phase LArTPC doped with xenon or photo-sensitive dopants that enhance the ionization yield may reach even lower energies.
By considering signals with S1 and S2, the dynamic range of such a detector can be extended to the MeV scale and beyond.
While the S2-only channel has absolute timing resolution limited by the maximum drift time of TPC, scintillation light at higher energies can enable nanosecond-scale timing resolution, and higher-multiplicity signals (e.g. those with coincident $\gamma$-rays) can be identified by the number of S2 pulses. In DarkSide-50, S2 pulses arising from interactions separated by more than \SI{4.7}{\milli\meter} were efficiently identified.
In addition to accessing a large dynamic range, extending to sub-keV thresholds, a LArTPC benefits from a scalable design. However, larger detectors may be overrun by cosmogenic backgrounds and require several meters of overburden.

%% file: Contributions/pip2bd.tex
\begin{figure}[htbp]
    \centering
    \includegraphics[width=0.75\textwidth]{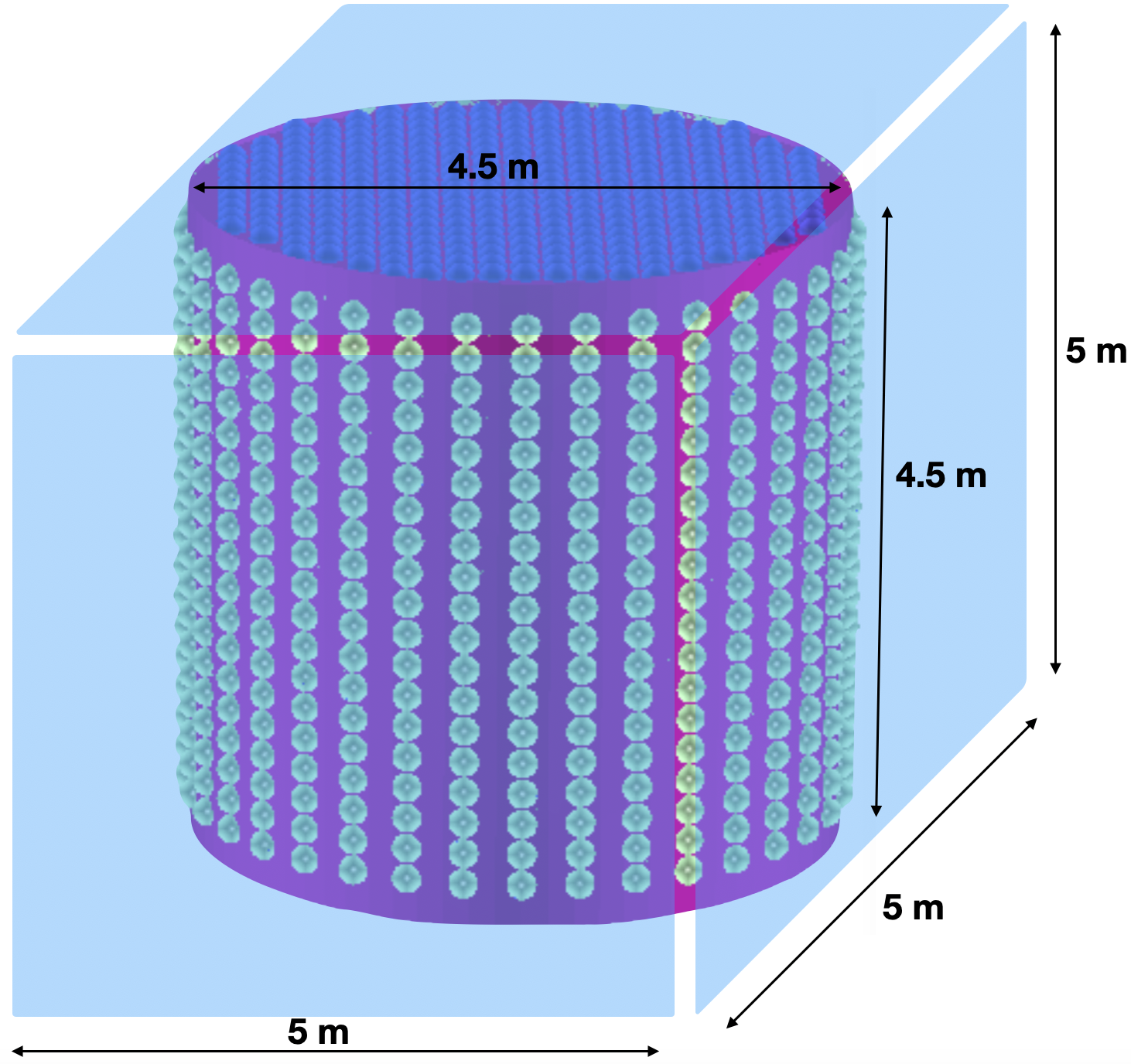}
    \caption{Geant4 rendering of the PIP2-BD detector. The active volume of the detector consists of a 4.5~m right cylinder inside of a 5~m per side cubic volume.}
    \label{fig:pip2bd-det}
\end{figure}

The reference PIP2-BD detector is a 100-tonne scale liquid argon (LAr) scintillation-only detector installed near the PIP-II target at a distance of $\sim$20~m. The active volume of the detector consists of a 2.5-m right cylinder inside a 5-m per side cubic volume which can also be instrumented as a active veto. The endcaps and side regions of the cylinder is covered in 8" PMTs cut out of a reflective teflon layer coated in the wavelength shifter tetraphenyl butadiene (TPB) that shifts the 128-nm LAr scintillation light to $\sim$400~nm which is detectable by the PMTs. A custom Geant4-based simulation using GDML input models the response of the detector. Figure~\ref{fig:pip2bd-det} shows the rendering of the detector within the simulation. The simulation uses a full optical photon generation and transport model in order to reconstruct the signals from neutrino and different dark sector signatures.         

A detector of this scale has leading sensitivity to various dark sector models when coupled to an accumulator ring. Current models explored in advance of the recent Snowmass process include dark photon decays, dark matter nucleus inelastic scattering, axion-like particles (ALPs). Possibilities also exist in the neutrino sector via sterile neutrino searches or probing non-standard neutrino interactions through the CEvNS interaction. The pulsed structure of even the minimal accumulator ring scenarios described in this white paper will greatly reduce the steady-state backgrounds, namely $^{39}$Ar, that are present when searching for keV-scale new physics such as the vector-portal light dark matter models. This detector also presents possibilities to search for new physics where the detectable signal is at the MeV-scale. A main next step for the detector response simulation is providing a reliable response model for MeV-scale physics

\begin{figure}[htbp]
    \centering
    \includegraphics[width=0.75\textwidth]{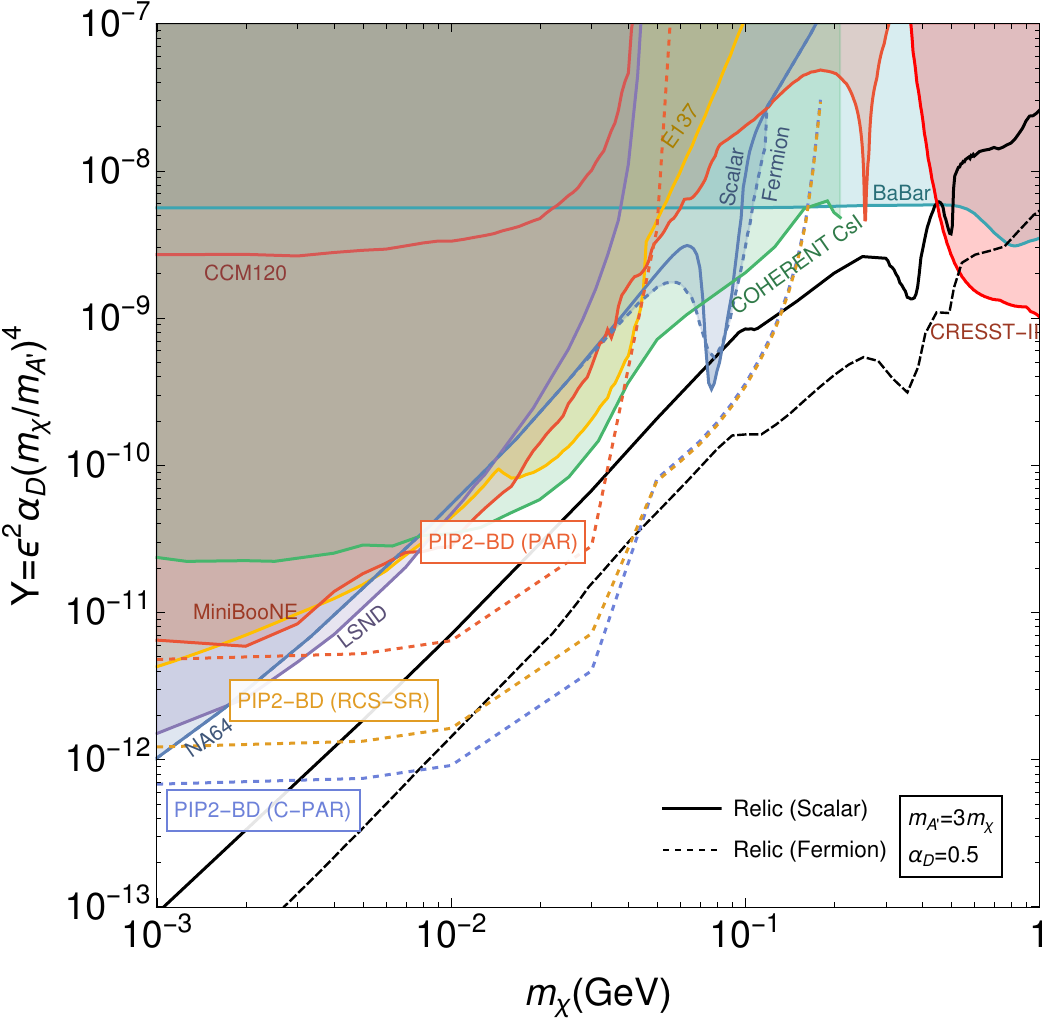}
    \caption{Sensitivity studies of PIP2-BD baseline detector concept to vector-portal light dark matter. The studies show the ability for leading probes on this model with the detector coupled to a PIP-II accumulator ring facility.}
    \label{fig:pip2bd-vpsens}
\end{figure}

A combination of light dark matter particles produced by the BdNMC code~\cite{deNiverville:2015mwa} and the Geant4 detector response simulation for background estimates provide a sensitivity study to the vector-portal light dark matter model. These studies assume that a 20~keV detector threshold is achievable in this detector which the current detector model simulations show with the addition of high purity liquid argon. These threshold have been achieved in other LAr scintillation detectors such as in COHERENT~\cite{COHERENT:2020iec} so the main question is the ease of the scalability of these thresholds. We explored three accelerator scenarios similar to those described in \ref{sec:pip2acc}; one similar to the PAR concept at 800~MeV proton energy, one similar to the C-PAR concept at 1.2~GeV proton energy, and one Rapid Cycling Synchrotron scenario at 2~GeV proton energy. The timing provided by the C-PAR scenario is found to be a powerful discriminator of the CEvNS background and provides the best sensitivity to the light dark matter models over a 5-year run of the PIP2-BD detector. Fig.~\ref{fig:pip2bd-vpsens} shows the results of the PIP2-BD sensitivity studies using the detector model described in this section.

Additionally, the produced dark matter can inelastically scatter within the detector where at low momentum transfer the inelastic scattering cross section is dominated by the Gamow-Teller transitions. Including the inelastic component improves the sensitivity at the lower dark matter mass regime. A background-free study showing the dark matter sensitivity similar to the RCS-SR scenario in Fig.~\ref{fig:pip2bd-vpsens} for the inelastic channel is shown in Fig.~4 of Ref.~\cite{Dutta:2023fij} and described further in Sec.~\ref{subsubsec:inelastics}.

\begin{figure}[htbp]
\centering
\includegraphics[width=0.5\textwidth]{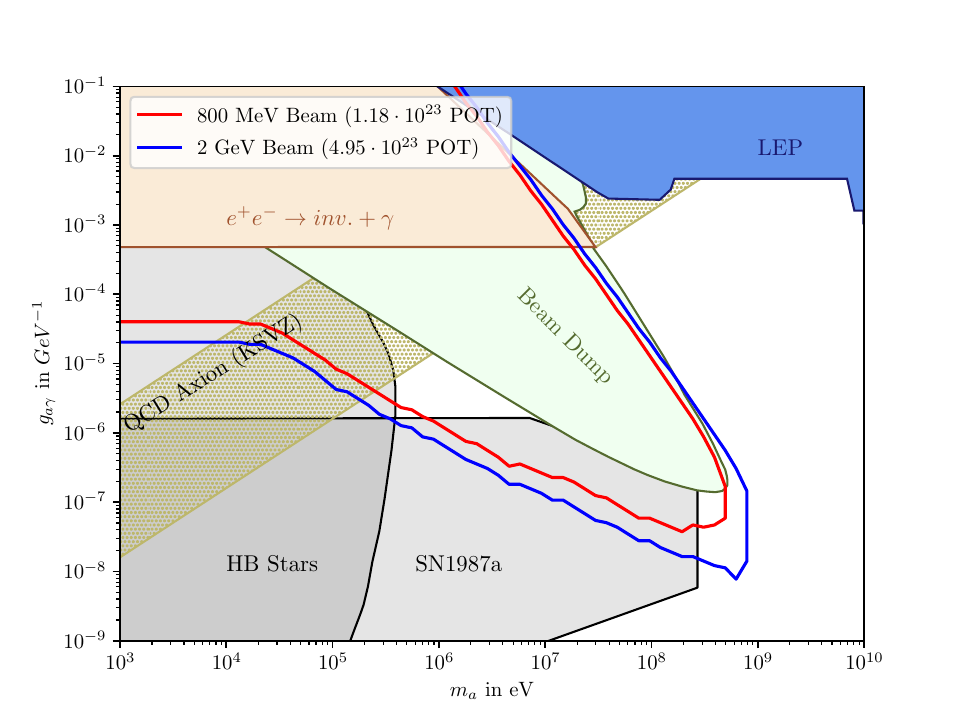}
  \includegraphics[width=0.46\textwidth]{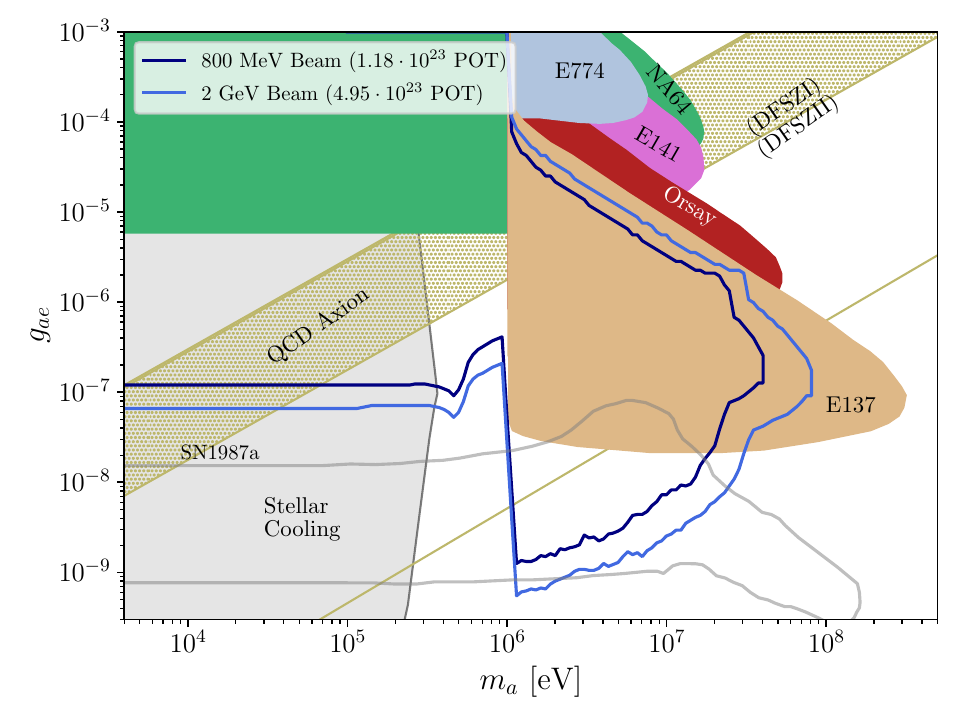}    \caption{Statistics-only, background-free, 90\% C.L. sensitivity estimates for the axion-like particle (ALP) model. In five years of running, PIP2-BD can scan the $g_{a\gamma}$ ``cosmological triangle" region a large part of an untested parameter region for $g_{a e}$.}
  \label{fig:pip2bd-alp}
\end{figure}

The baseline PIP2-BD detector provides powerful probes of ALPs coupling to both electrons and photons. The photon, electron, and positron flux above 100~keV is extracted from the beam dump simulation producing the proton collisions with the fixed target. After calculating the probability that a generated ALP decays in our detector, the 3-event, statistics-only, background-free sensitivity to $g_{a\gamma}$ and $g_{ae}$ scanning over the $m_a$ parameter space assuming a 75\% efficiency above 100~keV with a 5-year run of the PIP2-BD detector, shown in Fig.~\ref{fig:pip2bd-alp}. For the photon couplings, the PIP2-BD detector explores the ''cosmological triangle" region of the $g_{a\gamma}$ coupling parameter space. 

\begin{figure}[htbp]
\centering
  \includegraphics[width=0.32\textwidth]{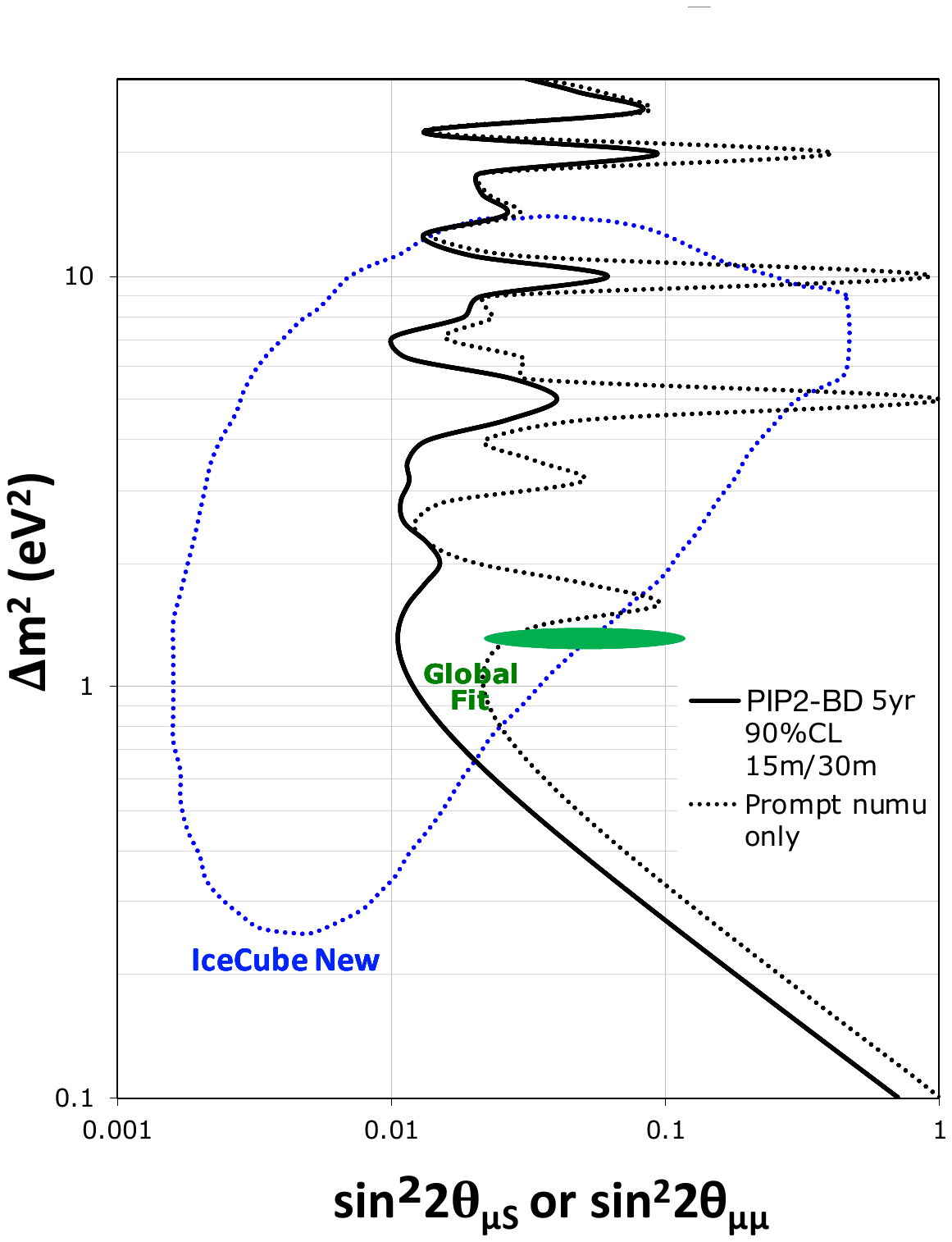}  \includegraphics[width=0.32\textwidth]{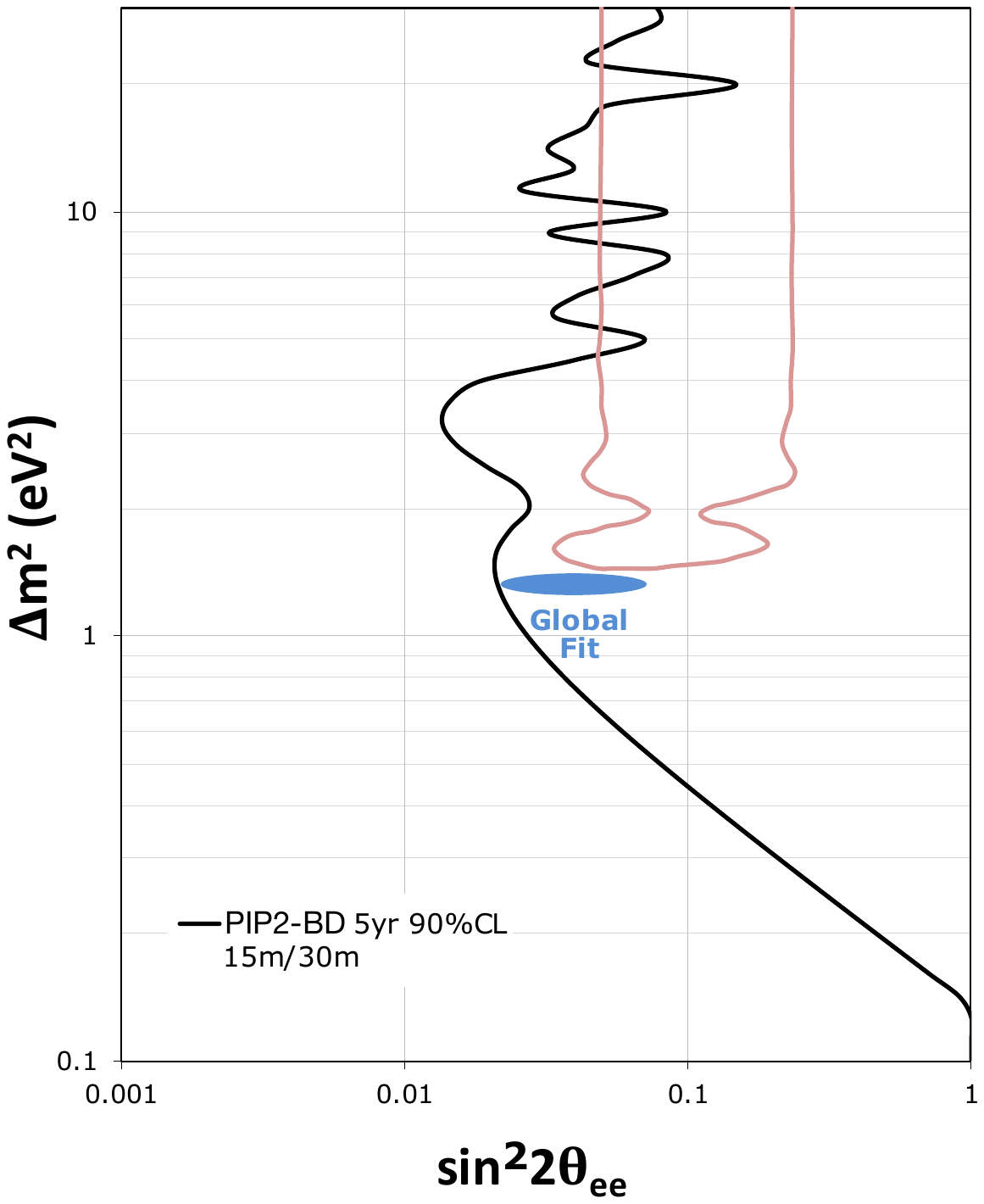}  \includegraphics[width=0.32\textwidth]{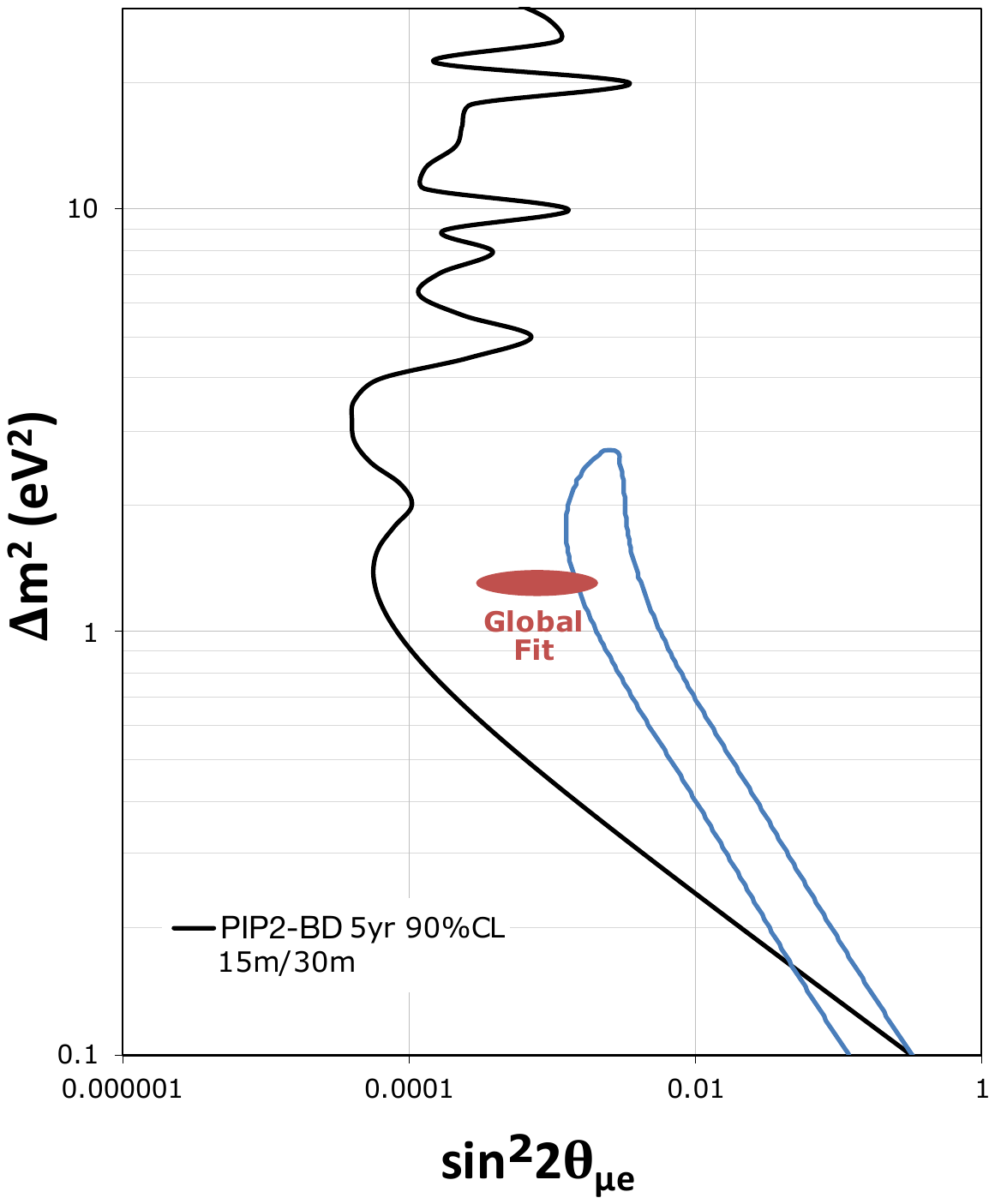}
  \caption{  PIP2-BD 90\% confidence limits on active-to-sterile neutrino mixing compared to existing $\nu_{\mu}$ disappearance limits from IceCube~\cite{Aartsen:2020iky} and a recent global fit~\cite{Diaz:2019fwt}, assuming a 5~year run (left). Also shown are the 90\% confidence limits for $\nu_{\mu}$ disappearance (left), $\nu_e$ disappearance (middle), and $\nu_e$ appearance (right), assuming the $\bar{\nu_{\mu}}$ and $\nu_e$ can be detected with similar assumptions as for the $\nu_{\mu}$. }
  \label{fig:pip2bd-sterilenu}
\end{figure}

Additionally, the PIP2-BD detector coupled to an accumulator ring is a powerful tool for exploring the neutrino sector. The monoenergetic muon neutrino disappearance together with the summed disappearance of $\nu_{\mu}$, $\nu_e$, and $\bar{\nu_{\mu}}$ allows for a definitive sterile neutrino search through the CEvNS interaction. With the addition of a second, identical, PIP2-BD detector at distances of 15 m and 30 m from the beam dump and assuming a 70\% efficiency. Calculating the rate-only 90\% confidence limits on the $\nu_{\mu}\rightarrow\nu_S$ mixing parameters including a 9\% normalization systematic, the PIP2-BD detector setup sets strong limits on the sterile neutrino over a 5 year run with the C-PAR setup described in~\ref{sec:accrings}, shown in Fig.~\ref{fig:pip2bd-sterilenu}.

%% file: Contributions/DAMSA.tex
\subsubsection{DAMSA Concept and Physics Motivation}~\label{subsubsec:damsa-concept}
DAMSA stands for Dump-produced Aboriginal Matter Search at an Accelerator and means deep thoughts, rumination, or reflection in Korean.
Its goal is to search for dark matter in the low mass regime, using the high-intensity 600~MeV proton beam accelerator capability for the rare isotope production, focusing on the Axion-Like Particles (ALP)~\cite{PhysRevD.107.L031901}.
Such a facility is advantageous for ALP discoveries due to the greatly reduced neutrino produced from the dump, given the beam energy the smaller number of charged pions are produced, and high-energy ones get absorbed before the decay.
Therefore, the PIP-II facility fits well with DAMSA's physics goals.

\subsubsection{Beam Dump and the Decay Vacuum Chamber} ~\label{subsubsec:damsa-bd-decay}
In order to explore the higher mass and/or larger coupling region of ALP parameter space, we put the detector as close as possible to the target. This allows to capture of relatively short-lived ALPs as well and expands the sensitivity reach in the higher mass and/or larger coupling regions.

\begin{figure}[htbp]
    \centering
    \includegraphics[width=0.75\textwidth]{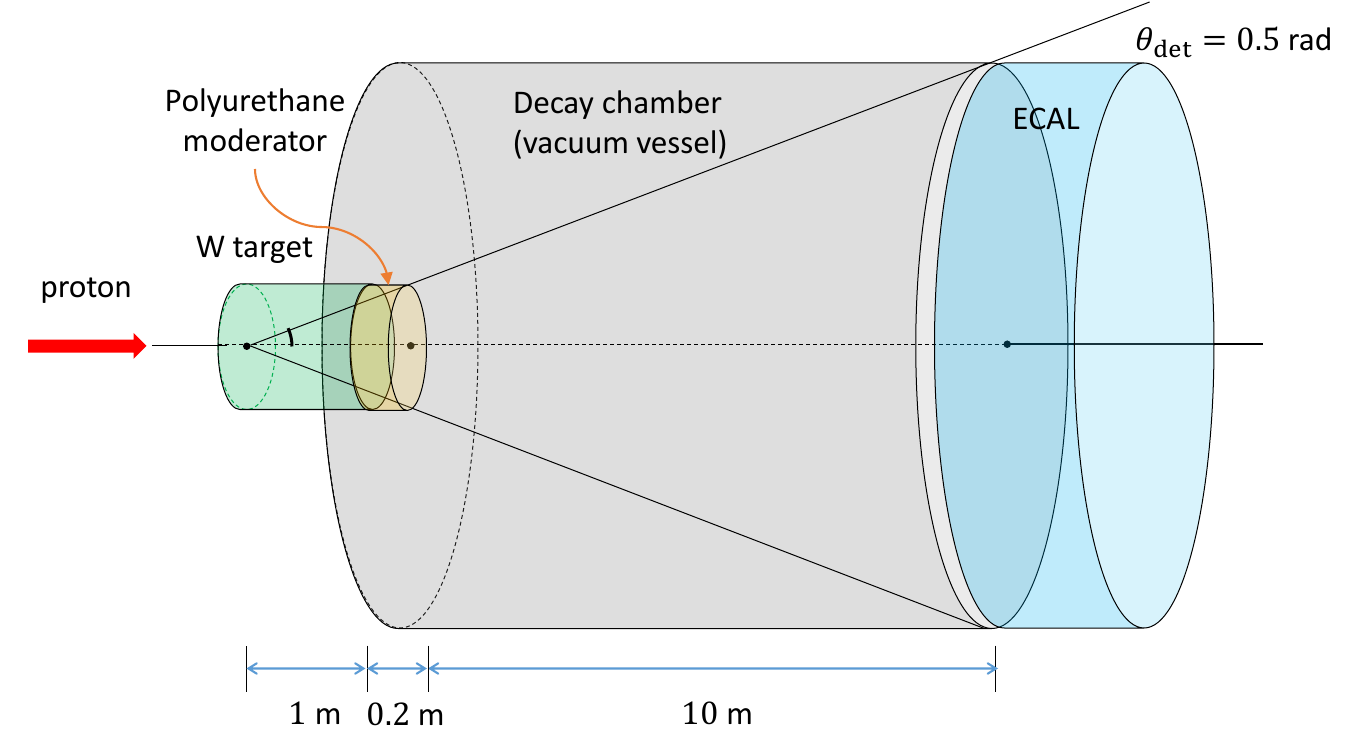}
    \caption{DAMSA Experimental Configuration}
    \label{fig:damsa-det}
\end{figure}

In the current version of the experiment design as shown in Fig.~\ref{fig:damsa-det}, we choose Tungsten(W) as the dump material and it has a cylindrical shape with 1 m diameter and 1 m deep. 1-meter deep tungsten contributes to creating lots of neutrons inside the dump but at the same time, 1 m is about 10 $\lambda_{\textrm{int.}}$ in terms of the nuclear interaction length of tungsten and so the dump absorbs the neutrons as well. Also, we have a 20 cm thick polyurethane neutron moderator surrounding it. The absorption or moderation capability as a function of its thickness has been studied with a Geant4 simulation and we have observed that the neutron moderation behavior follows a power law distribution and we have chosen 20 cm as an optimized thickness.

Remarkably, to handle the beam-induced neutrons, we placed a cylindrical vacuum chamber between the beam dump and the detector. and behind the moderator, we placed a vacuum decay chamber made of 0.6 cm thick, 10 m diameter, and 10 m long stainless steel plate.

\subsubsection{Neutron Background Mitigation Strategy} ~\label{subsubsec:damsa-n-mitigation}
This section discusses the neutron background mitigation strategy for the DAMSA experiment, which covered three energy levels for protons on target (POT): 600 MeV, 800 MeV, and 1 GeV. The motivation is to analyze the sources of background. An initial simulation was conducted using \texttt{GEANT4} with \texttt{QGSP\_BIC\_AllHP} physics list to determine the particles present, following the tungsten dump and polyurethane neutron moderator, and their counts relative to the POT (\SI{e6}{POT}). 

With the primary interest determined to be neutrons, due to spallation and excess photons, a neutron-only distribution was also simulated following the tungsten dump and polyurethane neutron moderator. A higher number of protons on target was used to create a high statistics neutron-only distribution (\SI{e8}{POT}). This distribution was then used in a subsequent simulation for the decay volume.

This simulation of the decay volume was fed via a magnified version of the high-statistics neutron distribution to provide the equivalent of \SI{e11}{POT}. Particle data was recorded following the decay volume.

Final photon pairs expected from a single beam pulse were analyzed based on their position, momentum, and time of arrival, and the appropriate/potential efficiency cuts that could be made were explored. Three cuts have been considered in the current analysis. A 15 MeV cut on individual photons was taken considering detector sensitivity threshold limitations. The Distance of Closest Approach (DCA) and Difference in Time of Arrival (TOA) were also considered, and efficiencies were graphed and tabled. The final part of the analysis considered the appropriate levels of energy, DCA, and TOA cuts.

For future work, it will be important to consider detector capabilities to provide tighter constraints on efficiency cuts. As well as conduct a sensitivity study to determine the level of sensitivity to Axion-like particles.

\subsubsection{Detector Requirements} ~\label{subsubsec:damsa-det-req}

\begin{table}
\centering
\begin{tabular}{ c | c | c | c} 
\hline
\hline
Description (per pulse)  & 600~MeV & 800~MeV & 1~GeV\\
\hline
Protons per pulse  ($n_{p}) $ & $6.87\times10^{14} $ &  \SI{6.87e14}{} &  \SI{6.87e14}{} \\ 
Beam-induced neutrons ($n_{n}$) & \SI{1.32e12}{} & \SI{2.35e12}{} & \SI{3.46e12}{}    \\ 
Neutron-induced photons ($n_{\gamma}$) & \SI{9.21e10}{} & \SI{1.66e11}{} & \SI{2.47e11}{}  \\
$\gamma$ after $E_\gamma>15$~MeV cut  ($n_{\gamma,E_{\gamma}}$) & \SI{4.87e6}{} & \SI{3.30e7}{} & \SI{7.19e7}{} \\
$\gamma$-pairs before DCA/$\Delta$TOA ($n_{\gamma\gamma}$) & \SI{2.37e13}{} & \SI{1.09e15}{} & \SI{5.17e15}{}  \\
$\gamma$-pairs after DCA $<$ \SI{1}{cm} ($n_{\gamma\gamma, \textrm{DCA}}$) & \SI{7.35e10}{} & \SI{3.25e12}{} & \SI{1.49e13}{} \\
$\gamma$-pairs after $\Delta$TOA $<$ \SI{0.1}{nsec} ($n_{\gamma\gamma, \textrm{DCA}, \Delta\textrm{TOA}}$) & \SI{6.48e8}{} & \SI{3.05e10}{} & \SI{1.38e11}{} \\
\hline 
\hline
Photon pair selection criteria & 600  MeV & 800 MeV & 1 GeV \\
\hline
DCA $<$ \SI{1}{cm} ($\epsilon_{\textrm{DCA}}$) & \SI{3.10e-3}{} & \SI{2.98e-3}{} & \SI{2.88e-3}{}    \\
$\Delta$TOA $<$ \SI{0.1}{nsec} ($\epsilon_{\Delta\textrm{TOA}}$) & \SI{8.82e-3}{} & \SI{9.38e-3}{} & \SI{9.29e-3}{} \\
Invariant Mass  4~MeV $ < M_{\gamma\gamma} < $ 6~MeV & \SI{1.63e-2}{} & \SI{1.62e-2}{} & \SI{1.75e-2}{} \\
Invariant Mass  9~MeV $ < M_{\gamma\gamma} < $ 11~MeV & \SI{1.88e-4}{} & \SI{2.07e-4}{} & \SI{2.61e-4}{}  \\
\hline \hline
\end{tabular}
\caption{A summary of the numbers related to the neutron-induced background mitigation and individual performance of the cuts. 
}
\label{tab: cuttable}
\end{table}

In the previous section, we discussed strategies to mitigate photons generated by beam-related neutrons, and the result of it is summarized in Table \ref{tab: cuttable}. Based on this result, we can establish a design goal for the detector. Assuming the energy threshold $E_{\text{thr.}}=15~\text{MeV}$, in order to discriminate neutron-induced photons without losing signals, we require the following detector performance characteristics:
\begin{itemize}
    \item The detector must have a good angular resolution in order to determine the vertex of two photons and the angular resolution depends on the position resolution of the detector, so we require a fine granularity electromagnetic calorimeter with position resolution, $\sigma_{x} < 1 ~\text{cm}$,
    \item The diphoton invariant mass depends on the energy resolution of the detector and in this study we assumed $\sigma_{E} < 1 ~\text{MeV}$, and
    \item In order to examine the temporal correlation between two photons, we need good timing resolution, and our requirement in this study is \SI{0.1}{nsec}.
\end{itemize}

When applying all independent cuts, for example with a 10 MeV ALP, we can expect background rejection factors for each proton energy configuration 600 MeV, 800 MeV, and 1000 MeV as $5.14\times10^{-9}$, $5.69\times 10^{-9}$, and $6.98\times 10^{-9}$, respectively. Based on further studies in \cite{PhysRevD.107.L031901}, we could expect an additional $O(2)$ background rejection factor which comes from the fiducial volume cut of the decay chamber and back-trace study of two photons momenta that selects photons emitted from the beam dump. Therefore, in the end, we expect $10^{-11}$ level of background rejection from the current strategy.

\subsubsection{Expected Sensitivity} ~\label{subsubsec:damsa-sensitivity}
\begin{figure}[htbp]
    \centering
    \includegraphics[width=0.75\textwidth]{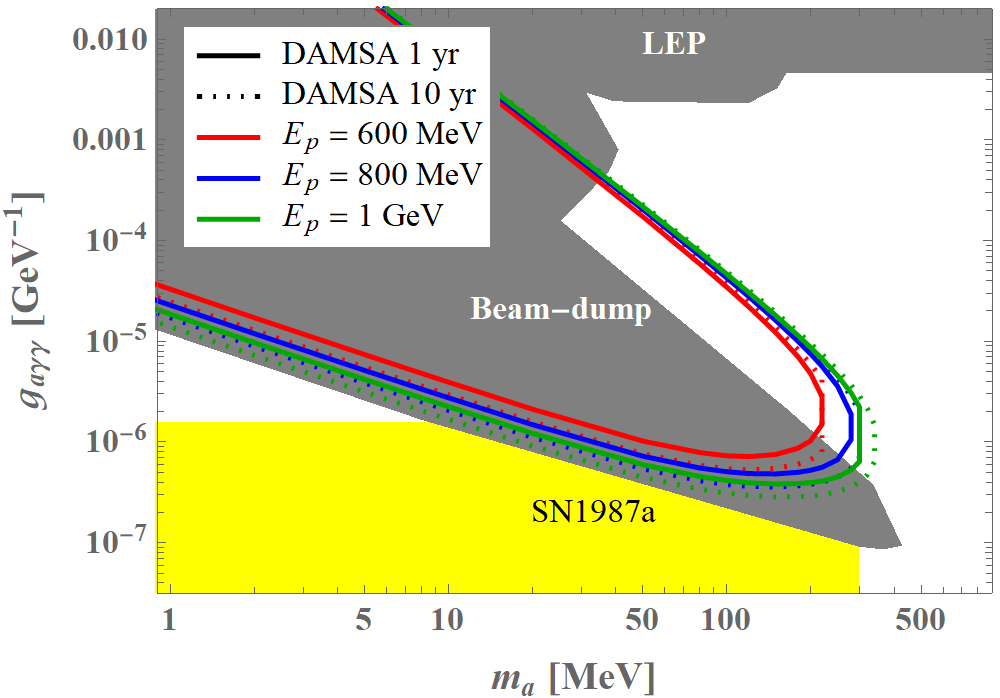}
    \caption{Expected DAMSA Sensitivity at RAON (600~MeV) and PIP-II energies (800~MeV and 1~GeV)}
    \label{fig:damsa-sensitivity}
\end{figure}

Taking into account the configuration of the experiment and the neutron-induced background estimate described in the previous sections, Fig.~\ref{fig:damsa-sensitivity} shows the sensitivity reach of the DAMSA experiment.
The sensitivity reach is compared between PIP-II 800~MeV (blue), PIP-II 1~GeV (green), and RAON, the nuclear rare isotope facility at 600~MeV (red).  
The solid lines represent the reach with one-year data taking while the dotted lines represent 10 years running.
Thanks to the 1~m proximity of the detector to the beam source including the decay volume, DAMSA can reach higher mass or higher coupling ALPs, which have not been accessible by other beam dump experiments within one year of data taking.

%% file: Contributions/conclusions.tex
The high intensity proton beams expected from PIP-II LINAC currently under construction with the completion near the end of the decade and potential upgrades from the accelerator complex evolution plan provides excellent opportunities for dark sector particle searches and exploring experimental concepts.
This white paper presents potential physics topics that fully take advantage of such powerful accelerator capabilities and clearly demonstrates that a significant level of interests and community within the field is building up for these new physics topics.
While some of the topics presented at the workshop have advanced significantly since then, this white paper presents an excellent snap shot of the community interests and its desire for pursuing the proposed topics at Fermilab.

To reiterate, PIP-II opens access to a wide range of kinematic phase space from eV to hundreds of MeV, that addresses different aspects of physics, providing complementarity to the reach of the Energy Frontier experiments. 
To fully leverage the accessibility, the most crucial element is the beam dump facility that could absorb as much of the proton beams as possible, generating a large number of dark sector particles by proton collisions within the dump.

The two proposed potential day one experiments at the PIP-II LINAC, DAMSA and PIP2-BD put an urgent timeline for developing the beam dump facility at the lab with a proposed name of F2D2, the Fermilab Facility for Dark Matter Discovery. The plan is to develop a facility in F2D2 that could meet the requirements for their physics topics, such as the accessibility to a very short distance (~1m) from the mean position of the photon production to the effective detector volume and other operational optimizations, such as the 40m.w.e. overburden.
Such a beam dump facility would provide an excellent opportunity for Fermilab to play a leadership role in dark sector particle searches at the low energy regime and enables the lab to transform itself into the facility of dark matter searches at an accelerator and attract the community to utilize it.

%% file: main.bbl
\begin{thebibliography}{100}

\bibitem{PIP2_CDR}
V.~Lebedev et~al.
\newblock {The PIP-II Conceptual Design Report}, 2017.

\bibitem{PIP2BD}
M.~Toups, R.~G. Van~de Water, Brian Batell, S.~J. Brice, Patrick deNiverville,
  Jeff Eldred, Roni Harnik, Kevin~J. Kelly, Doojin Kim, Tom Kobilarcik, Gordan
  Krnjaic, B.~R. Littlejohn, Bill Louis, Pedro A.~N. Machado, Z.~Pavlovic,
  William Pellico, Michael Shaevitz, P.~Snopok, Rex Tayloe, R.~T. Thornton,
  Jacob Zettlemoyer, and Bob Zwaska.
\newblock {PIP2-BD: GeV Proton Beam Dump at Fermilab's PIP-II Linac}, 2022.

\bibitem{BoosterRookie}
B.~Worthel et~al.
\newblock {The Booster Rookie Book}, 2009.

\bibitem{AMF}
Bertrand Echenard et~al.
\newblock {A New Charged Lepton Flavor Violation Program at Fermilab}.
\newblock In {\em {Snowmass 2021}}, 3 2022.

\bibitem{AMF_Workshop_23}
Bertrand Echenard et~al.
\newblock {Workshop on a Future Muon Program At Fermilab}, 3 2023.

\bibitem{COHERENT:2017ipa}
D.~Akimov et~al.
\newblock {Observation of Coherent Elastic Neutrino-Nucleus Scattering}.
\newblock {\em Science}, 357(6356):1123--1126, 2017.

\bibitem{CCM:2021leg}
A.~A. Aguilar-Arevalo et~al.
\newblock {First dark matter search results from Coherent CAPTAIN-Mills}.
\newblock {\em Phys. Rev. D}, 106(1):012001, 2022.

\bibitem{CCM:2021jmk}
A.~A. Aguilar-Arevalo et~al.
\newblock {Prospects for detecting axionlike particles at the Coherent
  CAPTAIN-Mills experiment}.
\newblock {\em Phys. Rev. D}, 107(9):095036, 2023.

\bibitem{COHERENT:2019kwz}
D.~Akimov et~al.
\newblock {Sensitivity of the COHERENT Experiment to Accelerator-Produced Dark
  Matter}.
\newblock {\em Phys. Rev. D}, 102(5):052007, 2020.

\bibitem{Dutta:2019eml}
Bhaskar Dutta, Shu Liao, Samiran Sinha, and Louis~E. Strigari.
\newblock {Searching for Beyond the Standard Model Physics with COHERENT Energy
  and Timing Data}.
\newblock {\em Phys. Rev. Lett.}, 123(6):061801, 2019.

\bibitem{Giunti:2019xpr}
C.~Giunti.
\newblock {General COHERENT constraints on neutrino nonstandard interactions}.
\newblock {\em Phys. Rev. D}, 101(3):035039, 2020.

\bibitem{Dutta:2020che}
Bhaskar Dutta, Rafael~F. Lang, Shu Liao, Samiran Sinha, Louis Strigari, and
  Adrian Thompson.
\newblock {A global analysis strategy to resolve neutrino NSI degeneracies with
  scattering and oscillation data}.
\newblock {\em JHEP}, 09:106, 2020.

\bibitem{Han:2019zkz}
Tao Han, Jiajun Liao, Hongkai Liu, and Danny Marfatia.
\newblock {Nonstandard neutrino interactions at COHERENT, DUNE, T2HK and LHC}.
\newblock {\em JHEP}, 11:028, 2019.

\bibitem{DeRomeri:2022twg}
V.~De~Romeri, O.~G. Miranda, D.~K. Papoulias, G.~Sanchez~Garcia, M.~T\'ortola,
  and J.~W.~F. Valle.
\newblock {Physics implications of a combined analysis of COHERENT CsI and LAr
  data}.
\newblock {\em JHEP}, 04:035, 2023.

\bibitem{Blanco:2019vyp}
Carlos Blanco, Dan Hooper, and Pedro Machado.
\newblock {Constraining Sterile Neutrino Interpretations of the LSND and
  MiniBooNE Anomalies with Coherent Neutrino Scattering Experiments}.
\newblock {\em Phys. Rev. D}, 101(7):075051, 2020.

\bibitem{Dutta:2019nbn}
Bhaskar Dutta, Doojin Kim, Shu Liao, Jong-Chul Park, Seodong Shin, and Louis~E.
  Strigari.
\newblock {Dark matter signals from timing spectra at neutrino experiments}.
\newblock {\em Phys. Rev. Lett.}, 124(12):121802, 2020.

\bibitem{Dutta:2020vop}
Bhaskar Dutta, Doojin Kim, Shu Liao, Jong-Chul Park, Seodong Shin, Louis~E.
  Strigari, and Adrian Thompson.
\newblock {Searching for dark matter signals in timing spectra at neutrino
  experiments}.
\newblock {\em JHEP}, 01:144, 2022.

\bibitem{GEANT4:2002zbu}
S.~Agostinelli et~al.
\newblock {GEANT4--a simulation toolkit}.
\newblock {\em Nucl. Instrum. Meth. A}, 506:250--303, 2003.

\bibitem{Alonso:2010fs}
J.~Alonso et~al.
\newblock {Expression of Interest for a Novel Search for CP Violation in the
  Neutrino Sector: DAEdALUS}.
\newblock 6 2010.

\bibitem{Tomalak:2021lif}
Oleksandr Tomalak.
\newblock {Radiative (anti)neutrino energy spectra from muon, pion, and kaon
  decays}.
\newblock {\em Phys. Lett. B}, 829:137108, 2022.

\bibitem{Pandey:2023arh}
V.~Pandey.
\newblock {Recent Progress in Low Energy Neutrino Scattering Physics and Its
  Implications for the Standard and Beyond the Standard Model Physics}.
\newblock 8 2023.

\bibitem{VanDessel:2020epd}
N.~Van~Dessel, V.~Pandey, H.~Ray, and N.~Jachowicz.
\newblock {Cross sections for coherent elastic and inelastic neutrino-nucleus
  scattering}.
\newblock {\em Universe}, 9:207, 2023.

\bibitem{Tomalak:2020zfh}
Oleksandr Tomalak, Pedro Machado, Vishvas Pandey, and Ryan Plestid.
\newblock {Flavor-dependent radiative corrections in coherent elastic
  neutrino-nucleus scattering}.
\newblock {\em JHEP}, 02:097, 2021.

\bibitem{Pandey:2014tza}
V.~Pandey, N.~Jachowicz, T.~Van~Cuyck, J.~Ryckebusch, and M.~Martini.
\newblock {Low-energy excitations and quasielastic contribution to
  electron-nucleus and neutrino-nucleus scattering in the continuum
  random-phase approximation}.
\newblock {\em Phys. Rev. C}, 92(2):024606, 2015.

\bibitem{Pandey:2016jju}
V.~Pandey, N.~Jachowicz, M.~Martini, R.~Gonz\'alez-Jim\'enez, J.~Ryckebusch,
  T.~Van~Cuyck, and N.~Van~Dessel.
\newblock {Impact of low-energy nuclear excitations on neutrino-nucleus
  scattering at MiniBooNE and T2K kinematics}.
\newblock {\em Phys. Rev. C}, 94(5):054609, 2016.

\bibitem{DUNE:2023rtr}
Adam Abed~Abud et~al.
\newblock {Impact of cross-section uncertainties on supernova neutrino spectral
  parameter fitting in the Deep Underground Neutrino Experiment}.
\newblock 3 2023.

\bibitem{Dutta:2023fij}
Bhaskar Dutta, Wei-Chih Huang, and Jayden~L. Newstead.
\newblock {Probing the dark sector with nuclear transition photons}.
\newblock 2 2023.

\bibitem{Dutta:2022tav}
Bhaskar Dutta, Wei-Chih Huang, Jayden~L. Newstead, and Vishvas Pandey.
\newblock {Inelastic nuclear scattering from neutrinos and dark matter}.
\newblock {\em Phys. Rev. D}, 106(11):113006, 2022.

\bibitem{Dent:2019ueq}
James~B. Dent, Bhaskar Dutta, Doojin Kim, Shu Liao, Rupak Mahapatra, Kuver
  Sinha, and Adrian Thompson.
\newblock {New Directions for Axion Searches via Scattering at Reactor Neutrino
  Experiments}.
\newblock {\em Phys. Rev. Lett.}, 124(21):211804, 2020.

\bibitem{alppip}
Bhaskar Dutta, Aparajitha Karthikeyan, and Adrian Thompson.
\newblock To appear, 2023.

\bibitem{Waites:2022tov}
Loyd Waites, Adrian Thompson, Adriana Bungau, Janet~M. Conrad, Bhaskar Dutta,
  Wei-Chih Huang, Doojin Kim, Michael Shaevitz, and Joshua Spitz.
\newblock {Axionlike particle production at beam dump experiments with distinct
  nuclear excitation lines}.
\newblock {\em Phys. Rev. D}, 107(9):095010, 2023.

\bibitem{Avignone:1988bv}
F.~T. Avignone, C.~Baktash, W.~C. Barker, F.~P. Calaprice, R.~W. Dunford, W.~C.
  Haxton, D.~Kahana, R.~T. Kouzes, H.~S. Miley, and D.~M. Moltz.
\newblock {Search for Axions From the 1115-kev Transition of $^{65}$Cu}.
\newblock {\em Phys. Rev. D}, 37:618--630, 1988.

\bibitem{absp}
Bhaskar Dutta, Wei-Chi Huang, and Jaydent Newstead.
\newblock To appear, 2023.

\bibitem{darkphoton}
Bhaskar Dutta, Aparajitha Karthikeyan, and Doojin Kim.
\newblock To appear, 2023.

\bibitem{deNiverville:2015mwa}
Patrick deNiverville, Maxim Pospelov, and Adam Ritz.
\newblock {Light new physics in coherent neutrino-nucleus scattering
  experiments}.
\newblock {\em Phys. Rev. D}, 92(9):095005, 2015.

\bibitem{Batell:2014mga}
Brian Batell, Rouven Essig, and Ze'ev Surujon.
\newblock {Strong Constraints on Sub-GeV Dark Sectors from SLAC Beam Dump
  E137}.
\newblock {\em Phys. Rev. Lett.}, 113(17):171802, 2014.

\bibitem{Maschuw:1998qh}
R.~Maschuw.
\newblock {Neutrino spectroscopy with KARMEN}.
\newblock {\em Prog. Part. Nucl. Phys.}, 40:183--192, 1998.

\bibitem{KARMEN:1998xmo}
B~Armbruster et~al.
\newblock {Measurement of the weak neutral current excitation C-12(nu(mu)
  nu'(mu))C*-12(1+,1,15.1-MeV) at E(nu(mu)) = 29.8-MeV}.
\newblock {\em Phys. Lett. B}, 423:15--20, 1998.

\bibitem{KARMEN:1991vkr}
G.~Drexlin et~al.
\newblock {First observation of the neutral current nuclear excitation C-12
  (nu, nu-prime) C-12* (1+, 1).}
\newblock {\em Phys. Lett. B}, 267:321--324, 1991.

\bibitem{MiniBooNE:2008yuf}
A.~A. Aguilar-Arevalo et~al.
\newblock {Unexplained Excess of Electron-Like Events From a 1-GeV Neutrino
  Beam}.
\newblock {\em Phys. Rev. Lett.}, 102:101802, 2009.

\bibitem{MiniBooNE:2018esg}
A.~A. Aguilar-Arevalo et~al.
\newblock {Significant Excess of ElectronLike Events in the MiniBooNE
  Short-Baseline Neutrino Experiment}.
\newblock {\em Phys. Rev. Lett.}, 121(22):221801, 2018.

\bibitem{MiniBooNE:2020pnu}
A.~A. Aguilar-Arevalo et~al.
\newblock {Updated MiniBooNE neutrino oscillation results with increased data
  and new background studies}.
\newblock {\em Phys. Rev. D}, 103(5):052002, 2021.

\bibitem{MiniBooNEDM:2018cxm}
A.~A. Aguilar-Arevalo et~al.
\newblock {Dark Matter Search in Nucleon, Pion, and Electron Channels from a
  Proton Beam Dump with MiniBooNE}.
\newblock {\em Phys. Rev. D}, 98(11):112004, 2018.

\bibitem{Jordan:2018qiy}
Johnathon~R. Jordan, Yonatan Kahn, Gordan Krnjaic, Matthew Moschella, and
  Joshua Spitz.
\newblock {Severe Constraints on New Physics Explanations of the MiniBooNE
  Excess}.
\newblock {\em Phys. Rev. Lett.}, 122(8):081801, 2019.

\bibitem{Dutta:2021cip}
Bhaskar Dutta, Doojin Kim, Adrian Thompson, Remington~T. Thornton, and
  Richard~G. Van~de Water.
\newblock {Solutions to the MiniBooNE Anomaly from New Physics in Charged Meson
  Decays}.
\newblock {\em Phys. Rev. Lett.}, 129(11):111803, 2022.

\bibitem{dscmd}
Bhaskar Dutta, Adrian Thompson, and Richard Van~de Water.
\newblock To appear, 2023.

\bibitem{Dirac:1931kp}
Paul Adrien~Maurice Dirac.
\newblock {Quantised singularities in the electromagnetic field,}.
\newblock {\em Proc. Roy. Soc. Lond. A}, 133(821):60--72, 1931.

\bibitem{Holdom:1985ag}
Bob Holdom.
\newblock {Two U(1)'s and Epsilon Charge Shifts}.
\newblock {\em Phys. Lett. B}, 166:196--198, 1986.

\bibitem{Holdom:1986eq}
Bob Holdom.
\newblock {Searching for $\epsilon$ Charges and a New U(1)}.
\newblock {\em Phys. Lett. B}, 178:65--70, 1986.

\bibitem{Chang:2018rso}
Jae~Hyeok Chang, Rouven Essig, and Samuel~D. McDermott.
\newblock {Supernova 1987A Constraints on Sub-GeV Dark Sectors, Millicharged
  Particles, the QCD Axion, and an Axion-like Particle}.
\newblock {\em JHEP}, 09:051, 2018.

\bibitem{Antypas:2022asj}
D.~Antypas et~al.
\newblock {New Horizons: Scalar and Vector Ultralight Dark Matter}.
\newblock 3 2022.

\bibitem{Pospelov:2007mp}
Maxim Pospelov, Adam Ritz, and Mikhail~B. Voloshin.
\newblock {Secluded WIMP Dark Matter}.
\newblock {\em Phys. Lett. B}, 662:53--61, 2008.

\bibitem{Gninenko:2006fi}
S.~N. Gninenko, N.~V. Krasnikov, and A.~Rubbia.
\newblock {Search for millicharged particles in reactor neutrino experiments: A
  Probe of the PVLAS anomaly}.
\newblock {\em Phys. Rev. D}, 75:075014, 2007.

\bibitem{Liu:2019knx}
Hongwan Liu, Nadav~Joseph Outmezguine, Diego Redigolo, and Tomer Volansky.
\newblock {Reviving Millicharged Dark Matter for 21-cm Cosmology}.
\newblock {\em Phys. Rev. D}, 100(12):123011, 2019.

\bibitem{Agrawal:2021dbo}
Prateek Agrawal et~al.
\newblock {Feebly-interacting particles: FIPs 2020 workshop report}.
\newblock {\em Eur. Phys. J. C}, 81(11):1015, 2021.

\bibitem{SENSEI:2023gie}
Liron Barak et~al.
\newblock {SENSEI: Search for Millicharged Particles produced in the NuMI
  Beam}.
\newblock 5 2023.

\bibitem{Boyarsky:2009ix}
Alexey Boyarsky, Oleg Ruchayskiy, and Mikhail Shaposhnikov.
\newblock {The Role of sterile neutrinos in cosmology and astrophysics}.
\newblock {\em Ann. Rev. Nucl. Part. Sci.}, 59:191--214, 2009.

\bibitem{Drewes:2013gca}
Marco Drewes.
\newblock {The Phenomenology of Right Handed Neutrinos}.
\newblock {\em Int. J. Mod. Phys. E}, 22:1330019, 2013.

\bibitem{Dasgupta:2021ies}
Basudeb Dasgupta and Joachim Kopp.
\newblock {Sterile Neutrinos}.
\newblock {\em Phys. Rept.}, 928:1--63, 2021.

\bibitem{Abdullahi:2022jlv}
Asli~M. Abdullahi et~al.
\newblock {The present and future status of heavy neutral leptons}.
\newblock {\em J. Phys. G}, 50(2):020501, 2023.

\bibitem{Batell:2022xau}
Brian Batell et~al.
\newblock {Dark Sector Studies with Neutrino Beams}.
\newblock In {\em {Snowmass 2021}}, 7 2022.

\bibitem{Ema:2023buz}
Yohei Ema, Zhen Liu, Kun-Feng Lyu, and Maxim Pospelov.
\newblock {Heavy Neutral Leptons from Stopped Muons and Pions}.
\newblock 6 2023.

\bibitem{LSND:2001akn}
L.~B. Auerbach et~al.
\newblock {Measurement of electron - neutrino - electron elastic scattering}.
\newblock {\em Phys. Rev. D}, 63:112001, 2001.

\bibitem{Daum:1987bg}
M.~Daum, B.~Jost, R.~M. Marshall, R.~C. Minehart, W.~A. Stephens, and K.~O.~H.
  Ziock.
\newblock {Search for Admixtures of Massive Neutrinos in the Decay $\pi^+ \to
  \mu^+$ Neutrino}.
\newblock {\em Phys. Rev. D}, 36:2624, 1987.

\bibitem{Britton:1992xv}
D.~I. Britton et~al.
\newblock {Improved search for massive neutrinos in pi+ ---\ensuremath{>} e+
  neutrino decay}.
\newblock {\em Phys. Rev. D}, 46:R885--R887, 1992.

\bibitem{PIENU:2017wbj}
A.~Aguilar-Arevalo et~al.
\newblock {Improved search for heavy neutrinos in the decay $\pi\rightarrow
  e\nu$}.
\newblock {\em Phys. Rev. D}, 97(7):072012, 2018.

\bibitem{PIENU:2019usb}
A.~Aguilar-Arevalo et~al.
\newblock {Search for heavy neutrinos in $\pi \to \mu\nu$ decay}.
\newblock {\em Phys. Lett. B}, 798:134980, 2019.

\bibitem{Bryman:2019bjg}
D.~A. Bryman and R.~Shrock.
\newblock {Constraints on Sterile Neutrinos in the MeV to GeV Mass Range}.
\newblock {\em Phys. Rev. D}, 100:073011, 2019.

\bibitem{T2K:2019jwa}
K.~Abe et~al.
\newblock {Search for heavy neutrinos with the T2K near detector ND280}.
\newblock {\em Phys. Rev. D}, 100(5):052006, 2019.

\bibitem{Arguelles:2021dqn}
Carlos~A. Arg\"uelles, Nicol\`o Foppiani, and Matheus Hostert.
\newblock {Heavy neutral leptons below the kaon mass at hodoscopic neutrino
  detectors}.
\newblock {\em Phys. Rev. D}, 105(9):095006, 2022.

\bibitem{MicroBooNE:2021usw}
P.~Abratenko et~al.
\newblock {Search for a Higgs Portal Scalar Decaying to Electron-Positron Pairs
  in the MicroBooNE Detector}.
\newblock {\em Phys. Rev. Lett.}, 127(15):151803, 2021.

\bibitem{Kelly:2021xbv}
Kevin~James Kelly and Pedro A.~N. Machado.
\newblock {MicroBooNE experiment, NuMI absorber, and heavy neutral leptons}.
\newblock {\em Phys. Rev. D}, 104(5):055015, 2021.

\bibitem{Berryman:2019dme}
Jeffrey~M. Berryman, Andre de~Gouvea, Patrick~J Fox, Boris~Jules Kayser,
  Kevin~James Kelly, and Jennifer~Lynne Raaf.
\newblock {Searches for Decays of New Particles in the DUNE Multi-Purpose Near
  Detector}.
\newblock {\em JHEP}, 02:174, 2020.

\bibitem{PIONEER:2022yag}
W.~Altmannshofer et~al.
\newblock {PIONEER: Studies of Rare Pion Decays}.
\newblock 3 2022.

\bibitem{PIONEER:2022alm}
W.~Altmannshofer et~al.
\newblock {Testing Lepton Flavor Universality and CKM Unitarity with Rare Pion
  Decays in the PIONEER experiment}.
\newblock In {\em {Snowmass 2021}}, 3 2022.

\bibitem{Fernandez-Martinez:2023phj}
Enrique Fern\'andez-Mart\'\i{}nez, Manuel Gonz\'alez-L\'opez, Josu
  Hern\'andez-Garc\'\i{}a, Matheus Hostert, and Jacobo L\'opez-Pav\'on.
\newblock {Effective portals to heavy neutral leptons}.
\newblock 4 2023.

\bibitem{SuperCDMS:2020hcc}
I.~Alkhatib et~al.
\newblock {Constraints on Lightly Ionizing Particles from CDMSlite}.
\newblock {\em Phys. Rev. Lett.}, 127(8):081802, 2021.

\bibitem{Petricca:2017zdp}
F.~Petricca et~al.
\newblock {First results on low-mass dark matter from the CRESST-III
  experiment}.
\newblock In {\em {15th International Conference on Topics in Astroparticle and
  Underground Physics (TAUP 2017) Sudbury, Ontario, Canada, July 24-28, 2017}},
  2017.

\bibitem{SENSEI:2020dpa}
{L. Barak {\it et al.} [SENSEI Collaboration]}.
\newblock {SENSEI: Direct-Detection Results on sub-GeV Dark Matter from a New
  Skipper-CCD}.
\newblock {\em Phys. Rev. Lett.}, 125(17):171802, 2020.

\bibitem{hawleyherrera2023sbcsnolab}
H.~Hawley-Herrera.
\newblock Sbc-snolab scintillation system and sipm implementation for dark
  matter searches, 2023.

\bibitem{mcpRoni2019}
Roni {Harnik}, Zhen {Liu}, and Ornella {Palamara}.
\newblock {Millicharged particles in liquid argon neutrino experiments}.
\newblock {\em Journal of High Energy Physics}, 2019(7):170, July 2019.

\bibitem{barak2023sensei}
Liron Barak, Itay~M. Bloch, Ana~M. Botti, Mariano Cababie, Gustavo Cancelo,
  Luke Chaplinsky, Michael Crisler, Alex Drlica-Wagner.~Rouven Essig, Juan
  Estrada, Erez Etzion, Stephen~E. Holland, Guillermo~Fernandez Moroni, Yaron
  Korn, Ian Lawson, Zhen Liu, Steffon Luoma, Sravan Munagavalasa, Aviv Orly,
  Dario Rodrigues, Aman Singal, Santiago~E. Perez, Ryan Plestid, Nathan~A.
  Saffold, Miguel~Sofo Haro, Roni Harnik, Leandro Stefanazzi, Kelly Stifter,
  Javier Tiffenberg, Sho Uemura, Tomer Volansky, and Tien-Tien Yu.
\newblock Sensei: Search for millicharged particles produced in the numi beam,
  2023.

\bibitem{platt2018}
M.~Platt, R.~Mahapatra, Raymond~A. Bunker, and John~L. Orrell.
\newblock {{SuperCDMS Underground Detector Fabrication Facility}}.
\newblock Technical Report PNNL--27319, 1424835, March 2018.

\bibitem{COHERENT:2020iec}
D.~Akimov et~al.
\newblock {First Measurement of Coherent Elastic Neutrino-Nucleus Scattering on
  Argon}.
\newblock {\em Phys. Rev. Lett.}, 126(1):012002, 2021.

\bibitem{Machado:2019xpc}
Pedro~A.N. Machado, Ornella Palamara, and David~W. Schmitz.
\newblock The short-baseline neutrino program at fermilab.
\newblock {\em Annual Review of Nuclear and Particle Science}, 69(1):363--387,
  2019.

\bibitem{Reichenbacher:2023ncb}
J.~Reichenbacher.
\newblock Backgrounds for supernova neutrino detection with underground
  detectors (liquid argon).
\newblock In {\em 3rd New Physics Opportunities at Neutrino Facilities
  Workshop: Astrophysical Neutrinos}. SLAC, 7 2023.

\bibitem{Parvu:2021ezc}
Mihaela Parvu and Ionel Lazanu.
\newblock {Radioactive background for ProtoDUNE detector}.
\newblock {\em JCAP}, 08:042, 2021.

\bibitem{Reichenbacher:1998ne}
J.~Reichenbacher.
\newblock {Investigation of the optical properties of large-area plastic
  scintillators for the KARMEN upgrade}.
\newblock Master's thesis, Karlsruhe Institute of Technology KIT, 12 1998.

\bibitem{KARMEN:2002zcm}
B.~Armbruster et~al.
\newblock {Upper limits for neutrino oscillations muon-anti-neutrino
  ---\ensuremath{>} electron-anti-neutrino from muon decay at rest}.
\newblock {\em Phys. Rev. D}, 65:112001, 2002.

\bibitem{Reichenbacher:2005nc}
J.~Reichenbacher.
\newblock {\em {Final KARMEN results on neutrino oscillations and neutrino
  nucleus interactions in the energy regime of supernovae}}.
\newblock Phd thesis, Karlsruhe Institute of Technology KIT, 6 2005.

\bibitem{Reichenbacher:2023nca}
J.~Reichenbacher.
\newblock Constraints on sterile neutrino evidence and lsnd with karmen1+2.
\newblock In {\em CETUP* 2023}. THE INSTITUTE for Underground Science at SURF,
  7 2023.

\bibitem{2020smartskipper}
Fernando {Chierchie}, Guillermo {Fernandez Moroni}, Leandro {Stefanazzi},
  Claudio {Chavez}, Eduardo {Paolini}, Gustavo {Cancelo}, Miguel {Sofo Haro},
  Javier {Tiffenberg}, Juan {Estrada}, and Sho {Uemura}.
\newblock {Smart-readout of the Skipper-CCD: Achieving Sub-electron Noise
  Levels in Regions of Interest}.
\newblock {\em arXiv e-prints}, page arXiv:2012.10414, December 2020.

\bibitem{Sofotalk2023}
M.~{Sofo Haro}.
\newblock {Development of new CCDs with non-destructive readout mode}, March
  2023.

\bibitem{DAMICM2023}
I.~Arnquist, N.~Avalos, D.~Baxter, X.~Bertou, N.~Castell\'o-Mor, A.~E.
  Chavarria, J.~Cuevas-Zepeda, J.~Cortabitarte Guti\'errez,
  J.~Duarte-Campderros, A.~Dastgheibi-Fard, O.~Deligny, C.~De~Dominicis,
  E.~Estrada, N.~Gadola, R.~Ga\"{\i}or, T.~Hossbach, L.~Iddir, L.~Khalil,
  B.~Kilminster, A.~Lantero-Barreda, I.~Lawson, S.~Lee, A.~Letessier-Selvon,
  P.~Loaiza, A.~Lopez-Virto, A.~Matalon, S.~Munagavalasa, K.~J. McGuire,
  P.~Mitra, D.~Norcini, G.~Papadopoulos, S.~Paul, A.~Piers, P.~Privitera,
  K.~Ramanathan, P.~Robmann, M.~Settimo, R.~Smida, R.~Thomas, M.~Traina,
  I.~Vila, R.~Vilar, G.~Warot, R.~Yajur, and J-P. Zopounidis.
\newblock First constraints from damic-m on sub-gev dark-matter particles
  interacting with electrons.
\newblock {\em Phys. Rev. Lett.}, 130:171003, Apr 2023.

\bibitem{Chierchie2023}
F.~Chierchie, C.R. Chavez, M.~Sofo Haro, G.~Fernandez Moroni, B.A.
  Cervantes-Vergara, S.~Perez, J.~Estrada, J.~Tiffenberg, S.~Uemura, and
  A.~Botti.
\newblock First results from a multiplexed and massive instrument with
  sub-electron noise skipper-ccds.
\newblock {\em Journal of Instrumentation}, 18(01):P01040, January 2023.

\bibitem{perez2023early}
Santiago Perez, Dario Rodrigues, Juan Estrada, Roni Harnik, Zhen Liu, Brenda~A.
  Cervantes-Vergara, Juan~Carlos DOlivo, Ryan~D. Plestid, Javier Tiffenberg,
  Tien-Tien Yu, Alexis Aguilar-Arevalo, Fabricio Alcalde-Bessia, Nicolas
  Avalos, Oscar Baez, Daniel Baxter, Xavier Bertou, Carla Bonifazi, Ana Botti,
  Gustavo Cancelo, Nuria Castello-Mor, Alvaro~E. Chavarria, Claudio~R. Chavez,
  Fernando Chierchie, Juan Manuel~De Egea, Cyrus Dreyer, Alex Drlica-Wagner,
  Rouven Essig, Ezequiel Estrada, Erez Etzion, Paul Grylls, Guillermo
  Fernandez-Moroni, Marivi Fernandez-Serra, Santiago Ferreyra, Stephen Holland,
  Agustin~Lantero Barreda, Andrew Lathrop, Ian Lawson, Ben Loer, Steffon Luoma,
  Edgar~Marrufo Villalpando, Mauricio~Martinez Montero, Kellie McGuire, Jorge
  Molina, Sravan Munagavalasa, Danielle Norcini, Alexander Piers, Paolo
  Privitera, Nathan Saffold, Richard Saldanha, Aman Singal, Radomir Smida,
  Miguel Sofo-Haro, Diego Stalder, Leandro Stefanazzi, Michelangelo Traina,
  Yu-Dai Tsai, Sho Uemura, Pedro Ventura, Rocio~Vilar Cortabitarte, and Rachana
  Yajur.
\newblock Early science with the oscura integration test, 2023.

\bibitem{Navarro2019}
Albert~Puig Navarro and Jonas Eschle.
\newblock phasespace: n-body phase space generation in python.
\newblock {\em Journal of Open Source Software}, 4(42):1570, 2019.

\bibitem{harnik2019millicharged}
Roni Harnik, Zhen Liu, and Ornella Palamara.
\newblock Millicharged particles in liquid argon neutrino experiments.
\newblock {\em Journal of High Energy Physics}, 2019(7):1--21, 2019.

\bibitem{FORMOSA}
Saeid {Foroughi-Abari}, Felix {Kling}, and Yu-Dai {Tsai}.
\newblock {FORMOSA: Looking Forward to Millicharged Dark Sectors}.
\newblock {\em arXiv e-prints}, page arXiv:2010.07941, October 2020.

\bibitem{FerMINI}
Kevin~J. Kelly and Yu-Dai Tsai.
\newblock {Proton fixed-target scintillation experiment to search for
  millicharged dark matter}.
\newblock {\em Phys. Rev. D}, 100(1):015043, 2019.

\bibitem{milliqan}
A.~Ball, J.~Brooke, C.~Campagnari, M.~Carrigan, M.~Citron, A.~De Roeck,
  M.~Ezeldine, B.~Francis, M.~Gastal, M.~Ghimire, J.~Goldstein, F.~Golf,
  A.~Haas, R.~Heller, C.{\hspace{0.167em} }S. Hill, L.~Lavezzo, R.~Loos,
  S.~Lowette, B.~Manley, B.~Marsh, D.{\hspace{0.167em}}W. Miller, B.~Odegard,
  R.~Schmitz, F.~Setti, H.~Shakeshaft, D.~Stuart, M.~Swiatlowski, J.~Yoo, and
  H.~Zaraket.
\newblock Sensitivity to millicharged particles in future proton-proton
  collisions at the {LHC} with the {milliQan} detector.
\newblock {\em Physical Review D}, 104(3), aug 2021.

\bibitem{PhysRevD.102.032002}
A.~Ball, G.~Beauregard, J.~Brooke, C.~Campagnari, M.~Carrigan, M.~Citron,
  J.~De~La~Haye, A.~De~Roeck, Y.~Elskens, R.~Escobar Franco, M.~Ezeldine,
  B.~Francis, M.~Gastal, M.~Ghimire, J.~Goldstein, F.~Golf, J.~Guiang, A.~Haas,
  R.~Heller, C.~S. Hill, L.~Lavezzo, R.~Loos, S.~Lowette, G.~Magill, B.~Manley,
  B.~Marsh, D.~W. Miller, B.~Odegard, F.~R. Saab, J.~Sahili, R.~Schmitz,
  F.~Setti, H.~Shakeshaft, D.~Stuart, M.~Swiatlowski, J.~Yoo, H.~Zaraket, and
  H.~Zheng.
\newblock Search for millicharged particles in proton-proton collisions at
  $\sqrt{s}=13\text{ }\text{ }\mathrm{TeV}$.
\newblock {\em Phys. Rev. D}, 102:032002, Aug 2020.

\bibitem{FPF}
Jonathan~L Feng, Felix Kling, Mary~Hall Reno, Juan Rojo, Dennis Soldin, Luis~A
  Anchordoqui, Jamie Boyd, Ahmed Ismail, Lucian Harland-Lang, Kevin~J Kelly,
  Vishvas Pandey, Sebastian Trojanowski, Yu-Dai Tsai, Jean-Marco Alameddine,
  Takeshi Araki, Akitaka Ariga, Tomoko Ariga, Kento Asai, Alessandro Bacchetta,
  Kincso Balazs, Alan~J Barr, Michele Battistin, Jianming Bian, Caterina
  Bertone, Weidong Bai, Pouya Bakhti, A~Baha Balantekin, Basabendu Barman,
  Brian Batell, Martin Bauer, Brian Bauer, Mathias Becker, Asher Berlin, Enrico
  Bertuzzo, Atri Bhattacharya, Marco Bonvini, Stewart~T Boogert, Alexey
  Boyarsky, Joseph Bramante, Vedran Brdar, Adrian Carmona, David~W Casper,
  Francesco~Giovanni Celiberto, Francesco Cerutti, Grigorios Chachamis, Garv
  Chauhan, Matthew Citron, Emanuele Copello, Jean-Pierre Corso, Luc Darm{\'{e}
  }, Raffaele~Tito D'Agnolo, Neda Darvishi, Arindam Das, Giovanni~De Lellis,
  Albert~De Roeck, Jordy de~Vries, Hans~P Dembinski, Sergey Demidov, Patrick
  deNiverville, Peter~B Denton, Frank~F Deppisch, P~S~Bhupal Dev, Antonia~Di
  Crescenzo, Keith~R Dienes, Milind~V Diwan, Herbi~K Dreiner, Yong Du, Bhaskar
  Dutta, Pit Duwentäster, Lucie Elie, Sebastian A~R Ellis, Rikard Enberg,
  Yasaman Farzan, Max Fieg, Ana~Luisa Foguel, Patrick Foldenauer, Saeid
  Foroughi-Abari, Jean-Fran{\c{c}}ois Fortin, Alexander Friedland, Elina Fuchs,
  Michael Fucilla, Kai Gallmeister, Alfonso Garcia, Carlos A~Garc{\'{\i}}a
  Canal, Maria~Vittoria Garzelli, Rhorry Gauld, Sumit Ghosh, Anish Ghoshal,
  Stephen Gibson, Francesco Giuli, Victor~P Gon{\c{c}}alves, Dmitry Gorbunov,
  Srubabati Goswami, Silvia Grau, Julian~Y Günther, Marco Guzzi, Andrew Haas,
  Timo Hakulinen, Steven~P Harris, Julia Harz, Juan Carlos~Helo Herrera,
  Christopher~S Hill, Martin Hirsch, Timothy~J Hobbs, Stefan Höche, Andrzej
  Hryczuk, Fei Huang, Tomohiro Inada, Angelo Infantino, Ameen Ismail, Richard
  Jacobsson, Sudip Jana, Yu~Seon Jeong, Tomas Je{\v{z}}o, Yongsoo Jho,
  Krzysztof Jod{\l}owski, Dmitry Kalashnikov, Timo~J Kärkkäinen, Cynthia
  Keppel, Jongkuk Kim, Michael Klasen, Spencer~R Klein, Pyungwon Ko, Dominik
  Köhler, Masahiro Komatsu, Karol Kova{\v{r}}{\'{\i}}k, Suchita Kulkarni,
  Jason Kumar, Karan Kumar, Jui-Lin Kuo, Frank Krauss, Aleksander Kusina, Maxim
  Laletin, Chiara~Le Roux, Seung~J Lee, Hye-Sung Lee, Helena Lefebvre, Jinmian
  Li, Shuailong Li, Yichen Li, Wei Liu, Zhen Liu, Mickael Lonjon, Kun-Feng Lyu,
  Rafal Maciula, Roshan~Mammen Abraham, Mohammad~R Masouminia, Josh McFayden,
  Oleksii Mikulenko, Mohammed M~A Mohammed, Kirtimaan~A Mohan, Jorge~G
  Morf{\'{\i}}n, Ulrich Mosel, Martin Mosny, Khoirul~F Muzakka, Pavel Nadolsky,
  Toshiyuki Nakano, Saurabh Nangia, Angel~Navascues Cornago, Laurence~J Nevay,
  Pierre Ninin, Emanuele~R Nocera, Takaaki Nomura, Rui Nunes, Nobuchika Okada,
  Fred Olness, John Osborne, Hidetoshi Otono, Maksym Ovchynnikov, Alessandro
  Papa, Junle Pei, Guillermo Peon, Gilad Perez, Luke Pickering, Simon Plätzer,
  Ryan Plestid, Tanmay~Kumar Poddar, Pablo Qu{\'{\i}}lez, Mudit Rai, Meshkat
  Rajaee, Digesh Raut, Peter Reimitz, Filippo Resnati, Wolfgang Rhode, Peter
  Richardson, Adam Ritz, Hiroki Rokujo, Leszek Roszkowski, Tim Ruhe, Richard
  Ruiz, Marta Sabate-Gilarte, Alexander Sandrock, Ina Sarcevic, Subir Sarkar,
  Osamu Sato, Christiane Scherb, Ingo Schienbein, Holger Schulz, Pedro
  Schwaller, Sergio~J Sciutto, Dipan Sengupta, Lesya Shchutska, Takashi
  Shimomura, Federico Silvetti, Kuver Sinha, Torbjörn Sjöstrand, Jan~T
  Sobczyk, Huayang Song, Jorge~F Soriano, Yotam Soreq, Anna Stasto, David
  Stuart, Shufang Su, Wei Su, Antoni Szczurek, Zahra Tabrizi, Yosuke Takubo,
  Marco Taoso, Brooks Thomas, Pierre Thonet, Douglas Tuckler, Agustin~Sabio
  Vera, Heinz Vincke, K~N Vishnudath, Zeren~Simon Wang, Martin~W Winkler,
  Wenjie Wu, Keping Xie, Xun-Jie Xu, Tevong You, Ji-Young Yu, Jiang-Hao Yu,
  Korinna Zapp, Yongchao Zhang, Yue Zhang, Guanghui Zhou, and Renata~Zukanovich
  Funchal.
\newblock The forward physics facility at the high-luminosity {LHC}.
\newblock {\em Journal of Physics G: Nuclear and Particle Physics},
  50(3):030501, jan 2023.

\bibitem{CDMS-II:2009ktb}
Z.~Ahmed et~al.
\newblock {Dark Matter Search Results from the CDMS II Experiment}.
\newblock {\em Science}, 327:1619--1621, 2010.

\bibitem{PICO:2019rsv}
C.~Amole et~al.
\newblock {Data-Driven Modeling of Electron Recoil Nucleation in PICO
  C$_3$F$_8$ Bubble Chambers}.
\newblock {\em Phys. Rev. D}, 100(8):082006, 2019.

\bibitem{Baxter:2017ozv}
D.~Baxter et~al.
\newblock {First Demonstration of a Scintillating Xenon Bubble Chamber for
  Detecting Dark Matter and Coherent Elastic Neutrino-Nucleus Scattering}.
\newblock {\em Phys. Rev. Lett.}, 118(23):231301, 2017.

\bibitem{BresslerPhD}
Matthew~John Bressler.
\newblock {\em {Operation and Calibration of Right-Side-Up Bubble Chambers at
  $\sim$keV Thresholds: Towards New Superheated Dark Matter Searches}}.
\newblock PhD thesis, Drexel University, 2022.

\bibitem{Alfonso-Pita:2022akn}
E.~Alfonso-Pita et~al.
\newblock {Snowmass 2021 Scintillating Bubble Chambers: Liquid-noble Bubble
  Chambers for Dark Matter and CE$\nu$NS Detection}.
\newblock In {\em {Snowmass 2021}}, 7 2022.

\bibitem{alfonso-pitaNewPhysicsSearches2022}
E.~{Alfonso-Pita}, L.~J. Flores, Eduardo Peinado, and
  E.~{V{\'a}zquez-J{\'a}uregui}.
\newblock New physics searches in a low threshold scintillating argon bubble
  chamber measuring coherent elastic neutrino-nucleus scattering in reactors.
\newblock {\em Physical Review D}, 105(11):113005, June 2022.

\bibitem{cennstheorygroupatif-unamPhysicsReachLow2021}
{CE{$\nu$}NS Theory Group at IF-UNAM}, L.~J. Flores, Eduardo Peinado, {SBC
  Collaboration}, E.~{Alfonso-Pita}, K.~Allen, M.~Baker, E.~Behnke,
  M.~Bressler, K.~Clark, R.~Coppejans, C.~Cripe, M.~Crisler, C.~E. Dahl, A.~{de
  St. Croix}, D.~Durnford, P.~Giampa, O.~Harris, P.~Hatch, H.~{Hawley-Herrera},
  C.~M. Jackson, Y.~Ko, C.~B. Krauss, N.~Lamb, M.~Laurin, I.~Levine, W.~H.
  Lippincott, R.~Neilson, S.~Pal, M.-C. Piro, Z.~Sheng,
  E.~{V{\'a}zquez-J{\'a}uregui}, T.~J. Whitis, S.~Windle, R.~Zhang, and
  A.~{Zu{\~n}iga-Reyes}.
\newblock Physics reach of a low threshold scintillating argon bubble chamber
  in coherent elastic neutrino-nucleus scattering reactor experiments.
\newblock {\em Physical Review D}, 103(9):L091301, May 2021.

\bibitem{darksidecollaborationDarkSide50532dayDark2018a}
{DarkSide Collaboration}, P.~Agnes, I.~F.~M. Albuquerque, T.~Alexander, A.~K.
  Alton, G.~R. Araujo, M.~Ave, H.~O. Back, B.~Baldin, G.~Batignani, K.~Biery,
  V.~Bocci, G.~Bonfini, W.~Bonivento, B.~Bottino, F.~Budano, S.~Bussino,
  M.~Cadeddu, M.~Cadoni, F.~Calaprice, A.~Caminata, N.~Canci, A.~Candela,
  M.~Caravati, M.~Cariello, M.~Carlini, M.~Carpinelli, S.~Catalanotti,
  V.~Cataudella, P.~Cavalcante, S.~Cavuoti, A.~Chepurnov, C.~Cical{\`o}, A.~G.
  Cocco, G.~Covone, D.~D'Angelo, M.~D'Incecco, D.~D'Urso, S.~Davini,
  A.~De~Candia, S.~De~Cecco, M.~De~Deo, G.~De~Filippis, G.~De~Rosa,
  M.~De~Vincenzi, A.~V. Derbin, A.~Devoto, F.~Di~Eusanio, G.~Di~Pietro,
  C.~Dionisi, M.~Downing, E.~Edkins, A.~Empl, A.~Fan, G.~Fiorillo, R.~S.
  Fitzpatrick, K.~Fomenko, D.~Franco, F.~Gabriele, C.~Galbiati, C.~Ghiano,
  S.~Giagu, C.~Giganti, G.~K. Giovanetti, O.~Gorchakov, A.~M. Goretti,
  F.~Granato, M.~Gromov, M.~Guan, Y.~Guardincerri, M.~Gulino, B.~R. Hackett,
  K.~Herner, B.~Hosseini, D.~Hughes, P.~Humble, E.~V. Hungerford, {\relax
  An}.~Ianni, V.~Ippolito, I.~James, T.~N. Johnson, K.~Keeter, C.~L. Kendziora,
  I.~Kochanek, G.~Koh, D.~Korablev, G.~Korga, A.~Kubankin, M.~Kuss,
  M.~La~Commara, M.~Lai, X.~Li, M.~Lissia, G.~Longo, Y.~Ma, A.~A. Machado,
  I.~N. Machulin, A.~Mandarano, L.~Mapelli, S.~M. Mari, J.~Maricic, C.~J.
  Martoff, A.~Messina, P.~D. Meyers, R.~Milincic, A.~Monte, M.~Morrocchi, B.~J.
  Mount, V.~N. Muratova, P.~Musico, A.~Navrer~Agasson, A.~O. Nozdrina,
  A.~Oleinik, M.~Orsini, F.~Ortica, L.~Pagani, M.~Pallavicini, L.~Pandola,
  E.~Pantic, E.~Paoloni, K.~Pelczar, N.~Pelliccia, A.~Pocar, S.~Pordes, S.~S.
  Poudel, D.~A. Pugachev, H.~Qian, F.~Ragusa, M.~Razeti, A.~Razeto,
  B.~Reinhold, A.~L. Renshaw, M.~Rescigno, Q.~Riffard, A.~Romani, B.~Rossi,
  N.~Rossi, D.~Sablone, O.~Samoylov, W.~Sands, S.~Sanfilippo, C.~Savarese,
  B.~Schlitzer, E.~Segreto, D.~A. Semenov, A.~Shchagin, A.~Sheshukov, P.~N.
  Singh, M.~D. Skorokhvatov, O.~Smirnov, A.~Sotnikov, C.~Stanford, S.~Stracka,
  Y.~Suvorov, R.~Tartaglia, G.~Testera, A.~Tonazzo, P.~Trinchese, E.~V.
  Unzhakov, M.~Verducci, A.~Vishneva, B.~Vogelaar, M.~Wada, T.~J. Waldrop,
  H.~Wang, Y.~Wang, A.~W. Watson, S.~Westerdale, M.~M. Wojcik, X.~Xiang,
  X.~Xiao, C.~Yang, Z.~Ye, C.~Zhu, and G.~Zuzel.
\newblock {{DarkSide-50}} 532-day dark matter search with low-radioactivity
  argon.
\newblock {\em Physical Review D}, 98(10):102006, November 2018.

\bibitem{darkside-50collaborationSearchLowmassDark2023}
{DarkSide-50 Collaboration}, P.~Agnes, I.~F.~M. Albuquerque, T.~Alexander,
  A.~K. Alton, M.~Ave, H.~O. Back, G.~Batignani, K.~Biery, V.~Bocci, W.~M.
  Bonivento, B.~Bottino, S.~Bussino, M.~Cadeddu, M.~Cadoni, F.~Calaprice,
  A.~Caminata, N.~Canci, M.~Caravati, N.~Cargioli, M.~Cariello, M.~Carlini,
  V.~Cataudella, P.~Cavalcante, S.~Cavuoti, S.~Chashin, A.~Chepurnov,
  C.~Cical{\`o}, G.~Covone, D.~D'Angelo, S.~Davini, A.~De~Candia, S.~De~Cecco,
  G.~De~Filippis, G.~De~Rosa, A.~V. Derbin, A.~Devoto, M.~D'Incecco,
  C.~Dionisi, F.~Dordei, M.~Downing, D.~D'Urso, G.~Fiorillo, D.~Franco,
  F.~Gabriele, C.~Galbiati, C.~Ghiano, C.~Giganti, G.~K. Giovanetti, A.~M.
  Goretti, G.~{Grilli di Cortona}, A.~Grobov, M.~Gromov, M.~Guan, M.~Gulino,
  B.~R. Hackett, K.~Herner, T.~Hessel, B.~Hosseini, F.~Hubaut, E.~V.
  Hungerford, {\relax An}.~Ianni, V.~Ippolito, K.~Keeter, C.~L. Kendziora,
  M.~Kimura, I.~Kochanek, D.~Korablev, G.~Korga, A.~Kubankin, M.~Kuss,
  M.~La~Commara, M.~Lai, X.~Li, M.~Lissia, G.~Longo, O.~Lychagina, I.~N.
  Machulin, L.~P. Mapelli, S.~M. Mari, J.~Maricic, A.~Messina, R.~Milincic,
  J.~Monroe, M.~Morrocchi, X.~Mougeot, V.~N. Muratova, P.~Musico, A.~O.
  Nozdrina, A.~Oleinik, F.~Ortica, L.~Pagani, M.~Pallavicini, L.~Pandola,
  E.~Pantic, E.~Paoloni, K.~Pelczar, N.~Pelliccia, S.~Piacentini, A.~Pocar,
  D.~M. Poehlmann, S.~Pordes, S.~S. Poudel, P.~Pralavorio, D.~D. Price,
  F.~Ragusa, M.~Razeti, A.~Razeto, A.~L. Renshaw, M.~Rescigno, J.~Rode,
  A.~Romani, D.~Sablone, O.~Samoylov, W.~Sands, S.~Sanfilippo, E.~Sandford,
  C.~Savarese, B.~Schlitzer, D.~A. Semenov, A.~Shchagin, A.~Sheshukov, M.~D.
  Skorokhvatov, O.~Smirnov, A.~Sotnikov, S.~Stracka, Y.~Suvorov, R.~Tartaglia,
  G.~Testera, A.~Tonazzo, E.~V. Unzhakov, A.~Vishneva, R.~B. Vogelaar, M.~Wada,
  H.~Wang, Y.~Wang, S.~Westerdale, M.~M. Wojcik, X.~Xiao, C.~Yang, and
  G.~Zuzel.
\newblock Search for low-mass dark matter {{WIMPs}} with 12 ton-day exposure of
  {{DarkSide-50}}.
\newblock {\em Physical Review D}, 107(6):063001, March 2023.

\bibitem{darksidecollaborationSearchDarkMatterNucleon2023}
{DarkSide Collaboration}, P.~Agnes, I.~F.~M. Albuquerque, T.~Alexander, A.~K.
  Alton, M.~Ave, H.~O. Back, G.~Batignani, K.~Biery, V.~Bocci, W.~M. Bonivento,
  B.~Bottino, S.~Bussino, M.~Cadeddu, M.~Cadoni, F.~Calaprice, A.~Caminata,
  M.~D. Campos, N.~Canci, M.~Caravati, N.~Cargioli, M.~Cariello, M.~Carlini,
  V.~Cataudella, P.~Cavalcante, S.~Cavuoti, S.~Chashin, A.~Chepurnov,
  C.~Cical{\`o}, G.~Covone, D.~D'Angelo, S.~Davini, A.~De~Candia, S.~De~Cecco,
  G.~De~Filippis, G.~De~Rosa, A.~V. Derbin, A.~Devoto, M.~D'Incecco,
  C.~Dionisi, F.~Dordei, M.~Downing, D.~D'Urso, M.~Fairbairn, G.~Fiorillo,
  D.~Franco, F.~Gabriele, C.~Galbiati, C.~Ghiano, C.~Giganti, G.~K. Giovanetti,
  A.~M. Goretti, G.~{Grilli di Cortona}, A.~Grobov, M.~Gromov, M.~Guan,
  M.~Gulino, B.~R. Hackett, K.~Herner, T.~Hessel, B.~Hosseini, F.~Hubaut, E.~V.
  Hungerford, {\relax An}.~Ianni, V.~Ippolito, K.~Keeter, C.~L. Kendziora,
  M.~Kimura, I.~Kochanek, D.~Korablev, G.~Korga, A.~Kubankin, M.~Kuss,
  M.~La~Commara, M.~Lai, X.~Li, M.~Lissia, G.~Longo, O.~Lychagina, I.~N.
  Machulin, L.~P. Mapelli, S.~M. Mari, J.~Maricic, A.~Messina, R.~Milincic,
  J.~Monroe, M.~Morrocchi, X.~Mougeot, V.~N. Muratova, P.~Musico, A.~O.
  Nozdrina, A.~Oleinik, F.~Ortica, L.~Pagani, M.~Pallavicini, L.~Pandola,
  E.~Pantic, E.~Paoloni, K.~Pelczar, N.~Pelliccia, S.~Piacentini, A.~Pocar,
  D.~M. Poehlmann, S.~Pordes, S.~S. Poudel, P.~Pralavorio, D.~D. Price,
  F.~Ragusa, M.~Razeti, A.~Razeto, A.~L. Renshaw, M.~Rescigno, J.~Rode,
  A.~Romani, D.~Sablone, O.~Samoylov, E.~Sandford, W.~Sands, S.~Sanfilippo,
  C.~Savarese, B.~Schlitzer, D.~A. Semenov, A.~Shchagin, A.~Sheshukov, M.~D.
  Skorokhvatov, O.~Smirnov, A.~Sotnikov, S.~Stracka, Y.~Suvorov, R.~Tartaglia,
  G.~Testera, A.~Tonazzo, E.~V. Unzhakov, A.~Vishneva, R.~B. Vogelaar, M.~Wada,
  H.~Wang, Y.~Wang, S.~Westerdale, M.~M. Wojcik, X.~Xiao, C.~Yang, and
  G.~Zuzel.
\newblock Search for {{Dark-Matter--Nucleon Interactions}} via {{Migdal
  Effect}} with {{DarkSide-50}}.
\newblock {\em Physical Review Letters}, 130(10):101001, March 2023.

\bibitem{darksidecollaborationSearchDarkMatter2023}
{DarkSide Collaboration}, P.~Agnes, I.~F.~M. Albuquerque, T.~Alexander, A.~K.
  Alton, M.~Ave, H.~O. Back, G.~Batignani, K.~Biery, V.~Bocci, W.~M. Bonivento,
  B.~Bottino, S.~Bussino, M.~Cadeddu, M.~Cadoni, F.~Calaprice, A.~Caminata,
  M.~D. Campos, N.~Canci, M.~Caravati, N.~Cargioli, M.~Cariello, M.~Carlini,
  V.~Cataudella, P.~Cavalcante, S.~Cavuoti, S.~Chashin, A.~Chepurnov,
  C.~Cical{\`o}, G.~Covone, D.~D'Angelo, S.~Davini, A.~De~Candia, S.~De~Cecco,
  G.~De~Filippis, G.~De~Rosa, A.~V. Derbin, A.~Devoto, M.~D'Incecco,
  C.~Dionisi, F.~Dordei, M.~Downing, D.~D'Urso, G.~Fiorillo, D.~Franco,
  F.~Gabriele, C.~Galbiati, C.~Ghiano, C.~Giganti, G.~K. Giovanetti, A.~M.
  Goretti, G.~{Grilli di Cortona}, A.~Grobov, M.~Gromov, M.~Guan, M.~Gulino,
  B.~R. Hackett, K.~Herner, T.~Hessel, B.~Hosseini, F.~Hubaut, E.~V.
  Hungerford, {\relax An}.~Ianni, V.~Ippolito, K.~Keeter, C.~L. Kendziora,
  M.~Kimura, I.~Kochanek, D.~Korablev, G.~Korga, A.~Kubankin, M.~Kuss,
  M.~La~Commara, M.~Lai, X.~Li, M.~Lissia, G.~Longo, O.~Lychagina, I.~N.
  Machulin, L.~P. Mapelli, S.~M. Mari, J.~Maricic, A.~Messina, R.~Milincic,
  J.~Monroe, M.~Morrocchi, X.~Mougeot, V.~N. Muratova, P.~Musico, A.~O.
  Nozdrina, A.~Oleinik, F.~Ortica, L.~Pagani, M.~Pallavicini, L.~Pandola,
  E.~Pantic, E.~Paoloni, K.~Pelczar, N.~Pelliccia, S.~Piacentini, A.~Pocar,
  D.~M. Poehlmann, S.~Pordes, S.~S. Poudel, P.~Pralavorio, D.~D. Price,
  F.~Ragusa, M.~Razeti, A.~Razeto, A.~L. Renshaw, M.~Rescigno, J.~Rode,
  A.~Romani, D.~Sablone, O.~Samoylov, W.~Sands, S.~Sanfilippo, E.~Sandford,
  C.~Savarese, B.~Schlitzer, D.~A. Semenov, A.~Shchagin, A.~Sheshukov, M.~D.
  Skorokhvatov, O.~Smirnov, A.~Sotnikov, S.~Stracka, Y.~Suvorov, R.~Tartaglia,
  G.~Testera, A.~Tonazzo, E.~V. Unzhakov, A.~Vishneva, R.~B. Vogelaar, M.~Wada,
  H.~Wang, Y.~Wang, S.~Westerdale, M.~M. Wojcik, X.~Xiao, C.~Yang, and
  G.~Zuzel.
\newblock Search for {{Dark Matter Particle Interactions}} with {{Electron
  Final States}} with {{DarkSide-50}}.
\newblock {\em Physical Review Letters}, 130(10):101002, March 2023.

\bibitem{darksidecollaborationCalibrationLiquidArgon2021}
{DarkSide Collaboration}, P.~Agnes, I.~F.~M. Albuquerque, T.~Alexander, A.~K.
  Alton, M.~Ave, H.~O. Back, G.~Batignani, K.~Biery, V.~Bocci, W.~M. Bonivento,
  B.~Bottino, S.~Bussino, M.~Cadeddu, M.~Cadoni, F.~Calaprice, A.~Caminata,
  N.~Canci, M.~Caravati, M.~Cariello, M.~Carlini, M.~Carpinelli,
  S.~Catalanotti, V.~Cataudella, P.~Cavalcante, S.~Cavuoti, A.~Chepurnov,
  C.~Cical{\`o}, A.~G. Cocco, G.~Covone, D.~D'Angelo, S.~Davini, A.~De~Candia,
  S.~De~Cecco, G.~De~Filippis, G.~De~Rosa, A.~V. Derbin, A.~Devoto,
  M.~D'Incecco, C.~Dionisi, F.~Dordei, M.~Downing, D.~D'Urso, G.~Fiorillo,
  D.~Franco, F.~Gabriele, C.~Galbiati, C.~Ghiano, C.~Giganti, G.~K. Giovanetti,
  O.~Gorchakov, A.~M. Goretti, A.~Grobov, M.~Gromov, M.~Guan, Y.~Guardincerri,
  M.~Gulino, B.~R. Hackett, K.~Herner, B.~Hosseini, F.~Hubaut, E.~V.
  Hungerford, {\relax An}.~Ianni, V.~Ippolito, K.~Keeter, C.~L. Kendziora,
  I.~Kochanek, D.~Korablev, G.~Korga, A.~Kubankin, M.~Kuss, M.~La~Commara,
  M.~Lai, X.~Li, M.~Lissia, G.~Longo, I.~N. Machulin, L.~P. Mapelli, S.~M.
  Mari, J.~Maricic, C.~J. Martoff, A.~Messina, P.~D. Meyers, R.~Milincic,
  M.~Morrocchi, X.~Mougeot, V.~N. Muratova, P.~Musico, A.~Navrer~Agasson, A.~O.
  Nozdrina, A.~Oleinik, F.~Ortica, L.~Pagani, M.~Pallavicini, L.~Pandola,
  E.~Pantic, E.~Paoloni, K.~Pelczar, N.~Pelliccia, E.~Picciau, A.~Pocar,
  S.~Pordes, S.~S. Poudel, P.~Pralavorio, F.~Ragusa, M.~Razeti, A.~Razeto,
  A.~L. Renshaw, M.~Rescigno, J.~Rode, A.~Romani, D.~Sablone, O.~Samoylov,
  W.~Sands, S.~Sanfilippo, C.~Savarese, B.~Schlitzer, D.~A. Semenov,
  A.~Shchagin, A.~Sheshukov, M.~D. Skorokhvatov, O.~Smirnov, A.~Sotnikov,
  S.~Stracka, Y.~Suvorov, R.~Tartaglia, G.~Testera, A.~Tonazzo, E.~V. Unzhakov,
  A.~Vishneva, R.~B. Vogelaar, M.~Wada, H.~Wang, Y.~Wang, S.~Westerdale, M.~M.
  Wojcik, X.~Xiao, C.~Yang, and G.~Zuzel.
\newblock Calibration of the liquid argon ionization response to low energy
  electronic and nuclear recoils with {{DarkSide-50}}.
\newblock {\em Physical Review D}, 104(8):082005, October 2021.

\bibitem{GlobalArgonDarkMatterCollaboration:2022czd}
P.~Agnes et~al.
\newblock {Sensitivity projections for a dual-phase argon TPC optimized for
  light dark matter searches through the ionization channel}.
\newblock {\em Phys. Rev. D}, 107(11):112006, 2023.

\bibitem{Aartsen:2020iky}
M.G. Aartsen et~al.
\newblock Searching for ev-scale sterile neutrinos with eight years of
  atmospheric neutrinos at the icecube neutrino telescope.
\newblock {\em Phys. Rev. D}, 102:052009, Sep 2020.

\bibitem{Diaz:2019fwt}
A.~Diaz, C.A. Argüelles, G.H. Collin, J.M. Conrad, and M.H. Shaevitz.
\newblock {Where Are We With Light Sterile Neutrinos?}
\newblock 5 2019.

\bibitem{PhysRevD.107.L031901}
Wooyoung Jang, Doojin Kim, Kyoungchul Kong, Youngjoon Kwon, Jong-Chul Park,
  Min~Sang Ryu, Seodong Shin, Richard~G. Van~de Water, Un-Ki Yang, and Jaehoon
  Yu.
\newblock Search prospects for axionlike particles at rare nuclear isotope
  accelerator facilities.
\newblock {\em Phys. Rev. D}, 107:L031901, Feb 2023.

\bibitem{CCM:2021yzc}
A.~A. Aguilar-Arevalo et~al.
\newblock {First Leptophobic Dark Matter Search from the
  Coherent\textendash{}CAPTAIN-Mills Liquid Argon Detector}.
\newblock {\em Phys. Rev. Lett.}, 129(2):021801, 2022.

\end{thebibliography}
